\documentclass[a4paper,11pt]{article}
\pdfoutput=1 

\usepackage{jheppub} 

\usepackage[T1]{fontenc} 

\usepackage{multirow}
 
\usepackage{rotating}
\usepackage{longtable}
\usepackage{dcolumn}  

\include{patch_amsmath}

\usepackage{booktabs}

\renewcommand{\arraystretch}{1.1}

\newcommand{\cm}{\mathrm{cm}}

\newcommand{\mev}{\mathrm{MeV}}
\newcommand{\mevc}{\mathrm{MeV}/c}
\newcommand{\mevm}{\mathrm{MeV}/c^2}
\newcommand{\gev}{\mathrm{GeV}}
\newcommand{\gevc}{\mathrm{GeV}/c}
\newcommand{\gevm}{\mathrm{GeV}/c^2}

\newcommand{\ee}{e^+e^-}
\newcommand{\uu}{\mu^+\mu^-}
\newcommand{\pp}{\pi^+\pi^-}
\newcommand{\kk}{K^+K^-}

\newcommand{\U}{\Upsilon}
\newcommand{\Ufo}{\Upsilon(4S)}
\newcommand{\Uf}{\Upsilon(5S)}
\newcommand{\Us}{\Upsilon(6S)}
\newcommand{\Uo}{\Upsilon(1S)}

\newcommand{\Un}{\Upsilon(nS)}

\newcommand{\hbn}{h_b(nP)}
\newcommand{\jp}{J/\psi}
\newcommand{\kp}{K^+}

\newcommand{\ks}{K^0_S}

\newcommand{\pip}{\pi^{+}}

\newcommand{\pim}{\pi^{-}}
\newcommand{\mbc}{M_{\rm bc}}
\newcommand{\mb}{m_B}

\newcommand{\dmbst}{\Delta m_{B^*}}

\newcommand{\DE}{\Delta E}
\newcommand{\fb}{\mathrm{fb}^{-1}}

\newcommand{\ecm}{E_\mathrm{cm}}
\newcommand{\eb}{E_B}
\newcommand{\pb}{p_B}
\newcommand{\ez}{E_{\rm cm0}}

\newcommand{\bb}{B\bar{B}}
\newcommand{\bbst}{B\bar{B}^*}
\newcommand{\bstbst}{B^*\bar{B}^*}
\newcommand{\bball}{B^{(*)}\bar{B}^{(*)}}
\newcommand{\opdi}{(1+\delta_\mathrm{ISR})}
\newcommand{\PB}{{\cal P}_B}
\newcommand{\rsb}{r_\mathrm{s.b.}}
\newcommand{\sh}{s}
\newcommand{\ff}{\phi}
\newcommand{\nr}{n}
\newcommand{\spread}{\sigma_{\ecm}}
\newcommand{\fsp}{f_{\sigma}}
\newcommand{\ipo}{I_{e^+}}
\newcommand{\deb}{\Delta E_\mathrm{BaBar}}
\newcommand{\D}{\Delta}
\newcommand{\ah}{a_\mathrm{h}}
\newcommand{\rafi}{R_\mathrm{5\,chan}^\mathrm{all}}
\newcommand{\rafilo}{R_\mathrm{5\,chan}}
\newcommand{\fra}{f_{B^+}/f_{B^0}}

\title{\boldmath Measurement of the energy dependence of the
  $e^+e^-\to{B}\bar{B}$, ${B}\bar{B}^*$ and ${B}^*\bar{B}^*$ exclusive
  cross sections}

\newcounter{AffiliationCounter}
\stepcounter{AffiliationCounter}\edef\instBilbao{\protect\theAffiliationCounter}
\stepcounter{AffiliationCounter}\edef\instBonn{\protect\theAffiliationCounter}
\stepcounter{AffiliationCounter}\edef\instBNL{\protect\theAffiliationCounter}
\stepcounter{AffiliationCounter}\edef\instBINP{\protect\theAffiliationCounter}
\stepcounter{AffiliationCounter}\edef\instCharles{\protect\theAffiliationCounter}
\stepcounter{AffiliationCounter}\edef\instCincinnati{\protect\theAffiliationCounter}
\stepcounter{AffiliationCounter}\edef\instDESY{\protect\theAffiliationCounter}
\stepcounter{AffiliationCounter}\edef\instFuJen{\protect\theAffiliationCounter}
\stepcounter{AffiliationCounter}\edef\instFudan{\protect\theAffiliationCounter}
\stepcounter{AffiliationCounter}\edef\instSokendai{\protect\theAffiliationCounter}
\stepcounter{AffiliationCounter}\edef\instGyeongsang{\protect\theAffiliationCounter}
\stepcounter{AffiliationCounter}\edef\instHanyang{\protect\theAffiliationCounter}
\stepcounter{AffiliationCounter}\edef\instHawaii{\protect\theAffiliationCounter}
\stepcounter{AffiliationCounter}\edef\instKEK{\protect\theAffiliationCounter}
\stepcounter{AffiliationCounter}\edef\instJPARC{\protect\theAffiliationCounter}
\stepcounter{AffiliationCounter}\edef\instHSE{\protect\theAffiliationCounter}
\stepcounter{AffiliationCounter}\edef\instJuelich{\protect\theAffiliationCounter}
\stepcounter{AffiliationCounter}\edef\instIKER{\protect\theAffiliationCounter}
\stepcounter{AffiliationCounter}\edef\instIISERM{\protect\theAffiliationCounter}
\stepcounter{AffiliationCounter}\edef\instIITB{\protect\theAffiliationCounter}
\stepcounter{AffiliationCounter}\edef\instIITH{\protect\theAffiliationCounter}
\stepcounter{AffiliationCounter}\edef\instIITM{\protect\theAffiliationCounter}
\stepcounter{AffiliationCounter}\edef\instIndiana{\protect\theAffiliationCounter}
\stepcounter{AffiliationCounter}\edef\instIHEP{\protect\theAffiliationCounter}
\stepcounter{AffiliationCounter}\edef\instProtvino{\protect\theAffiliationCounter}
\stepcounter{AffiliationCounter}\edef\instVienna{\protect\theAffiliationCounter}
\stepcounter{AffiliationCounter}\edef\instNapoli{\protect\theAffiliationCounter}
\stepcounter{AffiliationCounter}\edef\instTorino{\protect\theAffiliationCounter}
\stepcounter{AffiliationCounter}\edef\instJAEA{\protect\theAffiliationCounter}
\stepcounter{AffiliationCounter}\edef\instJSI{\protect\theAffiliationCounter}
\stepcounter{AffiliationCounter}\edef\instKarlsruhe{\protect\theAffiliationCounter}
\stepcounter{AffiliationCounter}\edef\instKAU{\protect\theAffiliationCounter}
\stepcounter{AffiliationCounter}\edef\instKitasato{\protect\theAffiliationCounter}
\stepcounter{AffiliationCounter}\edef\instKISTI{\protect\theAffiliationCounter}
\stepcounter{AffiliationCounter}\edef\instKorea{\protect\theAffiliationCounter}
\stepcounter{AffiliationCounter}\edef\instKyotoSangyo{\protect\theAffiliationCounter}
\stepcounter{AffiliationCounter}\edef\instKyungpook{\protect\theAffiliationCounter}
\stepcounter{AffiliationCounter}\edef\instLAL{\protect\theAffiliationCounter}
\stepcounter{AffiliationCounter}\edef\instLebedev{\protect\theAffiliationCounter}
\stepcounter{AffiliationCounter}\edef\instLjubljana{\protect\theAffiliationCounter}
\stepcounter{AffiliationCounter}\edef\instLMU{\protect\theAffiliationCounter}
\stepcounter{AffiliationCounter}\edef\instLuther{\protect\theAffiliationCounter}
\stepcounter{AffiliationCounter}\edef\instMNIT{\protect\theAffiliationCounter}
\stepcounter{AffiliationCounter}\edef\instMaribor{\protect\theAffiliationCounter}
\stepcounter{AffiliationCounter}\edef\instMPI{\protect\theAffiliationCounter}
\stepcounter{AffiliationCounter}\edef\instMississippi{\protect\theAffiliationCounter}
\stepcounter{AffiliationCounter}\edef\instMiyazaki{\protect\theAffiliationCounter}
\stepcounter{AffiliationCounter}\edef\instMEPhI{\protect\theAffiliationCounter}
\stepcounter{AffiliationCounter}\edef\instNagoya{\protect\theAffiliationCounter}
\stepcounter{AffiliationCounter}\edef\instNagoyaKMI{\protect\theAffiliationCounter}
\stepcounter{AffiliationCounter}\edef\instUNapoli{\protect\theAffiliationCounter}
\stepcounter{AffiliationCounter}\edef\instNara{\protect\theAffiliationCounter}
\stepcounter{AffiliationCounter}\edef\instNCU{\protect\theAffiliationCounter}
\stepcounter{AffiliationCounter}\edef\instNUU{\protect\theAffiliationCounter}
\stepcounter{AffiliationCounter}\edef\instTaiwan{\protect\theAffiliationCounter}
\stepcounter{AffiliationCounter}\edef\instKrakow{\protect\theAffiliationCounter}
\stepcounter{AffiliationCounter}\edef\instNihonDental{\protect\theAffiliationCounter}
\stepcounter{AffiliationCounter}\edef\instNiigata{\protect\theAffiliationCounter}
\stepcounter{AffiliationCounter}\edef\instNovaGorica{\protect\theAffiliationCounter}
\stepcounter{AffiliationCounter}\edef\instNovosibirsk{\protect\theAffiliationCounter}
\stepcounter{AffiliationCounter}\edef\instOsakaCity{\protect\theAffiliationCounter}
\stepcounter{AffiliationCounter}\edef\instPNNL{\protect\theAffiliationCounter}
\stepcounter{AffiliationCounter}\edef\instPanjab{\protect\theAffiliationCounter}
\stepcounter{AffiliationCounter}\edef\instPeking{\protect\theAffiliationCounter}
\stepcounter{AffiliationCounter}\edef\instPittsburgh{\protect\theAffiliationCounter}
\stepcounter{AffiliationCounter}\edef\instNPC{\protect\theAffiliationCounter}
\stepcounter{AffiliationCounter}\edef\instRIKENMSL{\protect\theAffiliationCounter}
\stepcounter{AffiliationCounter}\edef\instUSTC{\protect\theAffiliationCounter}
\stepcounter{AffiliationCounter}\edef\instSeoul{\protect\theAffiliationCounter}
\stepcounter{AffiliationCounter}\edef\instShoyaku{\protect\theAffiliationCounter}
\stepcounter{AffiliationCounter}\edef\instSoochow{\protect\theAffiliationCounter}
\stepcounter{AffiliationCounter}\edef\instSoongsil{\protect\theAffiliationCounter}
\stepcounter{AffiliationCounter}\edef\instSungkyunkwan{\protect\theAffiliationCounter}
\stepcounter{AffiliationCounter}\edef\instSydney{\protect\theAffiliationCounter}
\stepcounter{AffiliationCounter}\edef\instTabuk{\protect\theAffiliationCounter}
\stepcounter{AffiliationCounter}\edef\instTata{\protect\theAffiliationCounter}
\stepcounter{AffiliationCounter}\edef\instTUM{\protect\theAffiliationCounter}
\stepcounter{AffiliationCounter}\edef\instTelAviv{\protect\theAffiliationCounter}
\stepcounter{AffiliationCounter}\edef\instToho{\protect\theAffiliationCounter}
\stepcounter{AffiliationCounter}\edef\instTohoku{\protect\theAffiliationCounter}
\stepcounter{AffiliationCounter}\edef\instERI{\protect\theAffiliationCounter}
\stepcounter{AffiliationCounter}\edef\instTokyo{\protect\theAffiliationCounter}
\stepcounter{AffiliationCounter}\edef\instTIT{\protect\theAffiliationCounter}
\stepcounter{AffiliationCounter}\edef\instTMU{\protect\theAffiliationCounter}
\stepcounter{AffiliationCounter}\edef\instVPI{\protect\theAffiliationCounter}
\stepcounter{AffiliationCounter}\edef\instWayneState{\protect\theAffiliationCounter}
\stepcounter{AffiliationCounter}\edef\instYamagata{\protect\theAffiliationCounter}
\stepcounter{AffiliationCounter}\edef\instYonsei{\protect\theAffiliationCounter}

\collaboration{The Belle Collaboration}
  \author[\instLebedev,\instHSE]{R.~Mizuk,} 
  \author[\instBINP,\instNovosibirsk]{A.~Bondar,} 
  \author[\instKEK,\instSokendai]{I.~Adachi,} 
  \author[\instTokyo]{H.~Aihara,} 
  \author[\instTabuk,\instKAU]{S.~Al~Said,} 
  \author[\instBNL]{D.~M.~Asner,} 
  \author[\instCincinnati]{H.~Atmacan,} 
  \author[\instBINP,\instNovosibirsk]{V.~Aulchenko,} 
  \author[\instHSE]{T.~Aushev,} 
  \author[\instTabuk]{R.~Ayad,} 
% ...

\abstract{We report the first measurement of the exclusive cross
  sections $e^+e^-\to{B}\bar{B}$, $e^+e^-\to{B}\bar{B}^*$, and
  $e^+e^-\to{B}^*\bar{B}^*$ in the energy range from $10.63\,\gev$ to
  $11.02\,\gev$. The $B$ mesons are fully reconstructed in a large
  number of hadronic final states and the three channels are
  identified using a beam-constrained-mass variable. The shapes of
  the exclusive cross sections show oscillatory behavior with several
  maxima and minima. The results are obtained using data collected by
  the Belle experiment at the KEKB asymmetric-energy $e^+e^-$
  collider.}


\begin{document}
\maketitle
\flushbottom

\section{Introduction}

\begin{figure}[htbp]
\centering
\includegraphics[width=0.52\linewidth]{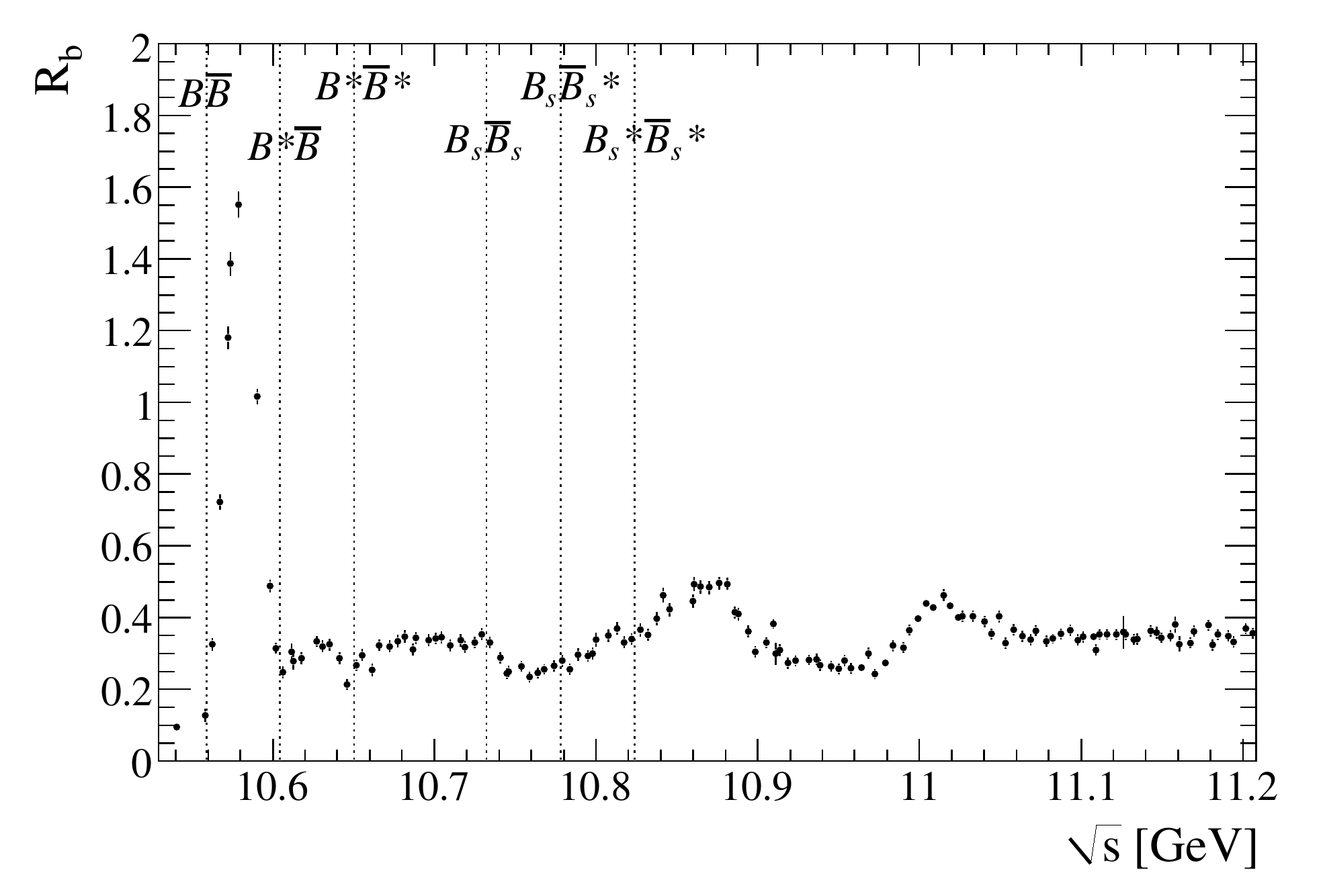}\hfill
\includegraphics[width=0.45\linewidth]{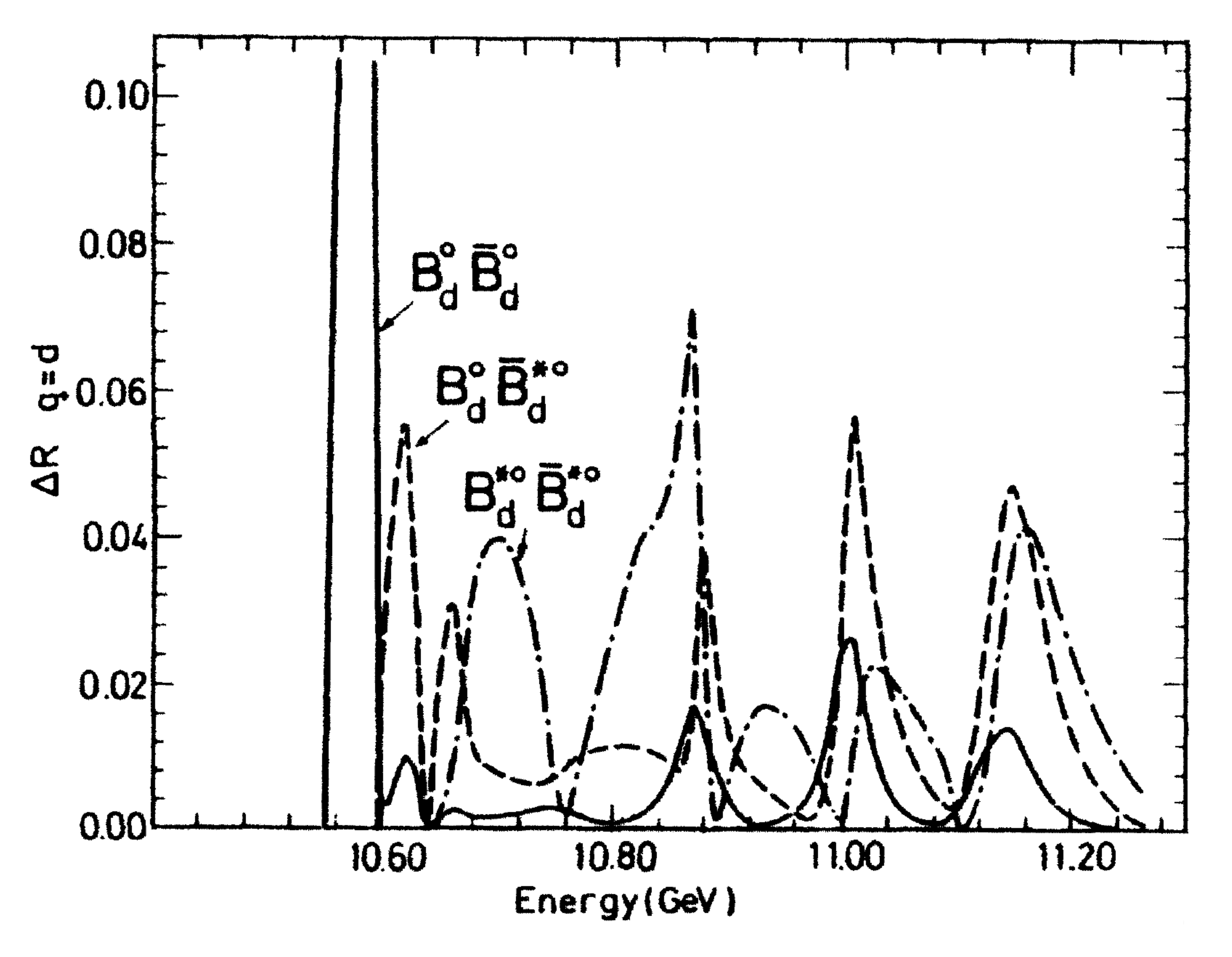}
\caption{ The $R_b$ scan results from BaBar~\cite{Aubert:2008ab}
  (left), and the expected in Unitarized Quark Model contributions of
  the $B\bar{B}$, $B\bar{B}^*$, and $B^*\bar{B}^*$
  channels~\cite{Ono:1985eu} (right). }
\label{xsecres}
\end{figure}

The total $\ee\to b\bar{b}$ cross section at various energies above
the $\bb$ threshold is shown in
Fig.~\ref{xsecres}~(left)~\cite{Aubert:2008ab}. It exhibits peaks of
$\Ufo$, $\U(10860)$, and $\U(11020)$, possibly a dip in the region of
$\U(10775)$, and also dips at the $\bbst$ and $\bstbst$ thresholds.
The exclusive two-body cross sections $\ee\to\bb$, $\ee\to\bbst$, and
$\ee\to\bstbst$, that saturate the total cross section below the
$\U(10860)$ peak and give a dominant contribution also at higher
energy, are expected to show much more pronounced behaviour, as shown
in Fig.~\ref{xsecres}~(right)~\cite{Ono:1985eu}.
The expected oscillatory behavior of the exclusive cross sections
might be due to the nodes of the $\Ufo$, $\U(10860)$, and $\U(11020)$
wave functions~\cite{Ono:1985eu}. These cross sections provide
important information about the interactions in this energy region
and, in particular, about the structure of the $\Ufo$, $\U(10860)$, and
$\U(11020)$ states. This topic is of special interest since the above
states show anomalies, primarily in the pattern of transitions to
lower bottomonium states, that are currently not well understood (for
a review see, e.g., Ref.~\cite{Bondar:2016hva}).

Here we report the first measurement of the energy dependence of the
$\ee\to\bb$, $\ee\to\bbst$, and $\ee\to\bstbst$ exclusive cross
sections.\footnote{$\bbst$ denotes the sum of $B\bar{B}^*$ and
  $B^*\bar{B}$.}
Our approach is to perform a full reconstruction of one $B$ meson in
hadronic channels, and then to identify the $\bb$, $\bbst$, and
$\bstbst$ signals using the $\mbc$ distribution,
$\mbc=\sqrt{(\ecm/2)^2-\pb^2}$, where $\ecm$ is the center-of-mass
(c.m.) energy and $\pb$ is the $B$-candidate momentum measured in the
c.m.\ frame. The $\mbc$ distribution for $\bb$ events peaks at the
nominal $B$-meson mass, $\mb$, while the distributions for $\bbst$ and
$\bstbst$ events peak approximately at $\mb-\frac{\dmbst}{2}$ and
$\mb-\dmbst$, respectively, where $\dmbst$ is the mass difference of
the $B^*$ and $B$ mesons~\cite{PDG}. If the $B$ meson originates from
a $B^*\to{B}\gamma$ decay, there is an additional broadening of the
signal due to the photon recoil momentum.

To reconstruct $B$ mesons in a large number of hadronic final states
we apply the Full Event Interpretation (FEI) package of the Belle~II
software that was developed primarily for tagging $B$ mesons in the
$\Ufo\to\bb$ decays~\cite{Keck:2018lcd}. This package uses
multivariate analysis for the event selection and provides high
flexibility in choosing the $B$ decay channels and the input variables
for the classifier.

Before going into details, we describe how the paper is organized and
give an overview of the analysis. In Section~\ref{sec:data_sets} we
briefly describe the Belle detector, the data samples and the
simulation. Selection of events is described in
Section~\ref{sec:selection}. For the FEI classifier, we choose input
variables that are not correlated with the $B$ candidate momentum,
which helps to avoid distortion of the background in the $\mbc$
distribution and to keep efficiency approximately independent of
$\ecm$. We do not include the energy of the $B$ candidate, $\eb$, into
the FEI training and use sidebands in the $\mbc$ {\it versus} $\DE$
plane to study background, where $\DE=\eb-\ecm/2$. We find that there
is a peaking background in the $\mbc$ distribution which is primarily
due to misreconstructed soft photons. We first apply FEI to the $\Ufo$
data sample (Section~\ref{sec:y4s}). We construct the $\mbc$ fit
function in which the signal shape is calculated based on the $\ecm$
spread, the cross section energy dependence, and the momentum
resolution. We also calibrate simulation of the peaking background and
determine the total $B$ meson yield which is later used to determine
the efficiency. We proceed with the analysis of the $\U(10860)$ data
sample (Section~\ref{sec:y5s}) with the aim to verify the fit
procedure and measure the signal yield. We also study the distribution
of $B$ mesons in the polar angle (Appendix~\ref{app:angular}). The
fits to the data samples at various energies are presented in
Section~\ref{sec:scan}. We measure the efficiency at the $\Ufo$ and
$\U(10860)$ energies (Section~\ref{sec:eff}). To determine the total
numbers of $B$ mesons in the $\U(10860)$ data sample, we use five
clean $B^+$ and $B^0$ decay channels reconstructed without
FEI. Determination of the cross sections, parameterization of the
cross section energy dependence, and estimation of the systematic
uncertainties are presented in Section~\ref{sec:xsec}. The results are
discussed in Section~\ref{sec:discussion}. As a byproduct, we measure
$f_s$, the fraction of the $B_s^{(*)}\bar{B}_s^{(*)}$ events at
$\U(10860)$ (Section~\ref{sec:fs}). We conclude in
Section~\ref{sec:concl}.

For brevity, in the following we denote $\U(10860)$ as $\Uf$ and
$\U(11020)$ as $\Us$.

\section{Belle detector and data sets}
\label{sec:data_sets}

The analysis is based on data collected by the Belle detector at the
KEKB asymmetric-energy $\ee$ collider~\cite{KEKB}.
The Belle detector is a large-solid-angle magnetic spectrometer that
consists of a silicon vertex detector (SVD), a 50-layer central drift
chamber (CDC), an array of aerogel threshold Cherenkov counters (ACC),
a barrel-like arrangement of time-of-flight scintillation counters
(TOF), and an electromagnetic calorimeter (ECL) comprised of CsI(Tl)
crystals located inside a superconducting solenoid that provides a
1.5~T magnetic field. An iron flux return located outside the coil is
instrumented to detect $K_L^0$ mesons and to identify muons (KLM).
Two different inner detector configurations were used. For the first
sample of $156\,\fb$, a $2.0\,$cm-radius beam pipe, and a 3-layer
silicon vertex detector were used; for the latter sample of
$833\,\fb$, a $1.5\,$cm-radius beam pipe, a 4-layer silicon vertex
detector (SVD2), and a small-cell inner drift chamber were used.
This analysis is based only on data collected with the SVD2
configuration.
Detailed description of the detector can be found in
Ref.~\cite{BELLE_DETECTOR,Brodzicka:2012jm}.

We use energy scan data with approximately $1\,\fb$ per point, six
points collected in 2007 and 16 points collected in 2010. We use also
the $\Uf$ on-resonance data with a total luminosity of $121\,\fb$
collected at five points with energies from $10.864\,\gev$ to
$10.868\,\gev$. The $\ecm$ calibration of these data is reported in
Ref.~\cite{Abdesselam:2019gth}. We combine the data samples with
similar energies so that finally we obtain 23 energy points. The
energies and integrated luminosities of these 23 data samples are
presented in Table~\ref{tab:xsec_results} below.
We use also the SVD2 part of the $\Ufo$ data sample with the primary
goal to calibrate the reconstruction efficiency; its integrated
luminosity is $571\,\fb$.

The signal $\ee\to B^{(*)}\bar{B}^{(*)}$ events and the background
$\ee\to B_s^{(*)}\bar{B}_s^{(*)}$, $\ee\to{q}\bar{q}$ ($q=u,d,s,c$)
events are generated using EvtGen~\cite{Lange:2001uf}.
The detector response is simulated using GEANT~\cite{GEANT}. The
Monte-Carlo (MC) simulation includes run-dependent variations in the
detector performance and background conditions.

\section{Event selection}
\label{sec:selection}

The event selection is performed primarily by FEI. Our strategy is not
to include the $\DE$ variable in the training of the FEI classifier so
that the $\DE$ sidebands (or, more precisely, the $\DE'$ sidebands
with the $\DE'$ variable defined as $\DE'\equiv\DE+\mbc-\mb$) are
available. We use the sidebands to study the smooth background
component. We perform the FEI training and then apply further
channel-dependent selection criteria on the FEI output variable and
the $\DE'$ variable, as described below.

We reconstruct $B^+$ and $B^0$ in the decay channels
$\bar{D}^{(*)}\pi^+(\pi^+\pi^-)$, $D_s^{(*)+}\bar{D}^{(*)}$,
$\jp\kp(\pim)$, $\jp\ks(\pip)$, $\jp\ks\pp$, $D^{(*)-}\pi^+\pi^+$, and
$D^{*-}K^+K^-\pi^+$, where $\bar{D}$ denotes $\bar{D}^0$ and $D^-$.
We do not use $B$-decay channels with $\pi^0$ as their energy
resolution is rather poor.
The $D^0$, $D^+$, and $D_s^+$ mesons are reconstructed in the final
states with $K^\pm$, $\ks$, $\pi^\pm$, up to one $\pi^0$, and
multiplicity up to five. The complete list of the $B$- and $D$-meson
decay channels is given in Appendix~\ref{app:decay_chan}. We
reconstruct $D^*$ in all possible decay modes: $D\pi$ and
$D\gamma$. $\jp$ are reconstructed in both $\uu$ and $\ee$
channels. To improve momentum resolution, we apply a mass-constrained
fit to $\pi^0$, $\jp$, and $D^*$; mass-vertex-constrained fit to $D$
and $D_s^+$; and vertex fit to $\ks$ and $B$.

In FEI, the Fast Boosted Decision Tree algorithm~\cite{Keck:2017gsv}
is used to discriminate signal and background events. First, the
final-state particles are classified, and then all the candidates for
unstable particles are obtained as combinations. Thus, FEI training is
performed in stages, and the results of the previous stage are used in
the training of the current one. The training is performed using MC
simulation at the $\Uf$ energy. The classifier output is the
probability that a given candidate is the signal.

As training variables of charged pions, kaons, and leptons, we use
particle identification information, momentum, and transverse
momentum. Prior to training we select the candidates that originate
from the interaction-point (IP) region: we require $dr<0.5\,\cm$ and
$dz<3\,\cm$, where $dr$ and $dz$ are cylindrical coordinates of
the point of the closest approach of the track to the beam. 
For kaons we apply an additional requirement of
$L(K)/(L(K)+L(\pi))>0.1$, where $L(K)$ and $L(\pi)$ are likelihoods of
the kaon and pion hypotheses formed from the measurements in the ACC
and TOF systems, as well as energy loss measurement in CDC. The
efficiency of this requirement is 98\% and the probability to
misidentify a pion as a kaon is about 20\%.

Photon candidates are clusters in ECL with the energies above 30 MeV
and without matching tracks. As training variables we use the number
of crystals in the cluster, the ratio of energy deposit in a
$3\times3$ matrix of crystals to that in a $5\times5$ matrix, cluster
energy, and polar angle.

For the $\pi^0\to\gamma\gamma$ candidates we use mass, momentum, and
decay angle, which is defined as the angle between the $\gamma$
momentum measured in the $\pi^0$ rest frame and the $\pi^0$ boost
direction from the laboratory frame. For the $\ks\to\pp$ candidates we
use mass and a set of parameters describing the displaced vertex of
$\ks$. These are the distance of closest approach between the two
daughter pions, the impact parameters of the daughter pions, the
distance between the IP and the $\ks$ vertex, and the angle between
the $\ks$ momentum and the direction from the IP to the $\ks$ vertex;
the latter three variables are measured in the plane perpendicular to
the beam direction.

The training variables for $\jp$, $D$, and $D^*$ candidates are the
signal probability of each daughter (thus the number of variables
varies with the channel) and the mass. In case of $D$, we use also the
$\chi^2$ of the mass-vertex fit; if the $D$ decay is a three-body
decay, we use invariant masses of pairs of its daughters to take into
account signals of $\rho$, $K^*$, and $\phi$.

For the $B$ meson candidates we use the signal probability of each
daughter and $\chi^2$ of the $B$ vertex fit. If there is a $D$ meson
in the decay, we include the distance between the $B$ and $D$
vertices, the significance of this distance, and the cosine of the
angle between the $D$ momentum and the direction from $B$ to $D$
vertices. In the case when there are several pions or pions and kaons
in the decay, we include invariant masses of the combinations in which
production of $\rho\,(\to\pi\pi)$, $a_1(\to3\pi)$ or $K^*(\to{K}\pi)$
is expected.

At the last stage when the training for the $B$ candidates is
performed, we include also variables that help to suppress continuum
production of light and charm quarks, $\ee\to{q}\bar{q}$. These are
the event-shape variable $R_2$ (the ratio of the second to zeroth
Fox-Wolfram moments~\cite{Fox:1978vu}) and the angle between the
thrust axis of the $B$ candidate and the rest of the event.
We also include two flags indicating the presence of a muon and an
electron, respectively, in the rest of the event.
We consider lepton candidates in the c.m.\ momentum windows
$1.0<p_\mu<2.6\,\gevc$ and $0.8<p_e<2.6\,\gevc$ where the contribution
of leptons from the semileptonic $B$ decays is enhanced. We require
that the leptons are well identified with a likelihood ratio above
0.9~\cite{BELLE_DETECTOR}. The efficiencies of this requirement are
71\% and 76\% for muons and electrons, respectively; the probabilities
to misidentify hadrons as leptons are 1\% and 0.1\%, respectively.

Although the training is performed individually for each decay channel
of every unstable particle in the decay chain, the signal probability
is defined in a universal way so that various channels can be
compared. Thus, the signal probability from the classifier is used to
rank multiple candidates. At the intermediate stage of the
reconstruction, we retain up to 10 best $D^0$, $D^+$, and $D_s^+$
candidates. At the final stage of the $B$ meson reconstruction, we
retain only one best candidate selected from all $B^+$ and $B^0$
candidates in the event.

The entire MC statistics corresponds to six times the statistics of
real data. We use half of the MC statistics to train the classifier
and the other half to determine the efficiency. The efficiency in the
part that was used for training is higher by a factor $1.025\pm0.006$,
which is an indication of a small overtraining.

\begin{figure}[htbp]
  \centering
  \includegraphics[width=8.5cm]{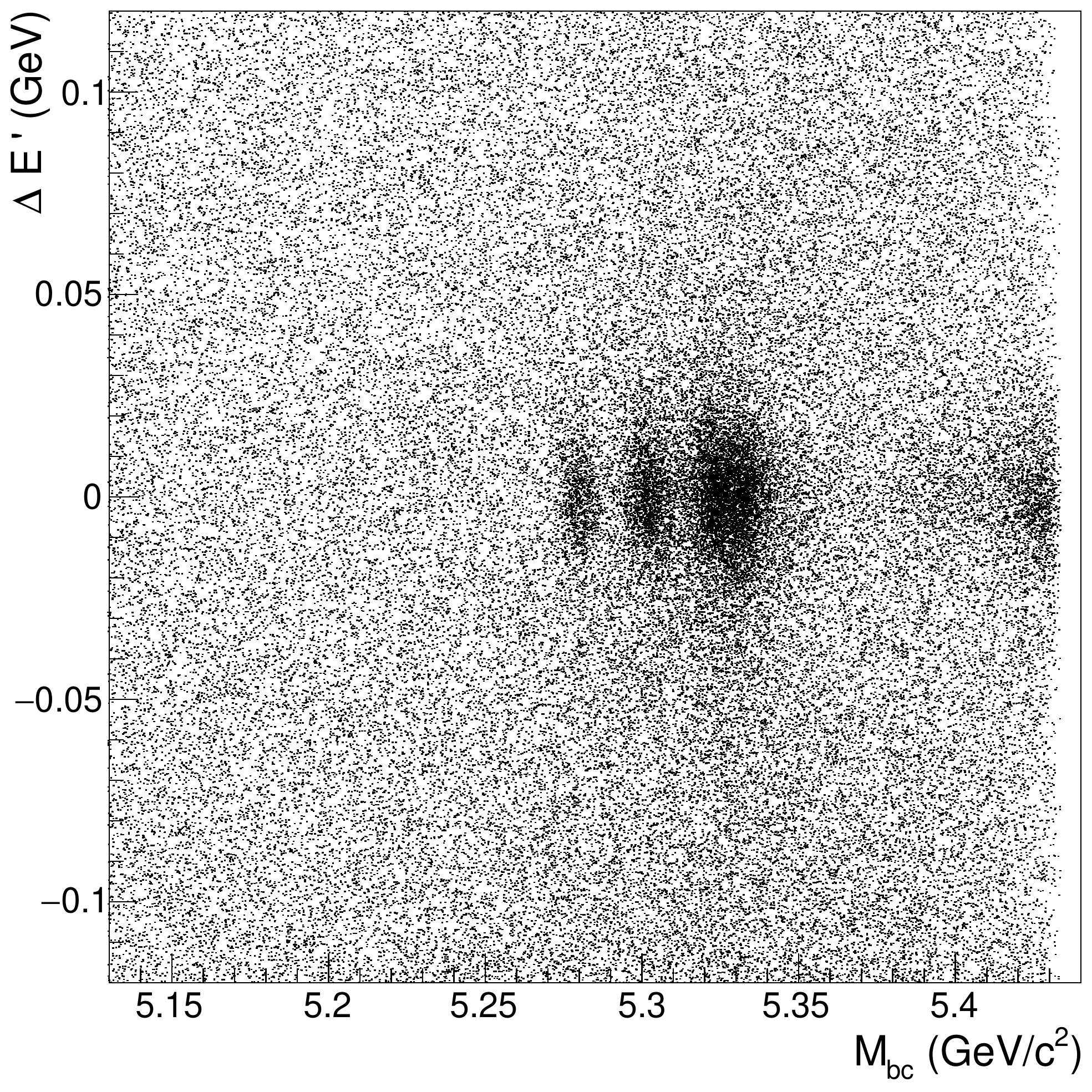}
  \caption{ The $\DE'$ {\it versus} $\mbc$ distribution for the $\Uf$
    data. }
  \label{de_vs_mbc_021020}
\end{figure}

The $\DE'$ {\it versus}\ $\mbc$ distribution for the $\Uf$ on-resonance
data is presented in Fig.~\ref{de_vs_mbc_021020}. Here we apply a
requirement on the $B$ meson signal probability from the FEI
classifier of $\PB>0.1$.
In the $\DE'$ projection, the signal events are concentrated near
zero. 

We optimize the requirements on the $\PB$ and $\DE'$ variables
individually for each $B$ decay channel. We evaluate the contribution
of each channel to an overall figure of merit (FoM) defined as
$N_\mathrm{S}/\sqrt{N_\mathrm{S}+N_\mathrm{B}}$, where $N_\mathrm{S}$
and $N_\mathrm{B}$ are the numbers of signal and background events,
respectively. This optimization is performed based on the $\Uf$
on-resonance data.
The requirements on $\PB$ are in the range 0.01 to 0.1; the $\DE'$
window size varies from $\pm10\,\mev$ to $\pm40\,\mev$ depending on
the channel. 

\begin{figure}[htbp]
  \centering
  \includegraphics[width=0.49\textwidth]{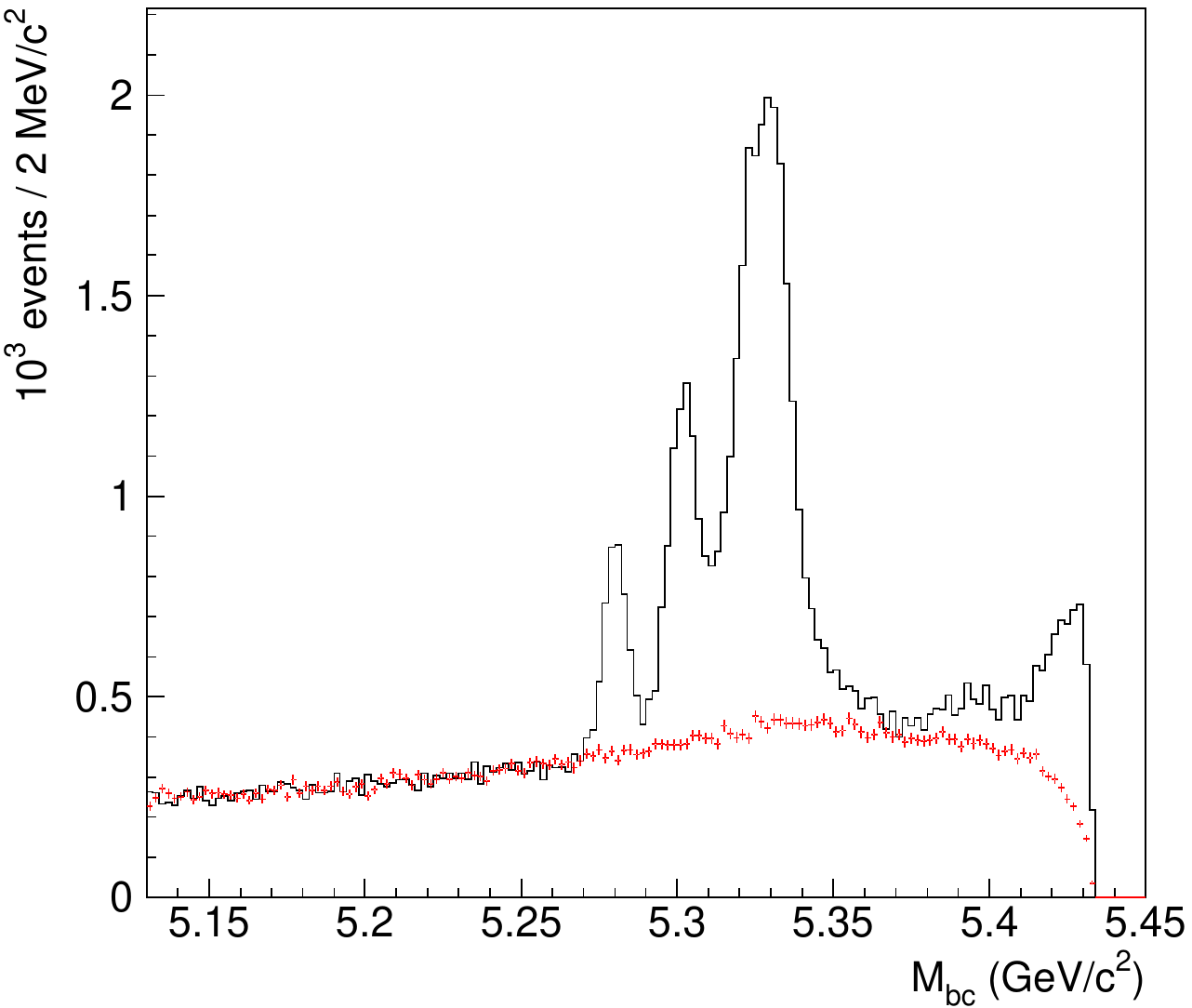}\hfill
  \includegraphics[width=0.49\textwidth]{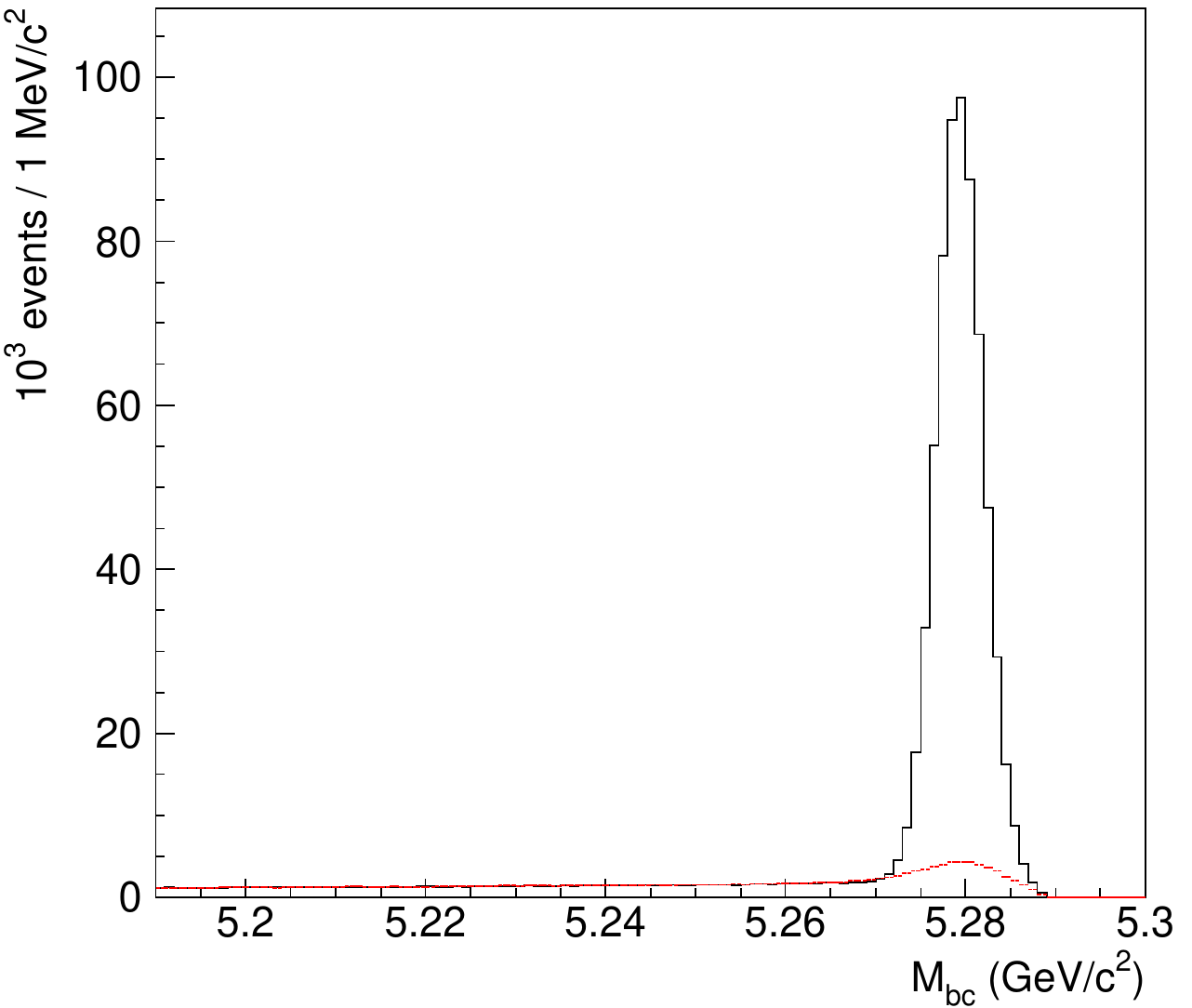}
  \caption{ The $\mbc$ distributions in the $\Uf$ (left) and $\Ufo$
    (right) data. Black solid histogram shows the $\DE'$ signal
    region, while red points with error bars show the normalized
    $\DE'$ sidebands. }
  \label{mbc_data_y5s_sig_sb_stat_220620}
\end{figure}

The $\mbc$ distributions in the $\Uf$ and $\Ufo$ data for the $\DE'$
signal region and sidebands are shown in
Fig.~\ref{mbc_data_y5s_sig_sb_stat_220620}.
The centers of the $\DE'$ sidebands are shifted by $\pm80\,\mev$ from
zero; the sizes of the high and low sidebands are the same as the size
of the signal region.
The three peaks in the $\Uf$ data from left to right are the signals
of $\bb$, $\bbst$, and $\bstbst$, respectively. The peaking structure
near the $\mbc$ threshold is due to the three-body processes
$\ee\to\bbst\pi$, $\bstbst\pi$, and the $\Ufo$ production in the
initial state radiation (ISR) process.
The $\DE'$ sidebands describe the combinatorial background outside the
peaks well.

The MC simulation at $\Uf$ shows that the background is dominated by
the $c\bar{c}$ production. The $B_s$ mesons do not produce a
significant contribution in the region of the $B^{(*)}\bar{B}^{(*)}$
signals.

In the optimization of the channel-dependent selection requirements,
$N_\mathrm{S}$ and $N_\mathrm{B}$ are determined using the $\DE'$
signal and sideband regions with an additional requirement of
$5.27<\mbc<5.34\,\gevm$.
The optimization for each channel is performed iteratively by scanning
FoM in turn in $\PB$ and $|\DE'|$.
Since we use the $\Uf$ on-resonance data for the optimization, the
measured $B$ meson yield could be biased.
We note, however, that the statistics in each $B$ decay channel is high
and we use a relatively large step in the $\PB$ and $|\DE'|$ scanning;
therefore, statistical fluctuations in the FoM dependence on $\PB$ and
$|\DE'|$ are small.
To further study this issue, we divide the $\Uf$ data sample into two
approximately equal parts. We then use part 1 for optimization and
reconstruct part 2 using the resulting selection
requirements. Similarly, we reconstruct part 1 using selection
requirements optimized with part 2. In this way we completely avoid
any bias in the yields due to statistical fluctuations in the value of
FoM during the optimization. We find a 1.3\% smaller ratio of the
yields at $\Uf$ and $\Ufo$ compared to the default procedure. We
consider this value as a symmetric systematic uncertainty due to the
optimization procedure.

\section{\boldmath Analysis of the $\Ufo$ data sample}
\label{sec:y4s}

Here our goal is to describe the $\mbc$ distribution in the $\Ufo$
data in terms of the $\ecm$ spread and the $B\bar{B}$ cross section
shape. This experience is essential for the analysis of the $\Uf$ and
energy scan data.

The information about the $B\bar{B}$ cross section shape in the $\Ufo$
region is rather limited. Current values for the $\Ufo$ mass and
width are dominated by the BaBar measurement in
2004~\cite{Aubert:2004pwa}. There is also a more precise scan
performed by BaBar in 2008~\cite{Aubert:2008ab}, however it was not
fitted in the original paper. We attempt to fit the 2008 scan results
using the $\Ufo$ parameterization from Ref.~\cite{Aubert:2004pwa},
which is based on the Quark Pair Creation
model~\cite{LeYaouanc:1972vsx}. However, the fit function
overestimates the measured values at high-mass side of $\Ufo$.
Since we need the cross section shape to calculate the signal shape in
the $\mbc$ distribution, we adopt the following strategy. We perform a
simultaneous fit to the $\mbc$ distributions and the scan results of
BaBar~\cite{Aubert:2008ab}. As a suitable model is missing, the cross
section is described by a high-order Chebyshev polynomial.

\subsection{$\mbc$ fit function}

The signal component of the $\mbc$ fit is calculated numerically as a
sequence of convolutions. It takes into account the beam-energy
spread, the energy dependence of the production cross section, the
ISR, and the momentum resolution.

The energy spread of the colliding beams is described by a single
Gaussian with the mean $\ez$, which is a nominal c.m.\ energy. The
distribution in $\ecm$ is multiplied by the energy dependence of the
cross section. We convolve the obtained function with the Kuraev-Fadin
radiation kernel~\cite{Kuraev:1985hb} to account for the ISR. We then
change the argument of the function from $\ecm$ to the $B$ meson
momentum $p_0$. At this step we account for the ISR recoil momentum of
the $B$ mesons.

We convolve the distribution in $p_0$ with the momentum resolution
functions. We use three resolution functions to describe the
candidates of three types: (1) MC truth-matched candidates in the
$\DE'$ signal window, (2) not MC truth-matched candidates in the
$\DE'$ signal window, and (3) candidates in the $\DE'$ sidebands.
The candidates of type 1 correspond to the signal, while candidates of
type 2 and 3 correspond to the peaking background.
Each of the three resolution functions, $f_i$, is a sum of two
Gaussians with special factors that account for the fact that $p$
can not be negative:
\begin{align}
  f_i(p-p_0) =  \frac{1-r_2(i)}{\sigma_1(i)}\; & \mathrm{exp}\{-\frac{(p-p_0-\mu_1(i))^2}{2\sigma_1(i)^2}\}\;
  (1-\mathrm{exp}\{-\frac{2p(p_0+\mu_1(i))}{\sigma_1(i)^2}\}) \nonumber \\
  +\; \frac{r_2(i)}{\sigma_2(i)}\; & \mathrm{exp}\{-\frac{(p-p_0-\mu_2(i))^2}{2\sigma_2(i)^2}\}\;
  (1-\mathrm{exp}\{-\frac{2p(p_0+\mu_2(i))}{\sigma_2(i)^2}\}).
  \label{eq:mom_res}
\end{align}
The special factors are found by considering the momentum resolution
function in three dimensions and analytically integrating out all
variables other than $p$. The parameters of the resolution function
are determined from the MC simulation (Table~\ref{tab:y4s_mom_res}).
\begin{table}[htbp]
  \caption{ Parameters of the momentum resolution functions of 
    Eq.~(\ref{eq:mom_res}). }
  \label{tab:y4s_mom_res}
  \centering
  \begin{tabular}{@{}cccc@{}} \toprule
    & $f_1$ & $f_2$ & $f_3$ \\
    \midrule
    $w$ & $-$ & $0.073$ & $0.041$ \\
    $\mu_1$ ($\mevc$) & $0.04$ & $-0.13$ &  $7.93$ \\
    $\sigma_1$ ($\mevc$) & $5.40$ &  $8.77$ &  $60.6$ \\
    $r_2$ & $0.52$ &  $0.43$ & $0.066$ \\
    $\mu_2$ ($\mevc$) & $0.36$ &  $7.27$ & $-3.79$ \\
    $\sigma_2$ ($\mevc$) & $10.6$ &  $48.8$ & $8.64$ \\
    \bottomrule
  \end{tabular}
\end{table}
The functions $f_2$ and $f_3$ are multiplied by additional factors
$w(i)$ which are weight factors for the peaking background in the
$\DE'$ signal and sideband regions.
The momentum distributions are transformed into the $\mbc$
distributions.

The smooth background is described by a threshold function
$\sqrt{\ecm/2-x}$ multiplied by a third-order Chebyshev polynomial. The
shape of the smooth background is the same in the $\DE'$ signal region
and sidebands while the normalizations are allowed to float
independently (the smooth component in the sidebands is multiplied by
a floated parameter $\rsb$).

When fitting the data, we introduce a shift and a width-correction
factor for each component of the momentum resolution function, $\sh_i$
and $\ff_i$. Using the $\DE'$ distributions we find that for the
signal component the shift is negligibly small, while the
width-correction factor is $\ff_1=1.187\pm0.012$.
We float the shift and the width-correction factor of the peaking background in
the sidebands, $\sh_3$ and $\ff_3$, and find that the values are
consistent with zero and one, respectively
(Table~\ref{tab:y4s_fit_results} below). Therefore, the shift and the
width-correction factor of the peaking background in the signal region, $\sh_2$
and $\ff_2$, that are poorly constrained by the $\mbc$ fit, are fixed
at $\sh_2=0$ and $\ff_2=1$.

The peaking background components in the $\DE'$ signal region and
sidebands are multiplied by a common normalization factor $\nr$ which
is floated in the fit. Thus, the floated parameters related to the
signal in the $\mbc$ distribution are the signal yield $N$ (integral
of the signal component; it does not include the integral of the
peaking background), the $\ecm$ spread $\spread$, the peaking
background normalization factor $\nr$, the shift $\sh_3$, and the
width-correction factor $\ff_3$.

\subsection{Cross section fit function}

We describe the energy dependence of the dressed cross
section\footnote{The dressed cross section differs from the Born cross
  section in that the vacuum polarization is accounted for.} by an
11th order Chebyshev polynomial (in case of the 10th order, $\chi^2$
of the fit is higher by about 20, while in case of the 12th order it
is almost unchanged). We require that the cross section is zero at the
$B\bar{B}$ threshold and is never negative. We then apply the ISR
correction and convolve with the BaBar energy spread of
$4.83\,\mev$~\cite{Aubert:2004pwa}. We use 9 BaBar points located
between the $B\bar{B}$ and $B\bar{B}^*$ thresholds. The accuracy of
the c.m.\ energy calibration at BaBar is $1.5\,\mev$. We introduce a
common shift for all nine points, $\deb$, and float it in the fit with
the uncertainty constraint adding a term $\deb^2/(1.5\,\mev)^2$ to
$\chi^2$.

\subsection{Results of the simultaneous fit}

\begin{figure}[htbp]
  \centering
  \includegraphics[width=0.49\textwidth]{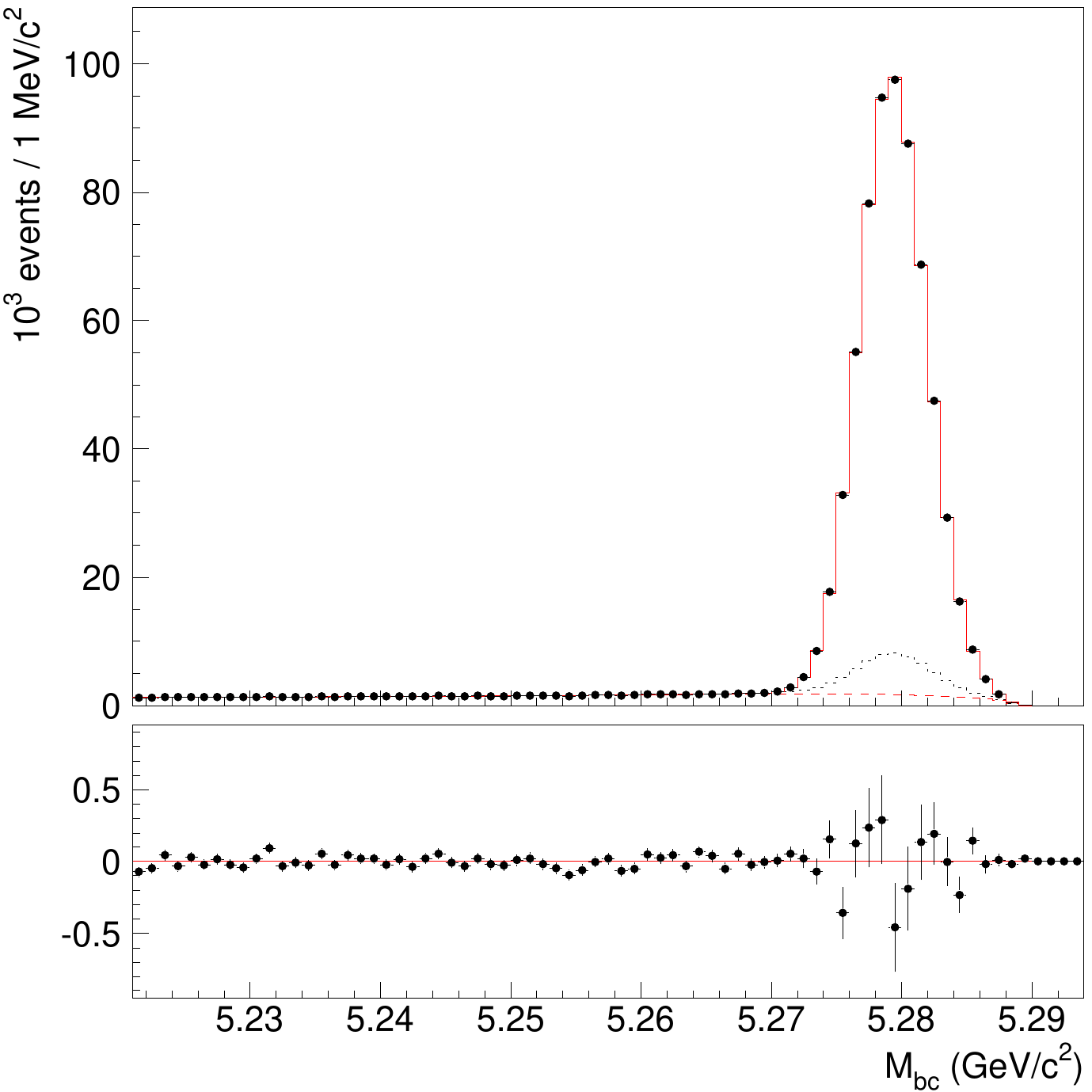}\hfill
  \includegraphics[width=0.49\textwidth]{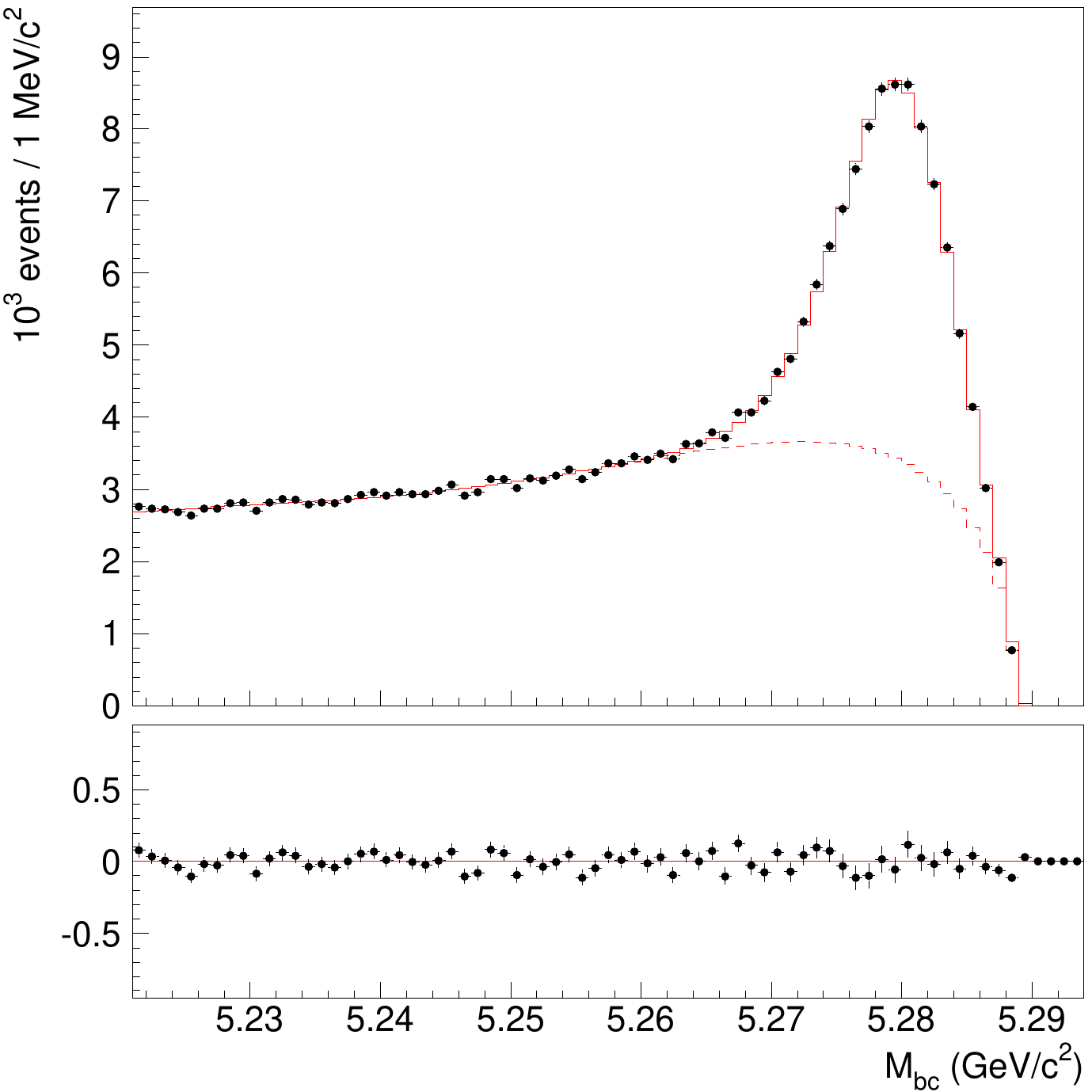}
  \caption{ The $\mbc$ distributions in the $\DE'$ signal region
    (left) and sidebands (right). Points with error bars are data,
    solid red histogram is the result of the simultaneous fit to these
    distributions and the cross section energy dependence
    (Fig.~\ref{xsec_fit_y4s_290820}), dashed red histogram is the
    smooth background, dotted black histogram is the peaking
    background in the $\DE'$ signal region. The lower panels show the
    residuals. }
  \label{mbc_fit_y4s_290820}
\end{figure}

\begin{figure}[htbp]
  \centering
  \includegraphics[width=0.5\textwidth]{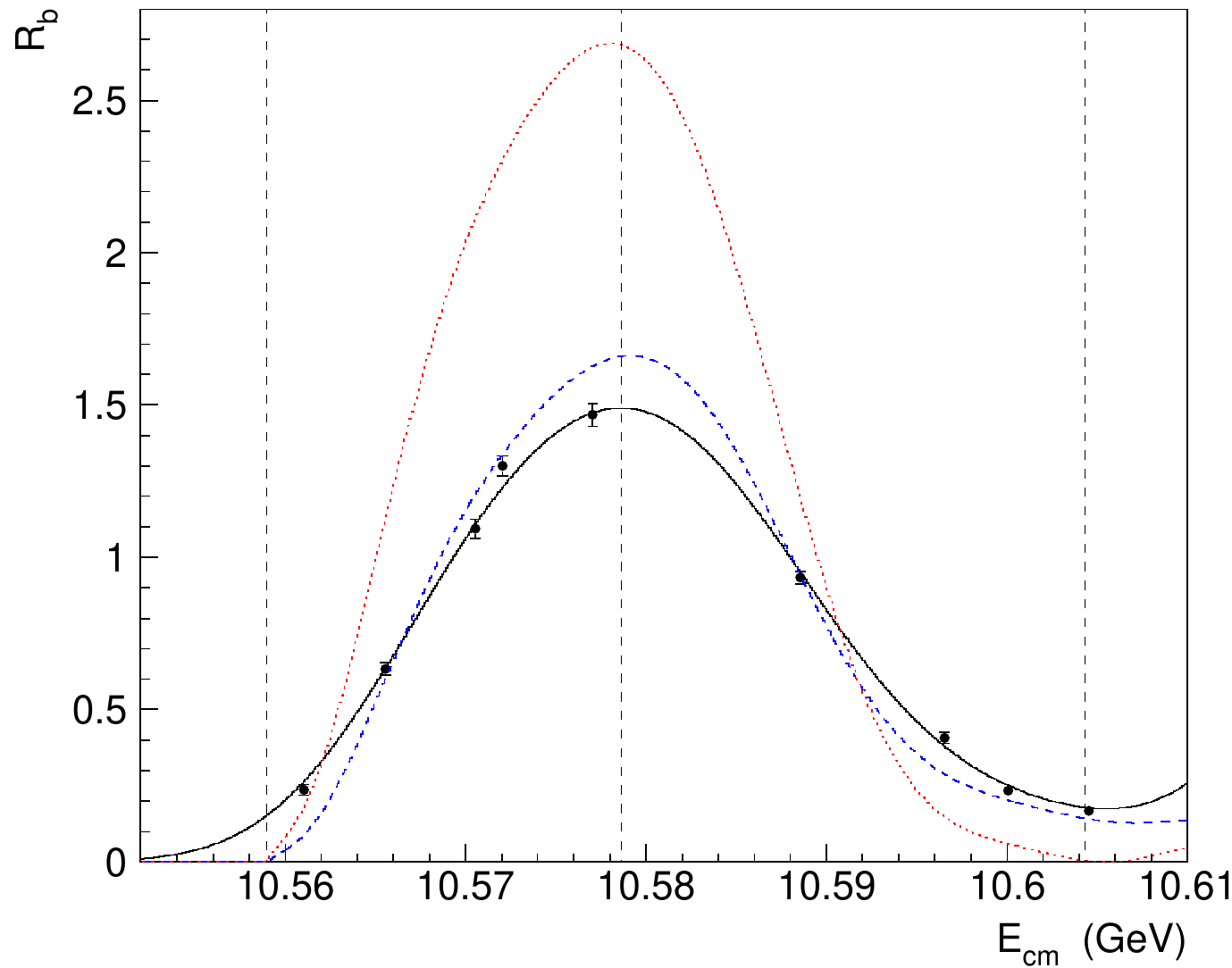}
  \caption{ Energy dependence of the total $b\bar{b}$ cross section
    ($R_\mathrm{b}$). Points with error bars are from
    Ref.~\cite{Aubert:2008ab}. Black solid curve is the result of the
    simultaneous fit to this distribution and the $\mbc$ distributions
    (Fig.~\ref{mbc_fit_y4s_290820}). Blue curve is the visible cross
    section before accounting for the $\ecm$ spread. Red curve is the
    dressed cross section. Vertical dashed lines indicate the
    $B\bar{B}$ threshold, the nominal Belle c.m.\ energy $\ez$, and
    the $B\bar{B}^*$ threshold, in increasing values of $\ecm$. }
  \label{xsec_fit_y4s_290820}
\end{figure}

The results of the simultaneous fit to the $\mbc$ distributions in the
$\DE'$ signal and sideband regions and to the cross section energy
dependence are presented in Figs.~\ref{mbc_fit_y4s_290820},
\ref{xsec_fit_y4s_290820} and Table~\ref{tab:y4s_fit_results}.
\begin{table}[htbp]
\caption{ Results of the simultaneous fit to the Belle $\mbc$
  distribution and the BaBar cross section scan
  results~\cite{Aubert:2008ab}. The first error is statistical, the
  second one (if present) is systematic.}
\renewcommand*{\arraystretch}{1.2}
\label{tab:y4s_fit_results}
\centering
\begin{tabular}{@{}cc@{}} \toprule
  $N$ & $(581.2\pm1.1\pm3.2)\times10^3$ \\
  $\spread$ & $(5.36\pm0.11\pm0.16)\,\mev$ \\
  $\deb$ & $(-1.75\pm0.14\pm0.67)\,\mev$ \\
  \midrule
  $\nr$ & $1.16\pm0.03$ \\
  $\sh_3$ & $(-0.2\pm0.6)\,\mevc$ \\
  $\ff_3$ & $1.00\pm0.02$ \\
  $\rsb$ & $1.017\pm0.005$ \\
\bottomrule
\end{tabular}
\end{table}
We perform the fit at several values of $\ez$, the average
c.m.\ energy of the Belle $\Ufo$ SVD2 data. For each fit, we determine
the difference between $\ez$ and the peak position of the visible
cross section, $\D\ecm$. We find that $\D\ecm$ is equal to zero at
$\ez=10.5787\,\gev$ and use this $\ez$ value in the default fit.
If $\ez$ is floated, we find $\ez=10.5791\pm0.0003\,\gev$, which is
$1.6\,\sigma$ away from the constraint. The p-value of the default fit
is 1.8\%.

To determine the systematic uncertainty, we consider a variation in
$\ez$ of $\pm0.5\,\mev$ that corresponds to a variation in $\D\ecm$ of
$\pm1.5\,\mev$. The decrease of the visible cross section at
$\D\ecm=\pm1.5\,\mev$ is about 1\%.
This variation produces a negligible change in the yield $N$, but is
a dominant uncertainty for $\spread$ and $\deb$.

We increase the order of the Chebyshev polynomial that describes the
cross section shape (11th order to 12th) and which is used in the
smooth background component (3rd order to 4th). In both cases we find
a negligible change in all fit results.

The normalization factor $\nr$ of the peaking background is found to
be 1.16 (Table~\ref{tab:y4s_fit_results}). This value is determined
primarily by the $\DE'$ sidebands. To estimate the systematic
uncertainty related to the peaking background in the $\DE'$ signal
region, we introduce a separate normalization factor for this
component, $\nr_2$, and repeat the fit fixing $\nr_2$ at 1.08 and
1.24. This source is dominant for the yield while for $\spread$ and
$\deb$ the changes are small.
The total systematic uncertainty for the yield $N$, $\spread$, and
$\deb$ is presented in Table~\ref{tab:y4s_fit_results}.
The value of the yield is used to determine the efficiency at the
$\Ufo$ energy (Section~\ref{sec:eff}).

\begin{figure}[htbp]
  \centering
  \includegraphics[width=0.47\textwidth]{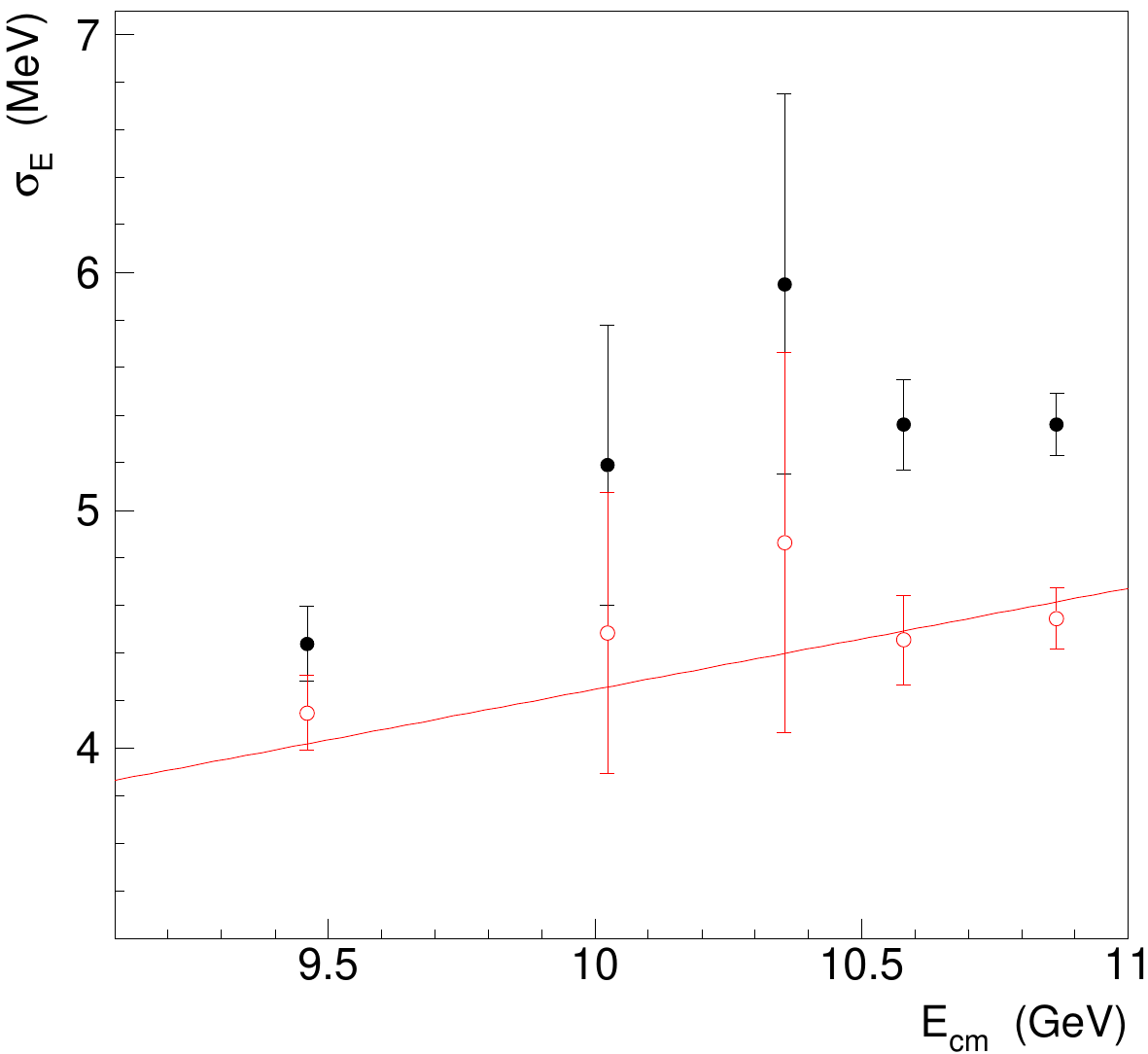}
  \caption{ Energy dependence of the $\ecm$ spread. Black dots with
    error bars are the measurements, red open dots are $\spread$
    corrected for the microwave instability effect, red line is the
    fit result.}
  \label{spread_vs_ecm_220221}
\end{figure}

The value of the $\ecm$ spread at $\Ufo$, $(5.36\pm0.19)\,\mev$, and
the measurements at other energies are shown in
Fig.~\ref{spread_vs_ecm_220221}.
The value at $\Uf$, $(5.36\pm0.13)\,\mev$, is measured in
Ref.~\cite{Abdesselam:2019gth} using the $\ee\to\Un\pp$ ($n=1,2,3$)
processes.
To find the values at $\U(1S,2S,3S)$, we use the visible cross
sections that are determined based on the event yields and
luminosities in Ref.~\cite{Brodzicka:2012jm}.
The procedure is the following. We determine the energy dependence of
the dressed cross section near the resonance using corresponding
parameters: mass, width, and electron width~\cite{PDG}. We then apply
the ISR correction by performing a convolution with the Kuraev-Fadin
radiation kernel~\cite{Kuraev:1985hb} and account for the $\ecm$
energy spread by performing another convolution with a Gaussian.
The integral of the cross section depends on the total and electron
widths, while the shape of the cross section is determined by the
energy spread, since the $\U(1S,2S,3S)$ resonances are very
narrow. Thus, the maximum of the visible cross section is sensitive to
$\spread$. For each $\Un$ ($n=1,2,3$) we fit a single point: the
measured visible cross section at the resonance maximum. The value of
the fit function is the maximum of the calculated cross section. The
parameters of the resonance are floated in the fit within the
uncertainties of their world-average values~\cite{PDG}, thus their
contribution to the uncertainty in the spread is accounted for. We
find that the dominant contribution is the uncertainty in the electron
width. The spread values at $\U(1S)$, $\U(2S)$, and $\U(3S)$ are
$(4.439\pm0.157)\,\mev$, $(5.19\pm0.59)\,\mev$, and
$(5.95\pm0.80)\,\mev$, respectively.

If the bunch current of the positron beam, $\ipo$, is above 0.5\,mA,
there is a microwave instability that increases the $\ecm$
spread~\cite{Cai:2009zz}. The increase factor, $\fsp$, depends
linearly on $\ipo$ reaching $\fsp=1.20$ at $\ipo=1.0$\,mA. The average
over data taking period value of $\fsp$ is determined based on $\ipo$
and integrated luminosity of each run. For the $\Un$ ($n=1,2,3,4,5$)
data samples, we find $\fsp=1.070$, 1.157, 1.223, 1.203, and 1.179,
respectively. The uncertainty in $\fsp$ is negligibly small compared
to the uncertainty in $\spread$. The corrected values
$\spread^\mathrm{cor}=\spread/\fsp$ at various $\ecm$ are shown in
Fig.~\ref{spread_vs_ecm_220221}. The dependence of
$\spread^\mathrm{cor}$ on $\ecm$ is consistent with the
proportionality hypothesis; from the fit we find:
\begin{equation}
  \spread^\mathrm{cor} = (4.247 \pm 0.073) \times 10^{-4} \times \ecm.
  \label{eq:spread_vs_ecm}
\end{equation}
The $\fsp$ values in the scan data samples are in the range
$1.11-1.20$. To determine $\spread$ of each scan data sample we use
its $\fsp$ factor and the $\spread^\mathrm{cor}$ value from
Eq.~(\ref{eq:spread_vs_ecm}). 

The shift in $\ecm$ of BaBar, $(-1.75\pm0.68)\,\mev$, is $1.2\,\sigma$
away from zero in terms of the BaBar accuracy of $1.5\,\mev$. This
shift could be used in future phenomenological analyses of the BaBar
scan results.

As a consistency check we estimate the ISR correction factor to be
$(1+\delta_\mathrm{ISR}) = 0.626\pm0.012$. The uncertainty here is a
systematic one due to variation of $\ez$. This value agrees
with the result of Ref.~\cite{Dong:2020tdw} of $0.622\pm0.018$.

\section{\boldmath Analysis of the $\Uf$ data sample}
\label{sec:y5s}

To fit the $\Uf$ data we include also the $\ee\to\bbst$ and
$\ee\to\bstbst$ signals. The decay $B^*\to{B}\gamma$ leads to
additional smearing of the $B$ momentum that we take into account in
the fit function by performing additional convolution; relativistic
kinematics is used in this calculation.
The distribution in the helicity angle of the $B^*\to{B}\gamma$ decay,
defined as the angle between the $B$ momentum measured in the $B^*$
rest frame and the boost direction of the $B^*$ from the c.m.\ frame, is
expected to be $1+\cos^2\theta_\mathrm{h}$ for $\ee\to\bbst$ and
$1+\ah\cos^2\theta_\mathrm{h}$ with $-1\leq\ah\leq1$ for
$\ee\to\bstbst$. In the fit, we float the parameter $\ah$.

\begin{figure}[htbp]
  \centering
  \includegraphics[width=0.57\textwidth]{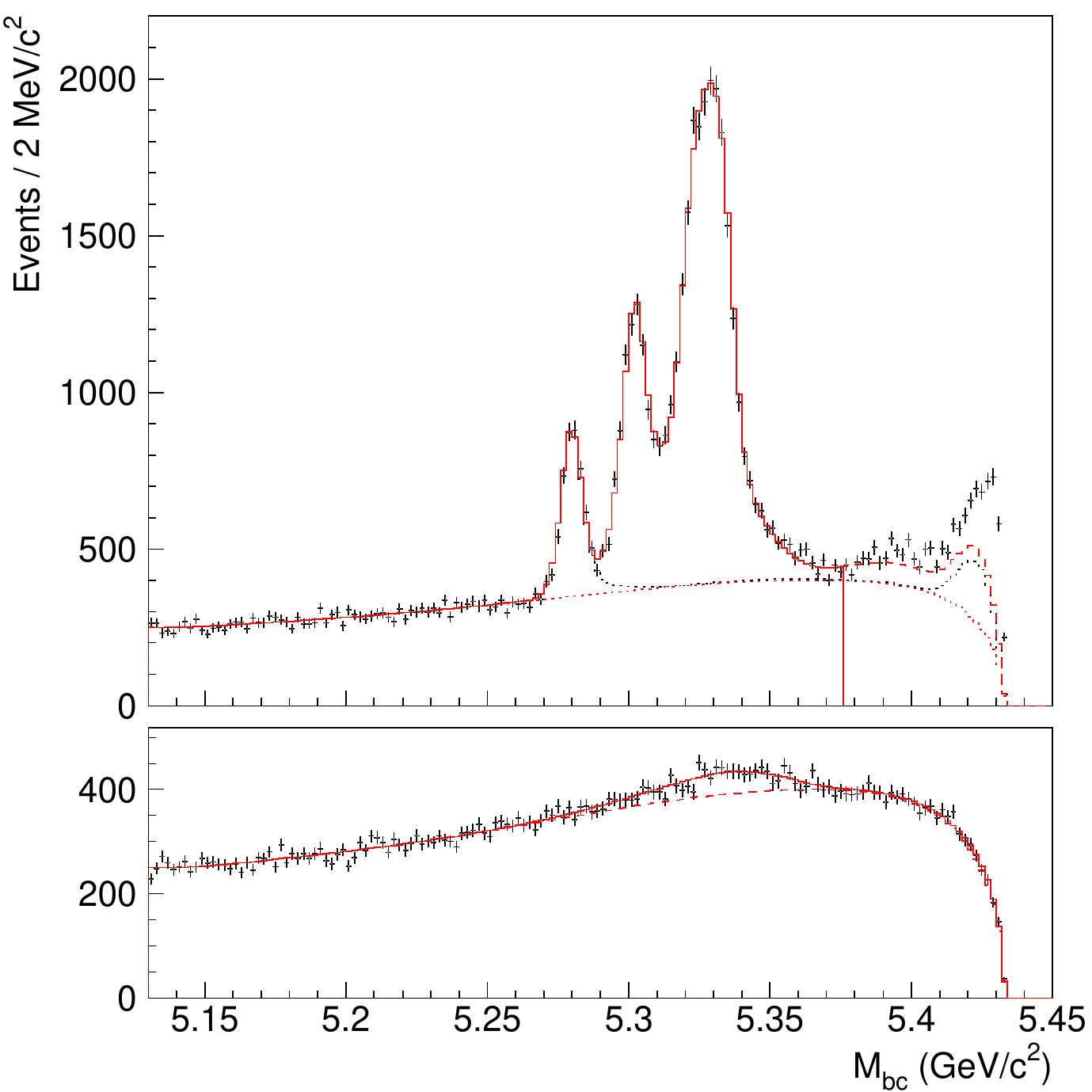}
  \caption{The $\mbc$ distributions in the $\DE'$ signal (top) and
    sideband (bottom) regions. Points with error bars are data, solid
    histogram is the result of the simultaneous fit, dashed histogram
    is the smooth background. Black dotted histogram indicates the
    contribution of the $B\bar{B}$ channel that includes a peak near
    threshold due to the ISR production of $\Ufo$. Vertical red line
    at $5.375\,\gevm$ in the top panel indicates the upper boundary of
    the fit interval. }
  \label{mbc_y5s_it10_140920}
\end{figure}

The energy dependence of the $\ee\to{B}^{(*)}\bar{B}^{(*)}$ cross
sections, that is needed for the fit function, is taken from the
measurements described below; the analysis is performed using an
iterative procedure. The parameters of the momentum resolution
function are determined from the $\Uf$ simulation. They are found to
be close to those at $\Ufo$. The factor $\nr$ for the normalization of
the peaking background is taken to be the same as in the $\Ufo$ data
(Table~\ref{tab:y4s_fit_results}).
The $\ecm$ spread is fixed to the fitted value
(Eq.~(\ref{eq:spread_vs_ecm})) multiplied by the microwave instability
correction factor; the result is $\spread=(5.44 \pm 0.09)\,\mev$.
The smooth background is described by a threshold function
$\sqrt{\ecm/2-x}$ multiplied by a 6th order Chebyshev polynomial; the
order is higher than at $\Ufo$ because we use a broader fit interval.
The result of the fit to the $\Uf$ data is shown in
Fig.~\ref{mbc_y5s_it10_140920} and Table~\ref{tab:y5s_fit_results}.
\begin{table}[htbp]
\caption{ Results of the fit to the $\mbc$ distribution at $\Uf$. The
  errors are statistical.}
\renewcommand*{\arraystretch}{1.2}
\label{tab:y5s_fit_results}
\centering
\begin{tabular}{@{}lc@{}} \toprule
  $N_\mathrm{total}$                            & $(23.66 \pm 0.22 \pm 0.34)\times10^3$ \\
  $N_\mathrm{B\bar{B}}\,/\,N_\mathrm{total}$     & $0.1121 \pm 0.0030$ \\
  $N_\mathrm{B\bar{B}^*}\,/\,N_\mathrm{total}$   & $0.3095 \pm 0.0045$ \\
  $N_\mathrm{B^*\bar{B}^*}\,/\,N_\mathrm{total}$ & $0.5784 \pm 0.0048$ \\
  \midrule
  $\ah$ & $-0.18 \pm 0.07$ \\
  \midrule
  $\sh_3$ & $(-39^{+16}_{-20})\,\mevc$ \\
  $\ff_3$ & $1.42^{+0.59}_{-0.33}$ \\
  $\rsb$ & $0.998\pm0.007$ \\
\bottomrule
\end{tabular}
\end{table}
We report the total yield $N_\mathrm{total}$ and the fractions of
various channels $N_{\bball}\,/\,N_\mathrm{total}$.
One of the fractions is not floated as its value is determined from
the constraint that the sum of the three fractions is equal to 1. To
find its statistical uncertainty, we repeat the fit with a different
choice of the fraction, which is not floated.
The yield is defined as the integral of the fit function in the region
$5.27<\mbc<5.35\,\gevm$.
The fit describes the data well, and its p-value is 87\%.
The $\mbc$ distribution in the $\DE'$ signal region is fitted only up
to $\mbc=5.375\,\mevm$ to avoid the region near the $\mbc$ threshold
where the contribution of the $\ee\to{B}\bar{B}^*\pi$ and
$\ee\to{B}^*\bar{B}^*\pi$ processes is
expected~\cite{Drutskoy:2010an}. We extrapolate the fit function
beyond the fit interval and indeed find an excess of events near the
threshold. The study of the three-body channels
$B^{(*)}\bar{B}^{(*)}\pi$ will be a subject of separate analysis.

The value of $N_\mathrm{total}$ is used to calculate the efficiency in
the $\Uf$ data (Section~\ref{sec:eff}), therefore we estimate its
systematic uncertainty (Table~\ref{tab:syst_y5s_yield}).
\begin{table}[htbp]
\caption{ Systematic uncertainties in the total signal yield
  $N_\mathrm{total}$ at $\Uf$ (in \%).}
\renewcommand*{\arraystretch}{1.2}
\label{tab:syst_y5s_yield}
\centering
\begin{tabular}{@{}lc@{}} \toprule
  Cross section shape               &      \\
  \;\;\;-- statistical uncertainty  & 0.37 \\
  \;\;\;-- parameterization         & 0.35 \\
  $\ecm$ spread                     & 0.12 \\
  Yield of peaking background       & 0.33 \\
  Shape of peaking background       & 0.13 \\
  Shape of smooth background        & 0.05 \\
  Optimization procedure            & 1.3  \\
  \midrule
  Total                             & 1.45 \\
  \bottomrule
\end{tabular}
\end{table}
The shape of the signal depends upon the energy dependence of the
$\bb$, $\bbst$, and $\bstbst$ cross sections. We find the corresponding
uncertainty as described in Section~\ref{sec:xsec}.
To estimate the uncertainty due to the $\ecm$ spread, we vary its
value within the uncertainty of $\pm0.09\,\mev$.
The normalization factor of the peaking background $\nr$ is varied
between 1.08 and 1.24.
In the fit to the $\Ufo$ data we find that the shift $\sh_3$ and the
width-correction factor $\ff_3$ of the peaking background in the $\DE'$ sidebands
are consistent with zero and one, respectively
(Table~\ref{tab:y4s_fit_results}), while in the $\Uf$ data they
deviate from the above values with a combined significance of
$2.6\,\sigma$ (Table~\ref{tab:y5s_fit_results}). To estimate the
uncertainty associated with the shape of the peaking background, we
repeat the fit fixing $\sh_3=0$ and $\ff_3=1$.
To account for the uncertainty in the shape of the smooth background
we change the order of the polynomial which is used in the
parameterization from 6th to 7th and 8th.
We account also for the systematic uncertainty due to the optimization
procedure of 1.3\% (Section~\ref{sec:selection}).
The deviations in the yield under the variations of the analysis are
considered as systematic uncertainties due to a given source. The
total systematic uncertainty (given both in
Tables~\ref{tab:y5s_fit_results} and \ref{tab:syst_y5s_yield}) is
estimated as a sum in quadrature of the individual contributions.

We study the distributions of the $\bb$, $\bbst$, and $\bstbst$ in the
polar angle of the $B$ meson and find that they agree with the
expectations. The details are provided in Appendix~\ref{app:angular}.

\section{Fits at various energies}
\label{sec:scan}

The fit function is the same as at $\Uf$, except for the normalization
of the signal. Here the integral of each signal component is not
normalized to unity in the range $5.27<\mbc<5.35\,\gevm$. Instead, it
is equal to the integral of the ISR kernel~\cite{Kuraev:1985hb}
multiplied by the relative change of the cross section with
energy. This normalization value is equal to the $\opdi$ correction
factor and thus the measured yields include the ISR correction and can
be used directly to determine the dressed cross sections. This
approach was used in previous energy scan
papers~\cite{Abdesselam:2015zza,Abdesselam:2019gth}.

We fix the shift $\sh_3$, the width-correction factor $\ff_3$ and the
relative normalization of the smooth background $\rsb$ to the fit
results at $\Uf$ (Table~\ref{tab:y5s_fit_results}). The angular
distribution parameter $\ah$ is floated within the allowed range
$-1\leq\ah\leq1$. In case of the smooth background only the
normalization is floated. The fits at various energies are shown in
Appendix~\ref{app:mbc_fits}.

Measured yields are used to calculate the dressed cross sections as
described in Section~\ref{sec:xsec} after we present the determination
of the efficiency in the next Section.
The values of $\ah$ in the scan data samples are poorly constrained by
the fit; their uncertainties are close to the size of the allowed
range.

\section{Determination of the efficiency}
\label{sec:eff}

To determine the efficiency at the $\Ufo$ energy we use the measured
$B$ meson yield $N(\Ufo)$ (Table~\ref{tab:y4s_fit_results}) and the
total number of the $B\bar{B}$ pairs in the $\Ufo$ SVD2 data,
$N_{\bb}(\Ufo)=(619.6\pm9.4)\times10^6$~\cite{Brodzicka:2012jm}. This
number is obtained by counting hadronic events at $\Ufo$ and
subtracting the continuum contribution, which is determined using data
collected $60\,\mev$ below $\Ufo$. The transitions from $\Ufo$ to
lower bottomonia have total branching fraction of 0.26\%~\cite{PDG}
and are neglected. The efficiency is calculated as
\begin{equation}
  \varepsilon_{\Ufo} = \frac{N(\Ufo)}{2\,N_{\bb}(\Ufo)} =
  (0.4690\pm0.0077)\times10^{-3}.
  \label{eq:eff_y4s}
\end{equation}

To determine precisely the ratio of the $B$ meson yields in the $\Ufo$
and $\Uf$ data samples, we use five final states with low multiplicity
for which the distribution over phase space is well known and which
can be reliably simulated:
\begin{enumerate}
\item $B^+\to\jp\,K^+$,
\item $B^0\to\jp\,K^*(892)^0$, $K^*(892)^0\to{K}^+\pi^-$,
\item $B^+\to\bar{D}^0\pi^+$, $\bar{D}^0\to{K}^+\pi^-$,
\item $B^+\to\bar{D}^0\pi^+$, $\bar{D}^0\to{K}^+\pi^+\pi^-\pi^-$,
\item $B^0\to D^-\pi^+$, $D^-\to{K}^+\pi^-\pi^-$.
\end{enumerate}
The signal-to-background ratio in these final states is high, and they
are reconstructed without application of FEI.
Selection requirements are taken to be the same as in
Ref.~\cite{Drutskoy:2010an}.
To minimize the sensitivity to the peaking background, we fit the
$\DE'$ spectra simultaneously in the $\Ufo$ and $\Uf$ data samples
applying a requirement $5.27<\mbc<5.35\,\gevm$. The fit is performed
separately for each channel. The fit function for the $\Ufo$ data is a
sum of two Gaussians to describe the signal and a first or second
order polynomial (depending on the channel) to describe the
background. The fit function for the $\Uf$ data is the same except
that the ratio of yields at $\Uf$ and $\Ufo$, shift, and
width-correction factor are introduced for the signal. The background
components in the $\Ufo$ and $\Uf$ data samples are floated
independently.
To estimate the systematic uncertainty, we consider variations of the
polynomial order and the fit interval. The systematic uncertainty is
calculated as the root-mean-square (RMS) of the deviations. 
We repeat the same fits for the MC samples and find that the
efficiency ratio is consistent with one for all the channels.
The efficiency-corrected yield ratios are $0.0393\pm0.0017$,
$0.0376\pm0.0023$, $0.0399\pm0.0021$, $0.0365\pm0.0033$, and
$0.0391\pm0.0022$ for the channels from one to five, respectively; the
uncertainties include the statistical and systematic contributions.
The average yield ratio is
\begin{equation}
  \rafilo = 0.03882\pm0.00097.
  \label{eq:r_5chan}
\end{equation}

In this calculation we implicitly assume that the ratio of the
$B^+B^-$ and $B^0\bar{B}^0$ production rates, $\fra$, is the same at
$\Ufo$ and $\Uf$. Since $\Uf$ is far from the $\bb$ threshold, one can
expect that isospin violation at $\Uf$ is small and $\fra=1$. At
$\Ufo$, $\fra$ was measured to be $1.058\pm0.024$~\cite{PDG}; thus, it
is shifted from 1 by 2.4 standard deviations. We repeat the
calculation taking into account the isospin non-conservation at
$\Ufo$, and find that $\rafilo$ increases by 0.53\%. Since the change
is very small, the isospin non-conservation is neglected.

The efficiency at the $\Uf$ energy is determined from the total $B$
meson yield $N_\mathrm{total}$ (Table~\ref{tab:y5s_fit_results}):
\begin{equation}
  \varepsilon_{\Uf} =
  \frac{N_\mathrm{total}}{2\,N_{\bb}(\Ufo)}\;
  \frac{1}{\rafilo}
  = (0.492\pm0.017)\times10^{-3}.
  \label{eq:eff_y5s}
\end{equation}
The ratio of the efficiencies at $\Uf$ and $\Ufo$, $1.049\pm0.032$,
agrees with the MC expectation of $1.028\pm0.004$. 

From MC simulation we find that the dependence of the efficiency on
the $B$ meson momentum is consistent with being linear.
Thus, for all energies and various $B^{(*)}\bar{B}^{(*)}$ final
states we determine the efficiency $\varepsilon$ based on the
corresponding average momentum and the values $\varepsilon_{\Ufo}$ and
$\varepsilon_{\Uf}$, assuming linear dependence on the $B$ meson
momentum.
To find the uncertainty in $\varepsilon$, we separate uncertainties in
$\varepsilon_{\Ufo}$ and $\varepsilon_{\Uf}$ into common and
uncorrelated parts; we then assume that the uncorrelated part varies
linearly with the $B$ meson momentum.

\section{Results for the cross sections}
\label{sec:xsec}

\begin{figure}[htbp]
  \centering
  \includegraphics[width=0.49\textwidth]{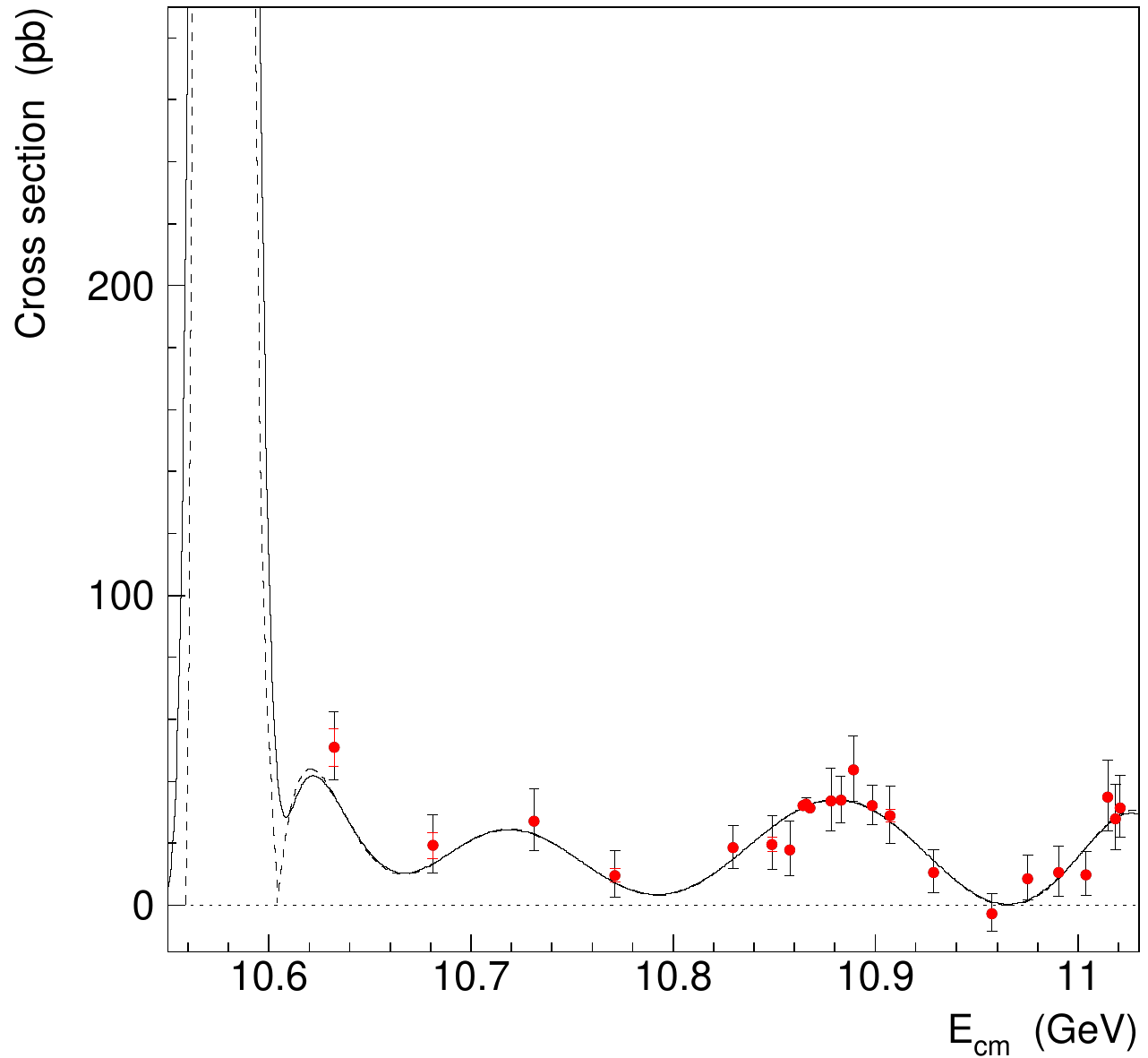}
  \includegraphics[width=0.49\textwidth]{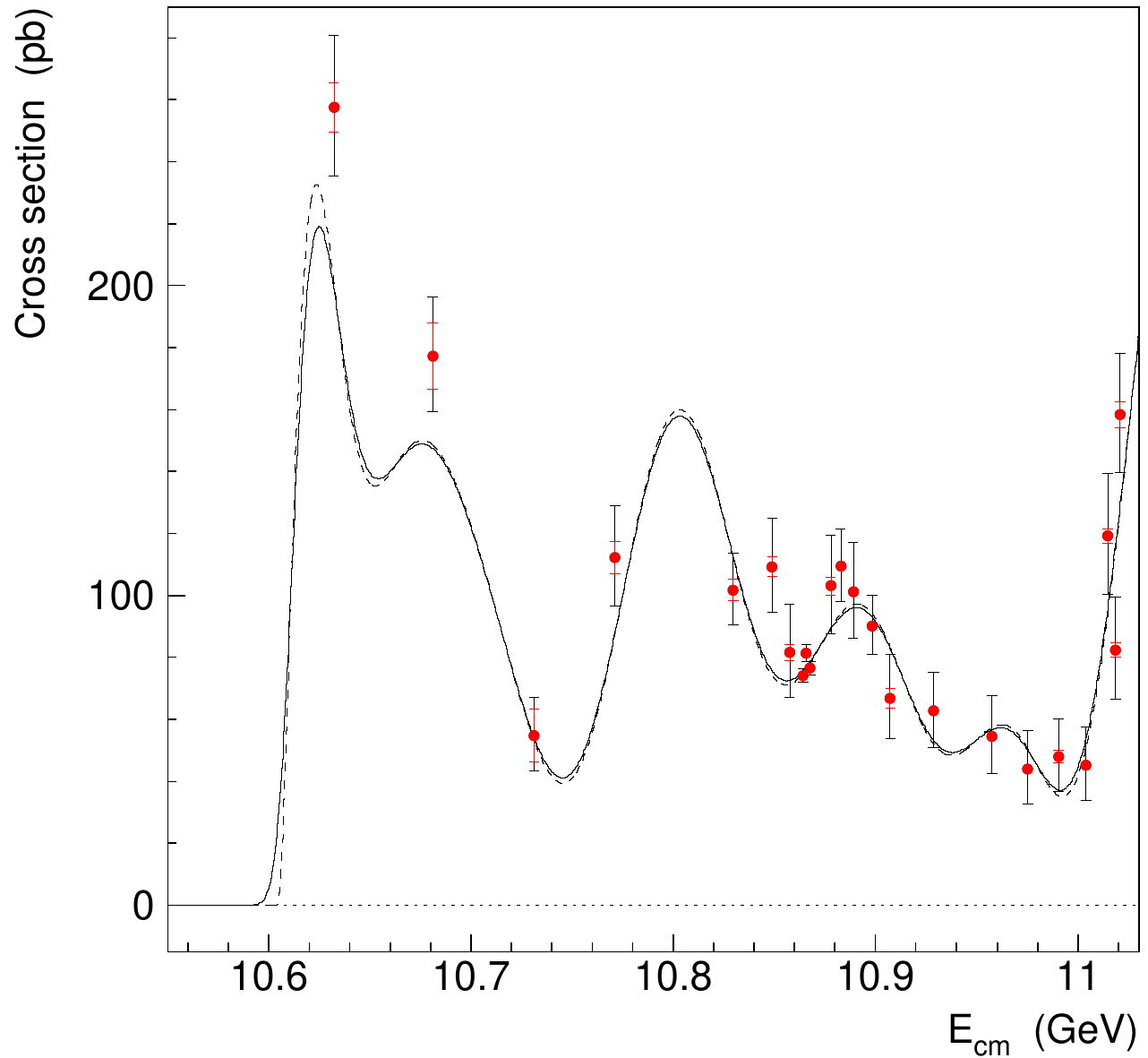}
  \includegraphics[width=0.49\textwidth]{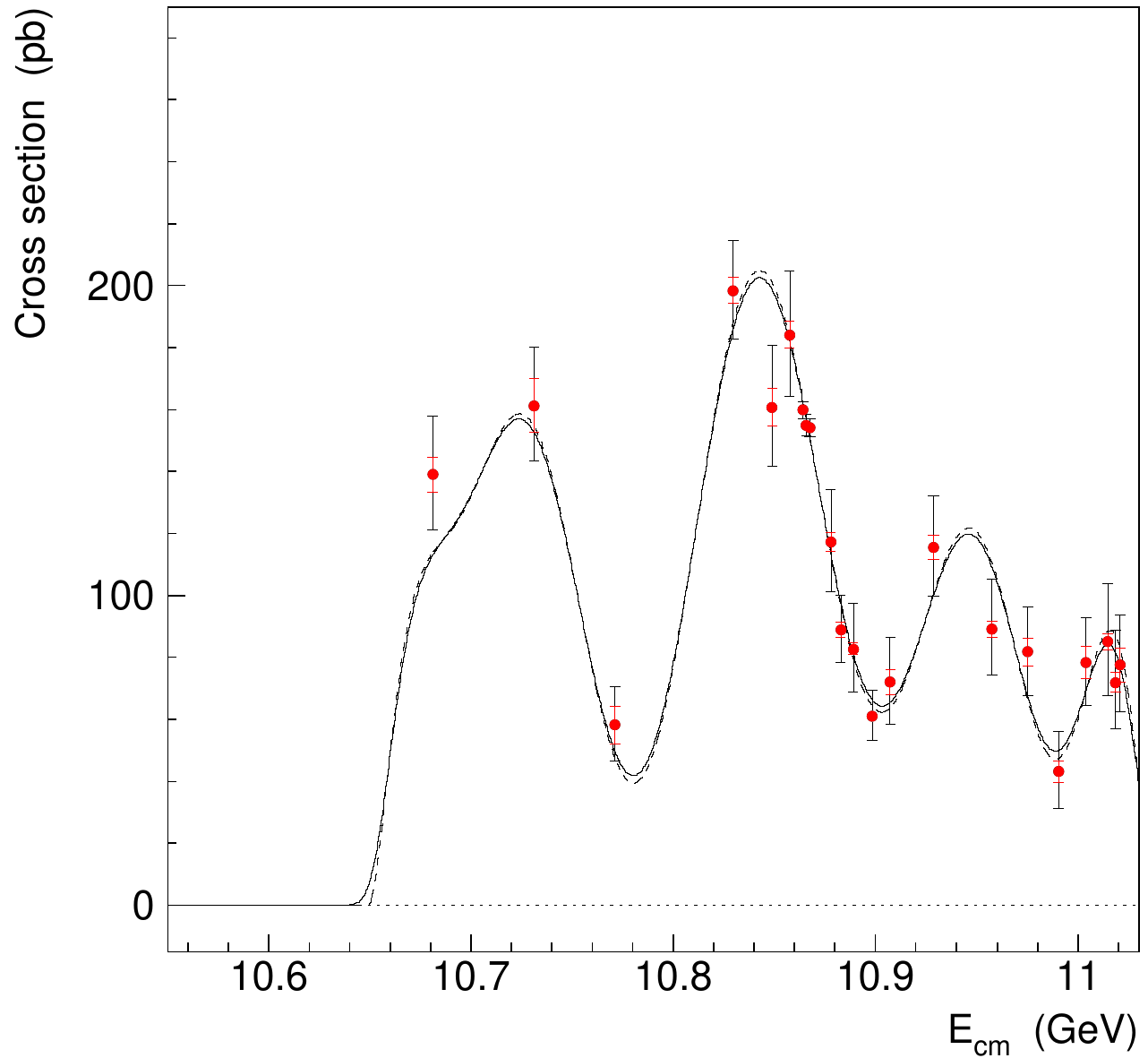}
  \caption{ Measured dressed cross sections at various energies for
    $\ee\to{B}\bar{B}$ (left), $\ee\to{B}\bar{B}^*$ (right), and
    $\ee\to{B}^*\bar{B}^*$ (bottom). The outer error bars indicate
    statistical uncertainties and inner red error bars indicate
    uncorrelated systematic uncertainties. Solid curves show the
    result of the simultaneous fit to these distributions and the
    total $b\bar{b}$ cross section energy dependence
    (Fig.~\ref{cmp_total_sum_exc_it13a_281020}). Dashed curves show
    the fit function before the convolution to account for the $\ecm$
    spread. }
  \label{bb_xs_vs_ecm_unc_it11_190920}
\end{figure}

The dressed cross sections are calculated as
\begin{equation}
  \sigma^\mathrm{dressed} = \frac{N}{\opdi\,L\,\varepsilon},
\end{equation}
where the ratio $N/\opdi$ is directly obtained from the fit. The cross
sections are presented in Table~\ref{tab:xsec_results} and in
Fig.~\ref{bb_xs_vs_ecm_unc_it11_190920}.
\begin{table}[htbp]
\caption{ Energies (in $\mev$), luminosities (in $\fb$) for various
  data samples and the results for the dressed cross sections (in
  pb). The first error in the cross section is statistical, the second
  is uncorrelated systematic, and the third is correlated systematic.}
\renewcommand*{\arraystretch}{1.4}
\label{tab:xsec_results}
\centering
\begin{tabular}{@{}ccccc@{}} \toprule
  $\ecm$ & $L$ & $\sigma(B\bar{B})$ & $\sigma(B\bar{B}^*)$ & $\sigma(B^*\bar{B}^*)$ \\
  \midrule
  $11020.8\pm1.4$ &  0.982 & $31.5\pm9.9\pm1.2\pm1.7$ & $158.4\pm19.3\pm4.2\pm7.7$ & $77.6\pm15.6\pm5.4\pm3.6$ \\
  $11018.5\pm2.0$ &  0.859 & $27.8\pm10.5\pm1.0\pm1.5$ & $82.4\pm16.5\pm2.3\pm4.0$ & $71.9\pm15.9\pm3.1\pm3.4$ \\
  $11014.8\pm1.4$ &  0.771 & $34.8\pm11.4\pm1.2\pm1.9$ & $119.2\pm19.5\pm2.4\pm5.8$ & $85.0\pm18.1\pm2.7\pm3.9$ \\
  $11003.9\pm1.0$ &  0.976 & $9.7\pm7.0\pm0.3\pm0.6$ & $45.2\pm11.8\pm1.3\pm2.2$ & $78.4\pm14.2\pm5.1\pm3.6$ \\
  $10990.4\pm1.3$ &  0.985 & $10.5\pm8.1\pm0.4\pm0.7$ & $47.9\pm11.7\pm2.0\pm2.3$ & $43.1\pm12.4\pm3.5\pm2.0$ \\
  $10975.3\pm1.4$ &  0.999 & $8.5\pm7.2\pm1.2\pm0.6$ & $44.0\pm11.9\pm0.8\pm2.1$ & $81.7\pm14.3\pm4.5\pm3.6$ \\
  $10957.5\pm1.5$ &  0.969 & $-2.8\pm6.0\pm0.1\pm0.3$ & $54.5\pm12.6\pm1.6\pm2.5$ & $89.2\pm15.5\pm2.5\pm3.8$ \\
  $10928.7\pm1.6$ &  1.149 & $10.5\pm6.9\pm0.9\pm0.6$ & $62.7\pm12.1\pm1.6\pm2.7$ & $115.6\pm16.2\pm3.8\pm4.7$ \\
  $10907.3\pm1.1$ &  0.980 & $28.8\pm9.1\pm2.0\pm1.4$ & $66.8\pm13.5\pm3.2\pm2.8$ & $72.1\pm14.0\pm4.0\pm2.8$ \\
  $10898.3\pm0.7$ &  2.408 & $32.2\pm6.3\pm0.5\pm1.4$ & $90.2\pm9.4\pm1.3\pm3.7$ & $61.1\pm8.0\pm1.4\pm2.3$ \\  
  $10888.9\pm0.8$ &  0.990 & $43.8\pm10.5\pm0.7\pm2.0$ & $101.2\pm15.6\pm1.0\pm4.1$ & $82.7\pm14.4\pm1.8\pm3.1$ \\
  $10882.8\pm0.7$ &  1.848 & $33.9\pm7.5\pm0.4\pm1.5$ & $109.6\pm11.7\pm1.5\pm4.4$ & $88.9\pm10.8\pm2.5\pm3.3$ \\  
  $10877.8\pm0.8$ &  0.978 & $33.7\pm10.1\pm1.7\pm1.5$ & $103.1\pm16.0\pm2.8\pm4.1$ & $117.3\pm16.6\pm3.0\pm4.3$ \\
  $10867.6\pm0.2$ &  45.28 & $31.3\pm1.5\pm0.0\pm1.3$ & $76.5\pm2.1\pm0.1\pm3.2$ & $154.1\pm2.7\pm0.2\pm6.2$ \\     
  $10865.8\pm0.3$ &  29.11 & $32.7\pm1.9\pm0.0\pm1.4$ & $81.3\pm2.7\pm0.1\pm3.4$ & $154.9\pm3.4\pm0.1\pm6.2$ \\     
  $10864.2\pm0.3$ &  47.65 & $32.2\pm1.4\pm0.0\pm1.4$ & $74.2\pm2.0\pm0.1\pm3.1$ & $159.9\pm2.7\pm0.3\pm6.3$ \\
  $10857.4\pm0.9$ &  0.988 & $17.8\pm8.8\pm1.2\pm0.8$ & $81.5\pm15.0\pm2.5\pm3.2$ & $184.1\pm20.4\pm4.4\pm6.5$ \\
  $10848.9\pm1.0$ &  0.989 & $19.6\pm8.7\pm2.3\pm0.9$ & $109.3\pm15.2\pm3.2\pm4.1$ & $160.8\pm19.4\pm6.2\pm5.6$ \\
  $10829.5\pm1.2$ &  1.697 & $18.6\pm7.0\pm0.7\pm0.8$ & $101.8\pm11.6\pm3.4\pm3.7$ & $198.4\pm16.0\pm4.2\pm6.6$ \\
  $10771.2\pm1.0$ &  0.955 & $9.7\pm7.6\pm2.2\pm0.5$ & $112.2\pm16.2\pm5.2\pm3.6$ & $58.2\pm12.1\pm6.1\pm1.7$ \\
  $10731.3\pm1.5$ &  0.946 & $27.0\pm10.1\pm1.4\pm1.0$ & $54.7\pm11.8\pm8.5\pm1.6$ & $161.3\pm18.4\pm8.7\pm4.2$ \\
  $10681.0\pm1.4$ &  0.949 & $19.2\pm9.3\pm4.1\pm0.7$ & $177.3\pm18.4\pm10.7\pm4.5$ & $139.0\pm18.4\pm5.7\pm3.1$ \\
  $10632.2\pm1.5$ &  0.989 & $51.0\pm11.1\pm6.0\pm1.4$ & $257.6\pm22.7\pm8.1\pm5.6$ & --- \\
  \bottomrule
\end{tabular}
\end{table}
The cross sections show a non-trivial behaviour with several maxima
and minima. There is no obvious signal of $\Uf$ that matches its mass
and width.

\begin{figure}[htbp]
  \centering
  \includegraphics[width=0.49\textwidth]{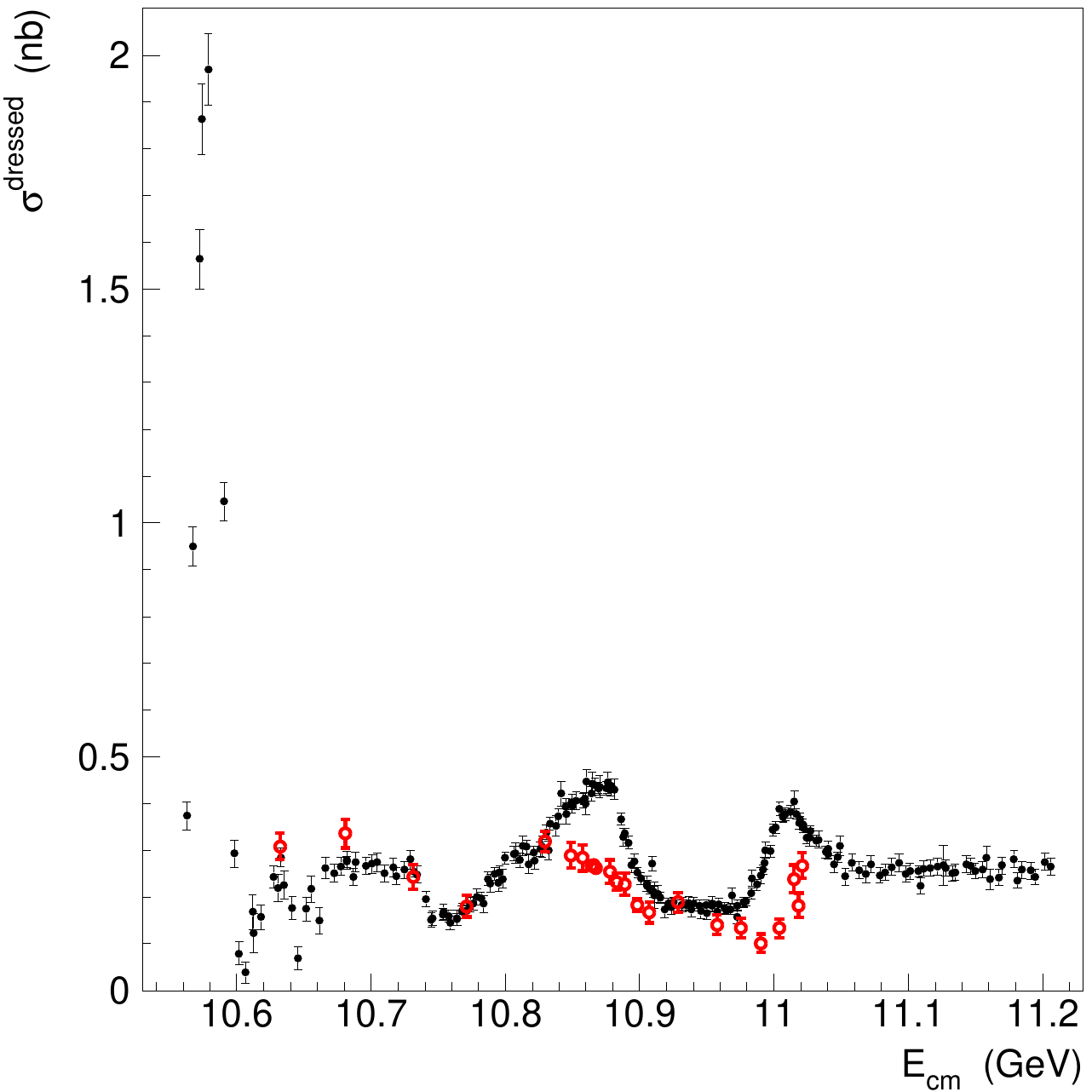}
  \includegraphics[width=0.49\textwidth]{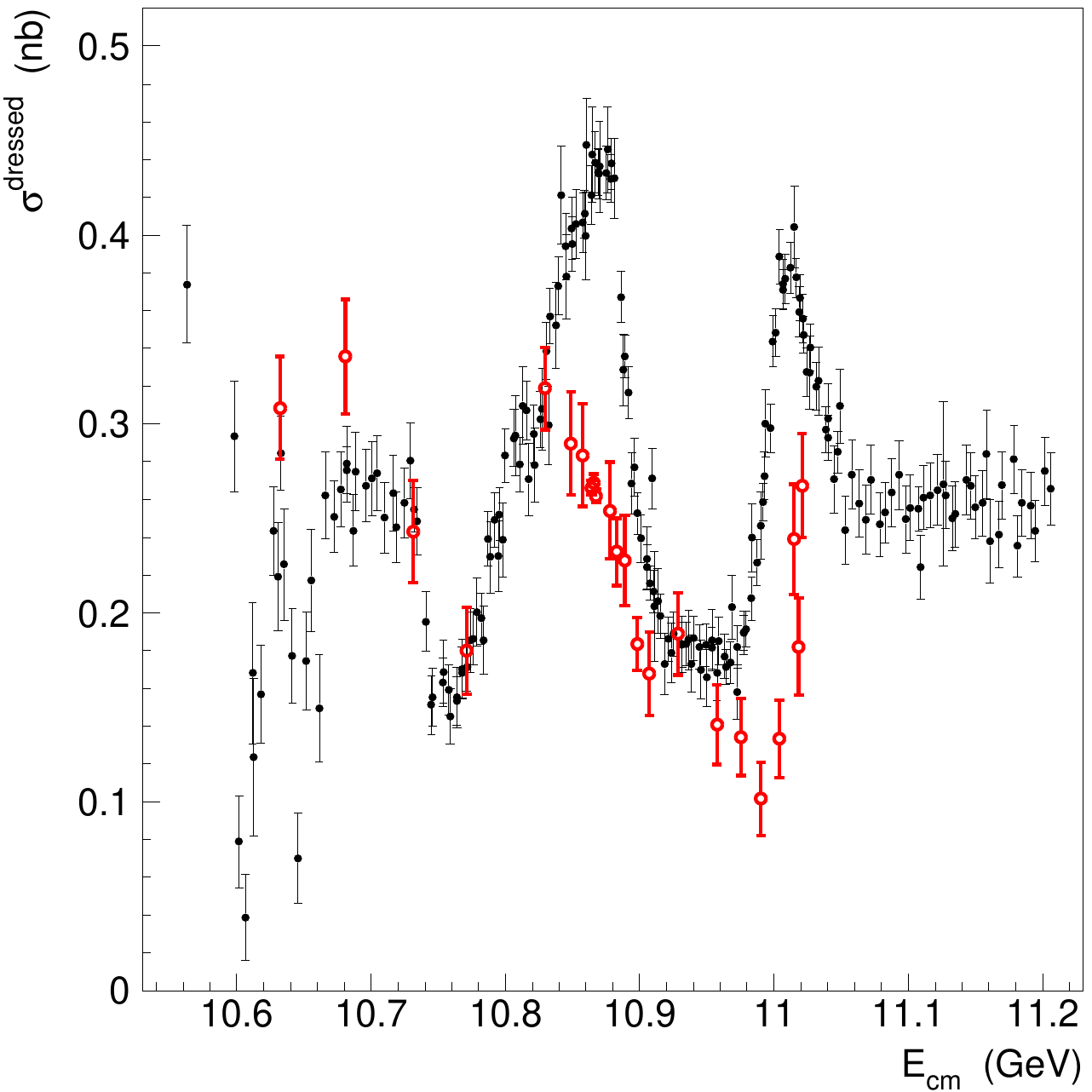}
  \caption{ Energy dependence of the total $b\bar{b}$ dressed cross
    section obtained in Ref.~\cite{Dong:2020tdw} from the visible
    cross sections measured by Belle~\cite{Santel:2015qga} and
    BaBar~\cite{Aubert:2008ab} (black dots). Open red circles
    represent the sum of the exclusive $\bb$, $\bbst$, and $\bstbst$
    cross sections measured in this work. Right panel is a zoom of the
    low cross section region. }
  \label{cmp_total_sum_exc_it13a_281020}
\end{figure}

In Fig.~\ref{cmp_total_sum_exc_it13a_281020} we plot the sum of the
exclusive $\bb$, $\bbst$, and $\bstbst$ cross sections superimposed on
the total $b\bar{b}$ dressed cross section that was obtained in
Ref.~\cite{Dong:2020tdw} from the visible cross sections measured by
Belle~\cite{Santel:2015qga} and BaBar~\cite{Aubert:2008ab}.
The sum is compatible with the total $b\bar{b}$ cross section up to
$\ecm=10.82\,\gev$; this value is close to the $B_s^*\bar{B}_s^*$
threshold. The deviation at higher energy is presumably due to the
contributions of $B_s$ mesons, multibody final states
$B^{(*)}\bar{B}^{(*)}\pi(\pi)$, and production of bottomonia with
light hadrons.

To calculate the shapes of the signals at various energies and to
determine the ISR corrections, we need to parameterize the energy
dependence of the cross sections. Since currently there is no suitable
phenomenological model for the cross section energy dependence, we fit
the cross sections using high-order Chebyshev polynomials.
The fit to the total $b\bar{b}$ visible cross section in the $\Ufo$
region (Fig.~\ref{xsec_fit_y4s_290820}) and the analysis in
Ref.~\cite{Dong:2020tdw} (Fig.~\ref{cmp_total_sum_exc_it13a_281020})
show that the dressed $\bb$ cross section goes to zero at the $\bbst$
threshold. Thus, for the $\bb$ channel we fit the cross section
starting from the $\bbst$ threshold, while below this threshold we use
the result of the fit shown in Fig.~\ref{xsec_fit_y4s_290820}. To
impose the requirement that the cross section is zero at the $\bbst$
threshold in the $\bb$ channel, as well as at the corresponding
thresholds in the $\bbst$ and $\bstbst$ channels, we add points at the
thresholds with zero values and small uncertainties.

\begin{figure}[htbp]
  \centering
  \includegraphics[width=0.49\textwidth]{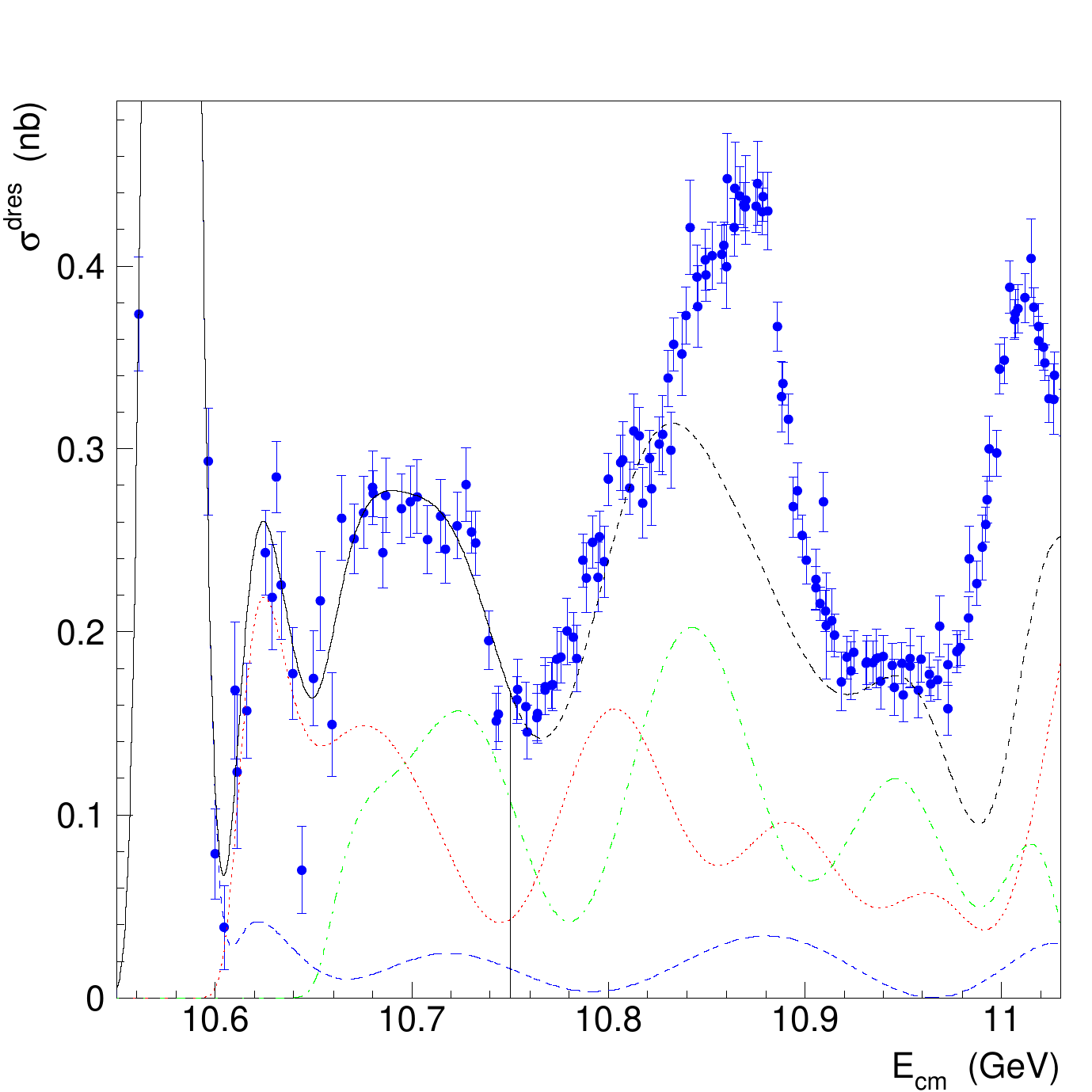}
  \caption{ Energy dependence of the total $b\bar{b}$ dressed cross
    section from Ref.~\cite{Dong:2020tdw} (blue dots). Solid black
    curve is the result of the simultaneous fit to this distribution
    and the exclusive $\bb$, $\bbst$, and $\bstbst$ cross section
    energy dependence
    (Fig.~\ref{bb_xs_vs_ecm_unc_it11_190920}). Vertical line at
    $10.75\,\gev$ indicates the upper boundary of the fit interval;
    dashed black curve is an extrapolation of the fit
    function. Also shown are the individual contributions of $\bb$
    (blue dashed curve), $\bbst$ (red dotted curve), and $\bstbst$
    (green dash-dotted curve). }
  \label{tot_xs_vs_ecm_it13a_281020}
\end{figure}

The energy dependence of the total $b\bar{b}$ dressed cross section
shows a dip at the $\bstbst$ threshold of $10.65\,\gev$
(Fig.~\ref{cmp_total_sum_exc_it13a_281020}). To take into account this
dip in the fits to the exclusive cross sections, we use the total
$b\bar{b}$ dressed cross section as an additional constraint.
We fit simultaneously the exclusive cross sections and the total
$b\bar{b}$ dressed cross section, the latter only in the region below
the $\bbst\pi$ threshold of $10.75\,\gev$. The fit function for the
total $b\bar{b}$ dressed cross section is just a sum of the individual
$\bb$, $\bbst$, and $\bstbst$ contributions. The orders of the
polynomials that we use for $\bb$, $\bbst$, and $\bstbst$ are 10, 17,
and 12, respectively. These orders provide sufficient flexibility to
describe the available data.
The polynomials are constrained to be positive by adding a penalty
term to the $\chi^2$ in case the polynomial becomes negative at any
energy.
To account for the $\ecm$ spread, we convolve the polynomials with the
Gaussian.
The results of the simultaneous fit are presented in
Figs.~\ref{bb_xs_vs_ecm_unc_it11_190920} and
\ref{tot_xs_vs_ecm_it13a_281020}.

\begin{figure}[htbp]
  \centering
  \includegraphics[width=0.49\textwidth]{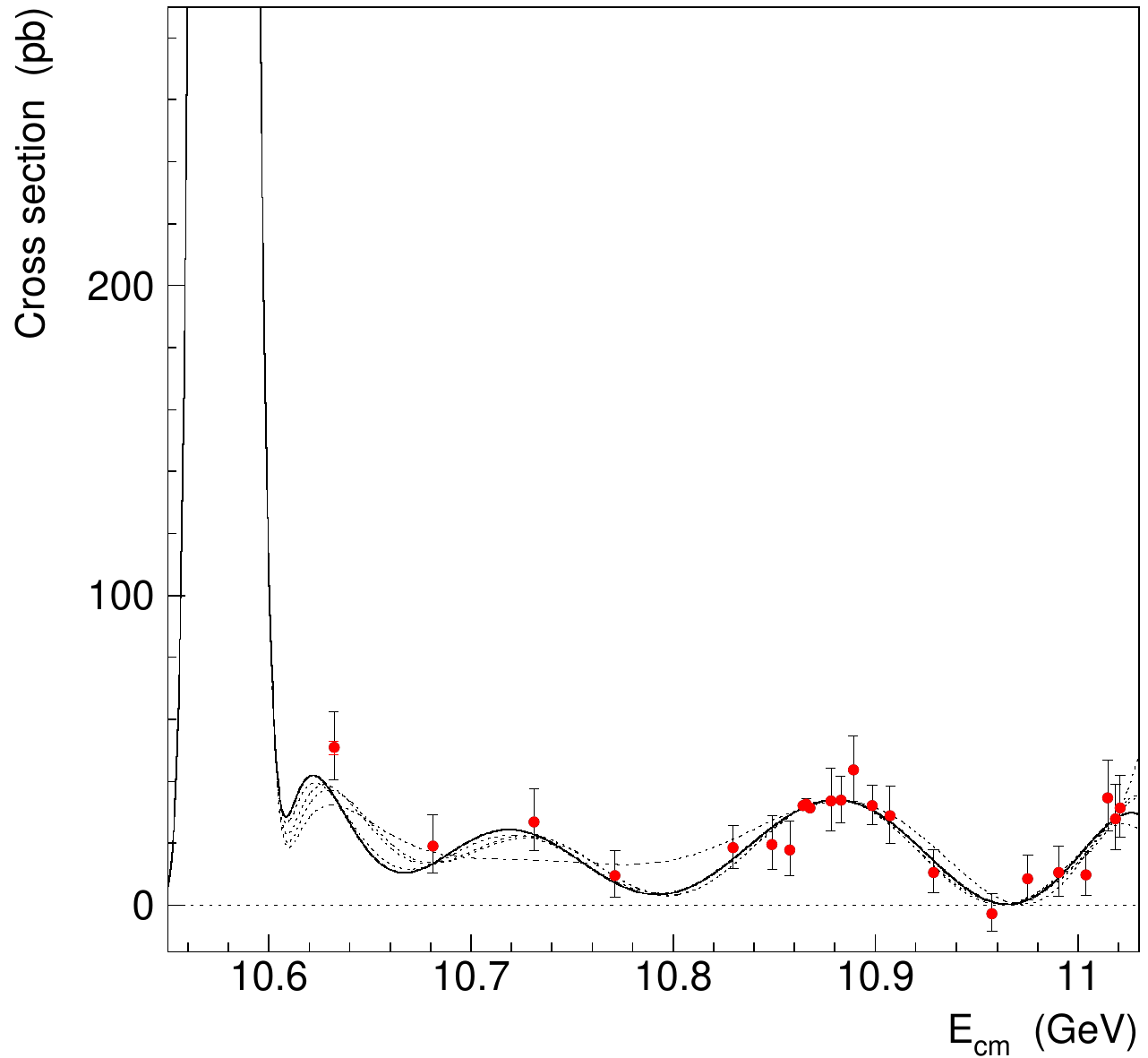}
  \includegraphics[width=0.49\textwidth]{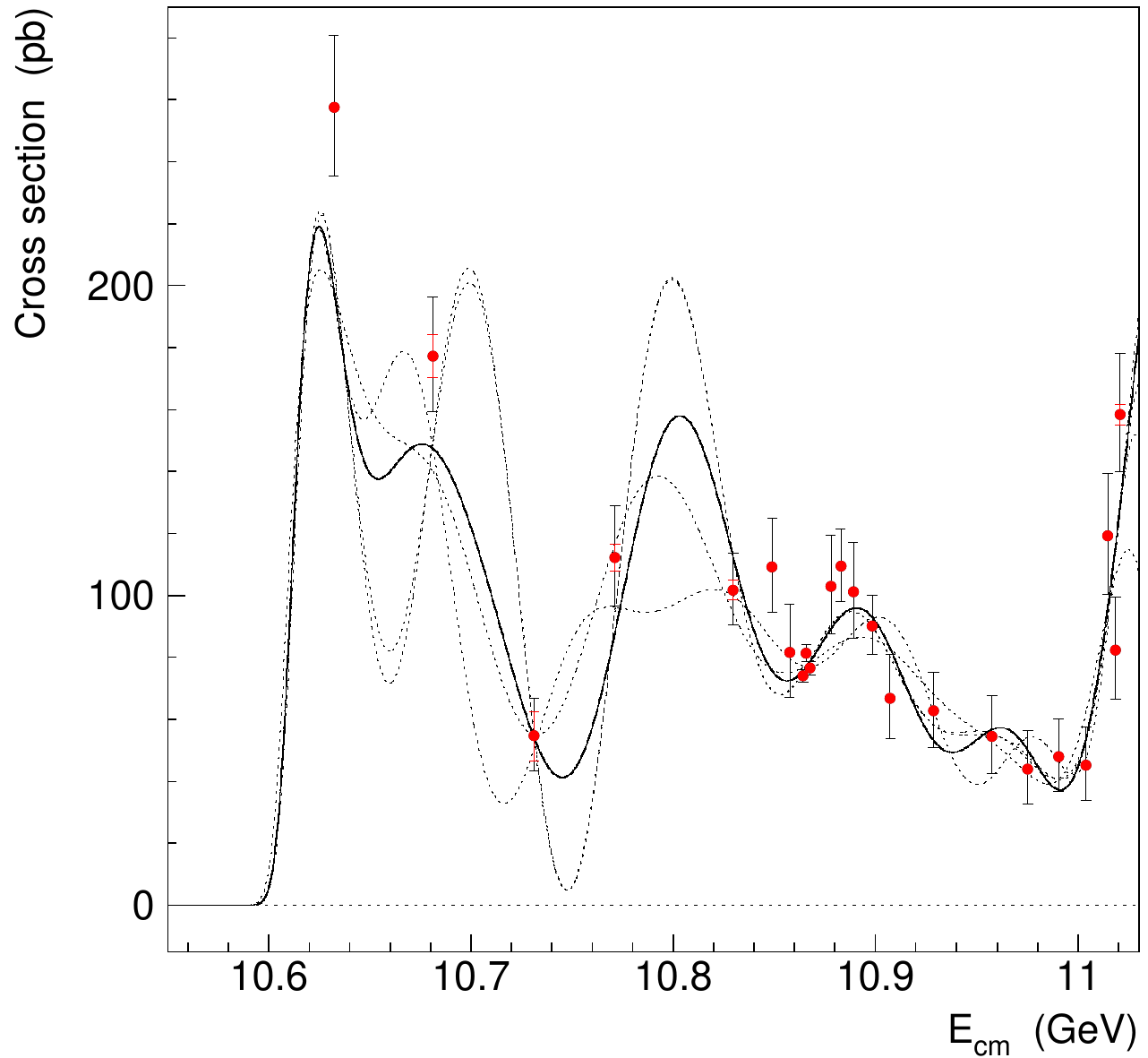}
  \includegraphics[width=0.49\textwidth]{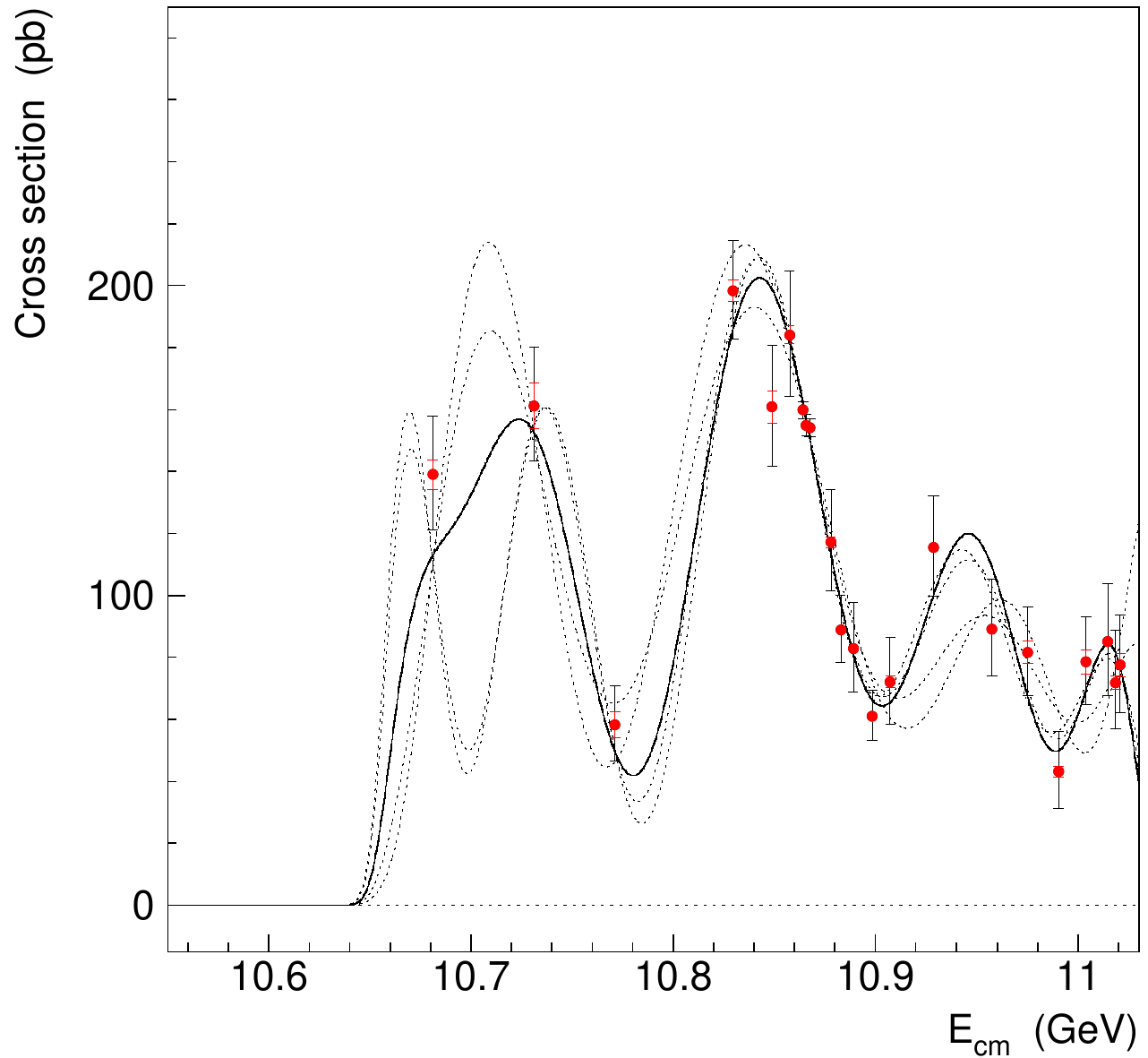}
  \caption{ Measured dressed cross sections at various energies for
    $\ee\to{B}\bar{B}$ (left), $\ee\to{B}\bar{B}^*$ (right), and
    $\ee\to{B}^*\bar{B}^*$ (bottom). The outer error bars indicate the
    statistical uncertainties and the inner red error bars indicate
    the systematic uncertainties due to the cross section
    parameterization. Solid curves show the fit results for the
    default set of polynomial orders. Dotted curves show the fit
    results for the polynomial orders varied by $\pm1$ and $\pm2$.}
  \label{bb_xs_vs_ecm_syst_xsec_param_it23_180221}
\end{figure}

The uncertainty in the shape of the cross section energy dependence
contributes to the systematic uncertainty in the cross section
measurements at various energies. The uncertainty in the shapes
originates from the parameterization and from the limited statistical
accuracy in the cross section measurements. In addition to the default
set of the polynomial orders of (10, 17, 12), we consider also sets
(11, 18, 13), (12, 19, 14), (9, 16, 11), and (8, 15, 10) that provide
a conservative estimation of the possible cross section behaviours.
Corresponding fit results are shown in
Fig.~\ref{bb_xs_vs_ecm_syst_xsec_param_it23_180221}. 

To estimate the influence of the statistical accuracy, we use toy
MC. We generate pseudoexperiments using the fitted cross sections as
central values and statistical uncertainties in data as standard
deviations. We fit the energy dependence of the cross sections in each
pseudoexperiment and, based on the fit results, determine the $\mbc$
signal shapes for all energies. We then refit the data using the new
shapes and repeat the measurement of the cross sections. The RMS of
the deviations is taken as a systematic uncertainty; the
parameterization and the statistical accuracy are considered as
separate sources. We find that for the three high-statistics $\Uf$
on-resonance points the deviations are strongly correlated, therefore
they are accounted for in the correlated uncertainty of these points;
for other points the uncertainties are considered to be uncorrelated.

To study the uncertainty due to the shape of the smooth background in
the $\mbc$ fits, we multiply the corresponding contribution by the
Chebyshev polynomial of the first or second order with floated
parameters. The RMS of the deviations of the yields are used to
calculate the uncertainties.

We vary the $\ecm$ values within their uncertainties, the deviations
in the yields are found to be negligible.
The cross section shape contributions and the smooth background shape
contribution are added in quadrature to obtain the total uncorrelated
systematic uncertainties. They are found to be small compared to
the statistical uncertainties as shown in
Fig.~\ref{bb_xs_vs_ecm_unc_it11_190920}.

Various contributions to the correlated systematic uncertainty,
estimated for the $\Uf$ high-statistics data, are presented in
Table~\ref{tab:cor_syst}. 
\begin{table}[htbp]
\caption{ Correlated systematic uncertainties in cross sections at
  $\Uf$ (in \%).}
\renewcommand*{\arraystretch}{1.2}
\label{tab:cor_syst}
\centering
\begin{tabular}{@{}lccc@{}} \toprule
  & $\sigma(B\bar{B})$ & $\sigma(B\bar{B}^*)$ & $\sigma(B^*\bar{B}^*)$ \\
  \midrule
  Cross section shape               &      &      &      \\ 
  \;\;\;-- statistical uncertainty  & 0.51 & 1.03 & 0.87 \\ 
  \;\;\;-- parameterization         & 0.17 & 1.30 & 1.55 \\ 
  $\ecm$ spread                     & 0.22 & 0.04 & 0.05 \\ 
  Yield of peaking background       & 0.12 & 0.04 & 0.05 \\ 
  Shape of peaking background       & 0.76 & 0.11 & 0.21 \\ 
  Efficiency                        & 3.40 & 3.40 & 3.40 \\
  Luminosity                        & 1.4  & 1.4  & 1.4  \\
  \midrule
  Total                             & 3.80 & 4.04 & 4.09 \\
  \bottomrule
\end{tabular}
\end{table}
The contribution of the cross-section shape is estimated as discussed
above.
The contributions of the $\ecm$ spread and the peaking background
yield and shape are estimated as described in Section~\ref{sec:y5s}
for the total $B$ meson yield. We account also for the uncertainty in
the efficiency (Eq.~(\ref{eq:eff_y5s})) and the uncertainty in the
integrated luminosity of 1.4\%. Total correlated uncertainty is
estimated as a sum in quadrature of the individual contributions.

The sources of the correlated systematic uncertainty at energies other
than $\Uf$ on-resonance are the same as listed in
Table~\ref{tab:cor_syst}, except for the cross section shape source
which is accounted for in the uncorrelated uncertainty. The
contributions of the $\ecm$ spread and the shape of the peaking
background are assumed to be the same as listed in
Table~\ref{tab:cor_syst}. The uncertainty in the efficiency varies
with the $B$ meson momentum as described in Section~\ref{sec:eff}.

To transform the multiplicative correlated uncertainty in the cross
section into the additive one, we use formula:
\begin{equation}
  (\sigma \pm \Delta_\sigma) \cdot (1 \pm \delta) =
  \sigma \pm \Delta_\sigma \pm 
  (\sigma \, \delta \oplus \Delta_\sigma \, \delta),
\end{equation}
where the symbol $\oplus$ denotes addition in quadrature.
The measured dressed cross sections at various energies with
statistical, uncorrelated systematic, and correlated systematic
uncertainties are presented in Table~\ref{tab:xsec_results}.

\section{Discussion}
\label{sec:discussion}

\begin{figure}[htbp]
  \centering
  \includegraphics[width=0.49\textwidth]{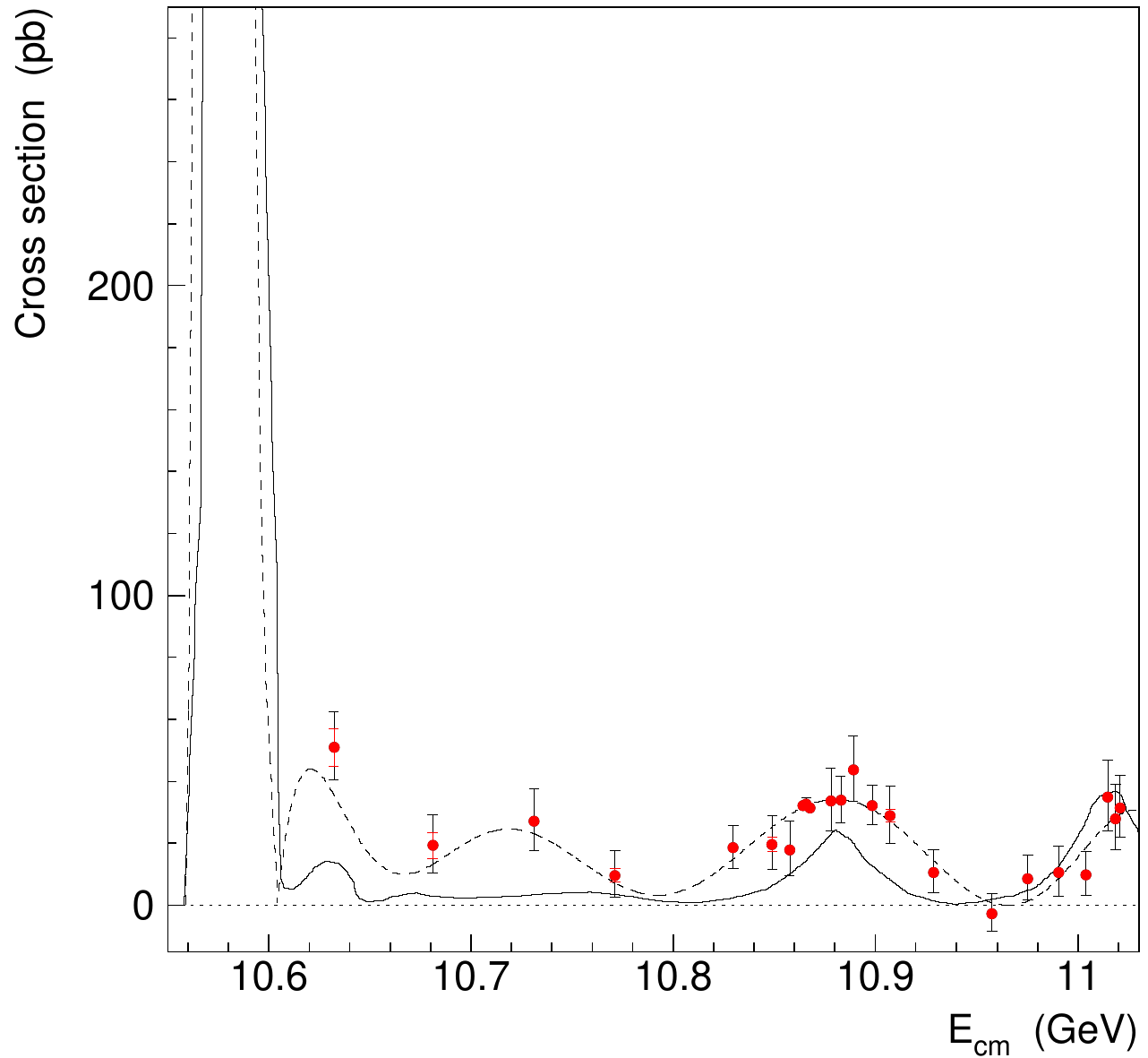}
  \includegraphics[width=0.49\textwidth]{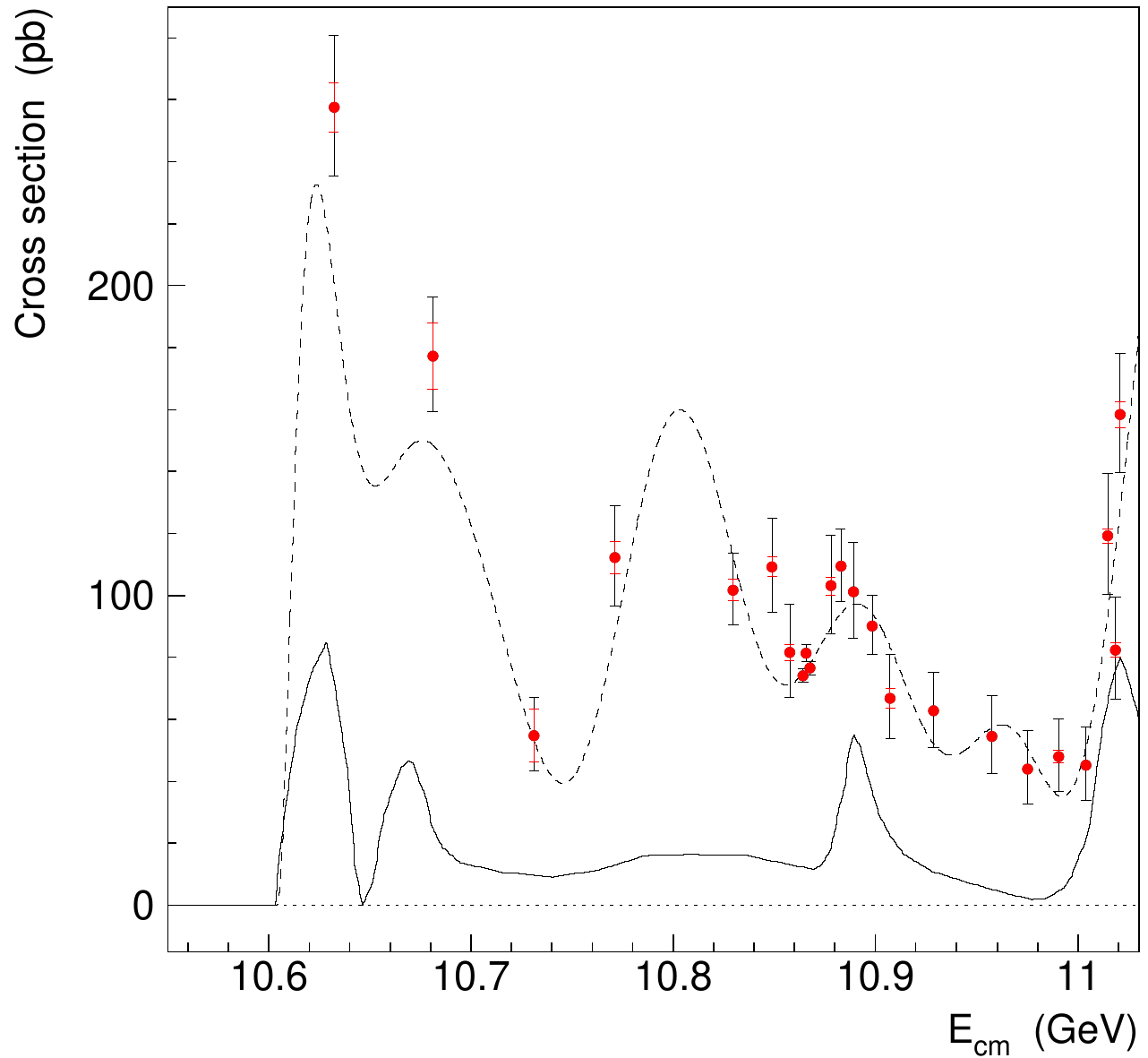}
  \includegraphics[width=0.49\textwidth]{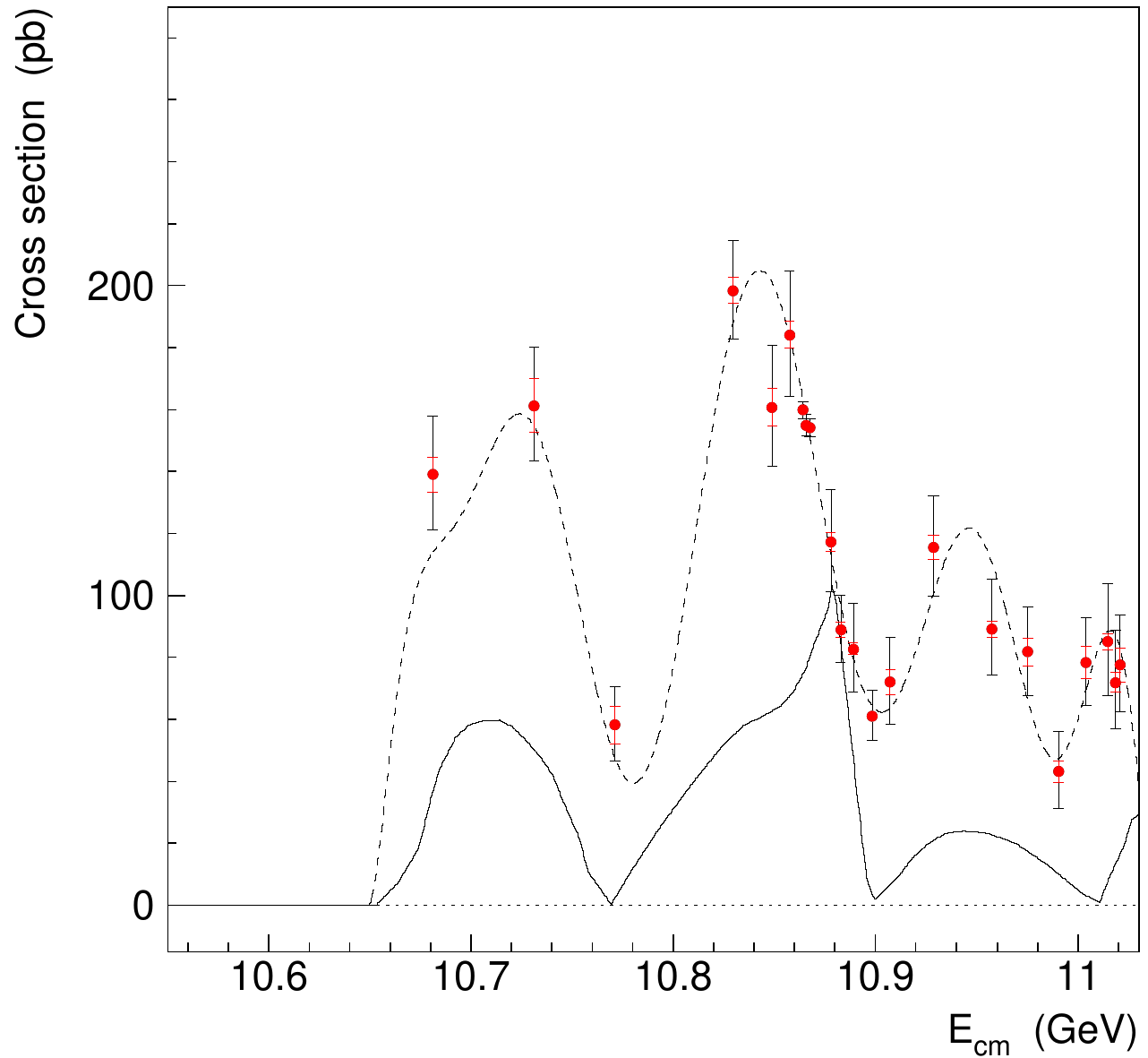}
  \caption{ Measured dressed cross sections at various energies for
    $\ee\to{B}\bar{B}$ (left), $\ee\to{B}\bar{B}^*$ (right), and
    $\ee\to{B}^*\bar{B}^*$ (bottom). Points and dashed curves are the
    same as in Fig.~\ref{bb_xs_vs_ecm_unc_it11_190920}. Solid curve
    shows the predictions of the Unitarized Quark
    Model~\cite{Ono:1985eu}. }
  \label{bb_xs_vs_ecm_theor_081120}
\end{figure}

Figure~\ref{bb_xs_vs_ecm_theor_081120} presents a comparison of the
measured exclusive cross sections with the predictions of the
Unitarized Quark Model (UQM)~\cite{Ono:1985eu}.
The data confirm the prediction that the cross sections show
oscillatory behaviour. Also, there is a rather good agreement in the
positions of the minima, that in the UQM are due to zeros in the
$\U(4S,5S,6S)$ wave functions. The UQM fails to describe the absolute
values of the cross sections. Contrary to the expectations, the cross
sections in the minima are not zero, which suggests that the UQM
misses some general non-resonant offset.

In the UQM there are narrow structures in all the channels that
correspond to the signals of $\Uf$. Data do not show such
structures. Thus, in the final states $\bb$, $\bbst$, and $\bstbst$ we
find no clear $\Uf$ signal. As follows from
Fig.~\ref{cmp_total_sum_exc_it13a_281020}, the narrow peak in the
$\Uf$ region is present in other $b\bar{b}$ final states,
$B_s^{(*)}\bar{B}_s^{(*)}$, $\bball\pi$, and the final states with
bottomonium and light hadrons. This finding contradicts to the
expectations of the potential models that the dominant decay channels
of $\Uf$ are $B^{(*)}\bar{B}^{(*)}$ (see, e.g.,
Ref.~\cite{Godfrey:2015dia}).

The sum of exclusive $B^{(*)}\bar{B}^{(*)}$ cross sections does not
saturate the total $b\bar{b}$ cross section for the energies above the
$B_s^*\bar{B}_s^*$ threshold, as shown in
Fig.~\ref{cmp_total_sum_exc_it13a_281020}. This sets goals for further
studies. It is of interest to measure the energy dependence of the
$\ee\to{B}_s^{(*)}\bar{B}_s^{(*)}$ and
$\ee\to{B}^{(*)}\bar{B}^{(*)}\pi(\pi)$ cross sections. These channels,
together with the $B^{(*)}\bar{B}^{(*)}$ channels measured herein,
should provide complete information for the coupled channel analysis
in the energy region under study.

The polarization of the $\bstbst$ channel is described by three
amplitudes, as discussed in Appendix~\ref{app:angular}. To measure
these amplitudes the reconstruction of a photon from the
$B^*\to{B}\gamma$ decay would be needed. Such a measurement requires
higher statistics than currently available $1\,\fb$ at the scan
points.

The separation between the points is rather large in the low-energy
region.
In particular, the model~\cite{Ono:1985eu} predicts an additional zero
in the $\bb$ and $\bbst$ cross sections which is in the gap between
two scan points. More scan data with smaller energy step sizes and
larger integrated luminosity in this region are needed to understand
reliably the shape of the exclusive cross sections. These data could
be collected by the Belle~II experiment.

\section{\boldmath Measurement of visible cross sections and event
  fractions at $\Uf$}
\label{sec:fs}

As a byproduct, we measure the visible cross sections $\ee\to\bb\,X$,
$\ee\to\bb$, $\ee\to\bbst$, and $\ee\to\bstbst$ at $\Uf$, as well as
corresponding fractions of events and the fraction of
$B_s^{(*)}\bar{B}_s^{(*)}$ events $f_s$.

To find the inclusive $\ee\to\bb\,X$ cross section, we use the same
method as for the measurement of the efficiency at $\Uf$
(Section~\ref{sec:eff}). The only difference is that the requirement
$\mbc<5.35\,\gevm$ is not applied; thus, we consider not only two-body
$\ee\to{B}^{(*)}\bar{B}^{(*)}$ but also multi-body
$\ee\to{B}^{(*)}\bar{B}^{(*)}\pi(\pi)$ processes.
The ratio of the $B$ meson yields in the $\Uf$ and $\Ufo$ SVD2 data
samples, averaged over the five low-multiplicity $B$-decay channels
that were used in Section~\ref{sec:eff}, is
\begin{equation}
  \rafi = 0.0503\pm0.0012,
\end{equation}
where the uncertainty includes statistical and systematic
contributions. 
For the cross section we find:
\begin{equation}
  \sigma^\mathrm{vis}(\ee\to\bb\,X) =
  \frac{\rafi\,\times\,N_{\bb}(\Ufo)}{L} = 
  (255.5\pm7.9)\,\mathrm{pb},
  \label{eq:sig_vis_bbx}
\end{equation}
where $L$ is the total integrated luminosity of the three
high-statistics points in Table~\ref{tab:xsec_results}. The fraction
of $\bb\,X$ events is
\begin{equation}
  f_{\bb\,X} = \frac{\sigma^\mathrm{vis}(\ee\to\bb\,X)}
  {\sigma_{b\bar{b}}} = 0.751\pm0.040,
\end{equation}
where $\sigma_{b\bar{b}}=(340\pm16)\,\mathrm{pb}$ is the total
$b\bar{b}$ cross section at $\Uf$~\cite{Esen:2012yz}.
Here and in the following we take into account that in the ratio of
cross sections the uncertainty due to the integrated luminosity
cancels.
The remaining events,
$f_{\mathrm{non}-\bb\,X}=1-f_{\bb\,X}=0.249\pm0.040$, contain $B_s$
mesons or bottomonia with light hadrons.

To estimate the fraction of bottomonium events, we consider all final
states with bottomonium listed in PDG~2020~\cite{PDG}. These are
$\U(nS)\pp$ ($n=1,2,3$), $\Uo\kk$, $\U(1D)\eta$, $\hbn\pp$ ($n=1,2$),
and $\chi_{bJ}\,\pp\pi^0$. The sum of corresponding fractions is
$(3.50^{+0.40}_{-0.42})\%$, where we assume that the uncertainties in
various fractions are uncorrelated.
Using isospin relations, we account also for the final states with
neutrals: the fractions for the $\pp$ transitions are multiplied by
1.5, while the fraction for the $\kk$ transition is multiplied by
2.0. The resulting sum is $(4.92^{+0.53}_{-0.56})\%$.
Belle reported preliminary results on the $\Un\eta$ ($n=1,2$) and
$\U(1D)\pp$ transitions~\cite{Krokovny:LaThuile2012} that show that
corresponding fractions are not large. There are also rather strict
upper limits on fractions of the $\eta_b(nS)\omega$
($n=1,2$)~\cite{Oskin:2020igx} and $\hbn\eta$
($n=1,2$)~\cite{Tamponi:2018cuf} transitions.
However, there are still channels for which no experimental
information is available. Among them are $4\pi$ transitions to $\Un$,
$\hbn$ and $\U(1D)$, as well as
$\Uf\to{Z}_b\,\pi\to\eta_b(1S)\rho\pi$.
To estimate the total fraction of bottomonium, we assume that all the
channels that are not in PDG 2020 contribute no more than already
measured channels and thus assign a large positive uncertainty:
\begin{equation}
  f_{\mathrm{bottomonium}} = (4.9^{+5.0}_{-0.6})\%.
\end{equation}

Finally, we estimate the fraction of the events with $B_s$ mesons: 
\begin{equation}
  f_s = f_{\mathrm{non}-\bb\,X} - f_{\mathrm{bottomonium}} = 0.200^{+0.040}_{-0.064},
\end{equation}
where the uncertainty includes statistical and systematic contributions. 
This result is consistent with the PDG 2020 value of
$0.201\pm0.031$~\cite{PDG} and the value that follows from the
measurement of the $\ee\to{B}_s^{(*)}\bar{B}_s^{(*)}$ cross section in
Ref~\cite{Oswald:2015dma}:
\begin{equation}
  f_s =
  \frac{\sigma(\ee\to{B}_s^{(*)}\bar{B}_s^{(*)})}{\sigma_{b\bar{b}}} =
  0.158 \pm 0.017.
\end{equation}
    
To measure the visible $\ee\to\bball$ cross sections, we use the
formula: 
\begin{equation}
\sigma^\mathrm{vis}(\ee\to\bball) = 
\frac{N_{\bball}\,\times\,k_{\bball}}
     {2\,\times\,\varepsilon_{\bball}\,\times\,L},
\end{equation}
where $N_{\bball}$ is the $\bball$ signal yield in the interval
$5.27<\mbc<5.35\,\mev$ (Table~\ref{tab:y5s_fit_results}). The factor
$k_{\bball}=1.042$, 1.120, and 1.089 for $\bb$, $\bbst$, and $\bstbst$,
respectively, accounts for the ISR tail in the $\mbc>5.35\,\mev$
region. For the $\bb$ channel, $k_{\bb}$ includes the ISR contribution
down to the $\bbst$ threshold, while the region between the $\bb$ and
$\bbst$ thresholds (the $\Ufo$ region) is accounted for
separately. For the $\bbst$ and $\bstbst$ channels, $k_{\bball}$
include the ISR contributions down to the corresponding thresholds.
The efficiency $\varepsilon_{\bball}$ is equal to
$\varepsilon_{\Uf}\,\times\,r_{\bball}$, where the factor
$r_{\bball}=1.010$, 1.004, and 0.998 for $\bb$, $\bbst$, and
$\bstbst$, respectively, accounts for the small variation of the
efficiency with the momentum of the $B$ meson
(Section~\ref{sec:eff}). Using Eq.~(\ref{eq:eff_y5s}), we obtain
\begin{equation}
  \sigma^\mathrm{vis}(\ee\to\bball) =
  \frac{N_{\bball}}{N_\mathrm{total}}\,
  k_{\bball}\,
  \frac{N_{\bb}(\Ufo)\,\times\,\rafilo}{r_{\bball}\,\times\,L},
  \label{eq:sig_vis_bball}
\end{equation}
here $N_{\bball}\,/\,N_\mathrm{total}$ are the fractions of various
channels given in Table~\ref{tab:y5s_fit_results}.

We study the systematic uncertainty in the values of the expression
$(N_{\bball}\,/\,N_\mathrm{total})\times{k}_{\bball}$ in the same way
as described in Section~\ref{sec:y5s} for $N_\mathrm{total}$. The
results are presented in Table~\ref{tab:syst_y5s_frac}.
\begin{table}[htbp]
\caption{ Systematic uncertainties in
  $(N_{\bball}\,/\,N_\mathrm{total})\times{k}_{\bball}$
  for various channels (in \%).} 
\renewcommand*{\arraystretch}{1.2}
\label{tab:syst_y5s_frac}
\centering
\begin{tabular}{@{}lccc@{}} \toprule
  & $\bb$ & $\bbst$ & $\bstbst$ \\
  \midrule
  Cross section shape:              &      &      &      \\
  \;\;\;-- statistical uncertainty  & 2.03 & 1.60 & 1.11 \\
  \;\;\;-- parameterization         & 1.56 & 2.83 & 2.38 \\
  $\ecm$ spread                     & 0.20 & 0.03 & 0.05 \\ 
  Yield of peaking background       & 0.12 & 0.04 & 0.05 \\ 
  Shape of peaking background       & 0.76 & 0.11 & 0.21 \\ 
  Shape of smooth background        & 0.28 & 0.09 & 0.10 \\
  \midrule
  Total                             & 2.70 & 3.25 & 2.64 \\
  \bottomrule
\end{tabular}
\end{table}

The contribution of the ISR events in the $\Ufo$ region is obtained by
integrating the ISR kernel multiplied by the dressed cross section
shown in Fig.~\ref{xsec_fit_y4s_290820} in the region between the $\bb$
and $\bbst$ thresholds. The resulting value is
$(10.53\pm0.31)\,\mathrm{pb}$, where the uncertainty includes the
contributions from the statistical (1.2\%) and systematic (2.6\%)
errors in $R_b$~\cite{Aubert:2008ab}, and the systematic uncertainty
due to the variation of $\ez$ (0.7\%).
The values of the measured visible cross sections are presented in
Table~\ref{tab:vis_frac}. We show also corresponding event fractions.
\begin{table}[htbp]
\caption{ Visible cross sections $\sigma^\mathrm{vis}$ (in pb) for
  various processes at $\Uf$ and corresponding
  $\sigma^\mathrm{vis}/\,\sigma_{b\bar{b}}$ fractions (in \%). The
  $\bb\,X$ final state includes $\bball$ and $\bball\pi(\pi)$. The
  errors contain the statistical and systematic contributions.}
\renewcommand*{\arraystretch}{1.2}
\label{tab:vis_frac}
\centering
\begin{tabular}{@{}ccc@{}} \toprule
  & $\sigma^\mathrm{vis}$ & $\sigma^\mathrm{vis}/\,\sigma_{b\bar{b}}$ \\ 
  \midrule
  $\ee\to\bb\,X$  & $255.5 \pm 7.9$ & $75.1 \pm 4.0$ \\
  $\ee\to\bb$     &  $33.3 \pm 1.2$ &  $9.8 \pm 0.5$ \\
  $\ee\to\bbst$   &  $68.0 \pm 3.3$ & $20.0 \pm 1.3$ \\
  $\ee\to\bstbst$ & $124.4 \pm 5.3$ & $36.6 \pm 2.2$ \\
  \bottomrule
\end{tabular}
\end{table}
Previous Belle measurement of these values~\cite{Drutskoy:2010an} did
not take into account the ISR tails of the signals. The results shown
in Table~\ref{tab:vis_frac} supersede those in
Ref.~\cite{Drutskoy:2010an}.

We also measure the ratio
$\sigma^\mathrm{vis}(\ee\to\bbst)/\sigma^\mathrm{vis}(\ee\to\bb\,X)$,
which is needed for the measurement of $f_s$ using lepton-charge
correlations in the dilepton events~\cite{Sia:2006cq}. Based on
Eqs.~(\ref{eq:sig_vis_bbx}) and (\ref{eq:sig_vis_bball}), we find
\begin{equation}
  \frac{\sigma^\mathrm{vis}(\ee\to\bbst)}{\sigma^\mathrm{vis}(\ee\to\bb\,X)}
  = \frac{N_{\bball}}{N_\mathrm{total}}\frac{1}{r_{\bball}}
  \frac{k_{\bball}}{R_\mathrm{5\,chan}^\mathrm{all}/R_\mathrm{5\,chan}}
  = 0.2664 \pm 0.0101,
  \label{eq:rat_bball_bbx}
\end{equation}
where the uncertainty includes statistical and systematic
contributions. While calculating the ratio
$R_\mathrm{5\,chan}^\mathrm{all}/R_\mathrm{5\,chan}=0.2962\pm0.0176$,
we take into account the correlation of the corresponding
uncertainties.

\section{Conclusions}
\label{sec:concl}

To conclude, we report the first measurement of the energy dependence
of the exclusive cross sections, $\ee\to\bb$, $\ee\to\bbst$, and
$\ee\to\bstbst$ in the region from $10.63$ to $11.02\,\gev$. The
results are presented in Table~\ref{tab:xsec_results} and
Fig.~\ref{bb_xs_vs_ecm_unc_it11_190920}. The cross sections show
non-trivial behavior with several maxima and minima.
They can be used in future phenomenological studies to shed light on
$b\bar{b}$-quark and $B^{(*)}\bar{B}^{(*)}$-meson interactions in this
energy region.

As a byproduct, we measure at $\Uf$ the fraction of the events
containing the $B_s$ mesons, $f_s=0.200^{+0.040}_{-0.064}$, where the
error contains statistical and systematic contributions. We measure also
the visible cross sections and corresponding fractions of the events for
the $\bb\,X$, $\bb$, $\bbst$ and $\bstbst$ final states
(Table~\ref{tab:vis_frac} and Eq.~(\ref{eq:rat_bball_bbx})).

\acknowledgments

We thank the KEKB group for the excellent operation of the
accelerator; the KEK cryogenics group for the efficient
operation of the solenoid; and the KEK computer group, and the Pacific Northwest National
Laboratory (PNNL) Environmental Molecular Sciences Laboratory (EMSL)
computing group for strong computing support; and the National
Institute of Informatics, and Science Information NETwork 5 (SINET5) for
valuable network support.  We acknowledge support from
the Ministry of Education, Culture, Sports, Science, and
Technology (MEXT) of Japan, the Japan Society for the 
Promotion of Science (JSPS), and the Tau-Lepton Physics 
Research Center of Nagoya University; 
the Australian Research Council including grants
DP180102629, 
DP170102389, 
DP170102204, 
DP150103061, 
FT130100303; 
Austrian Federal Ministry of Education, Science and Research (FWF) and
FWF Austrian Science Fund No.~P~31361-N36;
the National Natural Science Foundation of China under Contracts
No.~11435013,  
No.~11475187,  
No.~11521505,  
No.~11575017,  
No.~11675166,  
No.~11705209;  
Key Research Program of Frontier Sciences, Chinese Academy of Sciences (CAS), Grant No.~QYZDJ-SSW-SLH011; 
the  CAS Center for Excellence in Particle Physics (CCEPP); 
the Shanghai Pujiang Program under Grant No.~18PJ1401000;  
the Shanghai Science and Technology Committee (STCSM) under Grant No.~19ZR1403000; 
the Ministry of Education, Youth and Sports of the Czech
Republic under Contract No.~LTT17020;
Horizon 2020 ERC Advanced Grant No.~884719 and ERC Starting Grant No.~947006 ``InterLeptons'' (European Union);
the Carl Zeiss Foundation, the Deutsche Forschungsgemeinschaft, the
Excellence Cluster Universe, and the VolkswagenStiftung;
the Department of Atomic Energy (Project Identification No. RTI 4002) and the Department of Science and Technology of India; 
the Istituto Nazionale di Fisica Nucleare of Italy; 
National Research Foundation (NRF) of Korea Grant
Nos.~2016R1\-D1A1B\-01010135, 2016R1\-D1A1B\-02012900, 2018R1\-A2B\-3003643,
2018R1\-A6A1A\-06024970, 2018R1\-D1A1B\-07047294, 2019K1\-A3A7A\-09033840,
2019R1\-I1A3A\-01058933;
Radiation Science Research Institute, Foreign Large-size Research Facility Application Supporting project, the Global Science Experimental Data Hub Center of the Korea Institute of Science and Technology Information and KREONET/GLORIAD;
the Polish Ministry of Science and Higher Education and 
the National Science Center;
the Ministry of Science and Higher Education of the Russian Federation, Agreement 14.W03.31.0026, 
and the HSE University Basic Research Program, Moscow; 
University of Tabuk research grants
S-1440-0321, S-0256-1438, and S-0280-1439 (Saudi Arabia);
the Slovenian Research Agency Grant Nos. J1-9124 and P1-0135;
Ikerbasque, Basque Foundation for Science, Spain;
the Swiss National Science Foundation; 
the Ministry of Education and the Ministry of Science and Technology of Taiwan;
and the United States Department of Energy and the National Science Foundation.

\appendix

\section{\boldmath $B$ and $D$ decay channels}
\label{app:decay_chan}

The $B$- and $D$-meson decay channels that are used in this analysis
are listed in Tables~\ref{tab:fei_b_chan} and \ref{tab:fei_d_chan}. 
\begin{table}[htbp]
  \caption{ Decay channels of $B^+$ and $B^0$ used in FEI. }
  \label{tab:fei_b_chan}
  \centering
  \begin{tabular}{@{}ll@{}} \toprule
    $B^+\to$ & $B^0\to$ \\
    \midrule
    $\bar{D}^0\pi^+$ & $D^-\pi^+$  \\
    $\bar{D}^0\pi^+\pi^+\pi^-$ & $D^-\pi^+\pi^+\pi^-$ \\
    $\bar{D}^{*0}\pi^+$ & $D^{*-}\pi^+$ \\
    $\bar{D}^{*0}\pi^+\pi^+\pi^-$ & $D^{*-}\pi^+\pi^+\pi^-$ \\
    \midrule
    $D_s^{+}\bar{D}^{0}$ & $D_s^{+}D^{-}$ \\
    $D_s^{*+}\bar{D}^{0}$ & $D_s^{*+}D^{-}$ \\
    $D_s^{+}\bar{D}^{*0}$ & $D_s^{+}D^{*-}$ \\
    $D_s^{*+}\bar{D}^{*0}$ & $D_s^{*+}D^{*-}$ \\
    \midrule
    $J/\psi \, K^+$ & $J/\psi \, \ks$ \\
    $J/\psi \, \ks\,\pi^+$ & $J/\psi \, K^+\pi^-$ \\
    $J/\psi \, K^+\pi^+\pi^-$ & \\
    \midrule
    $D^-\pi^+\pi^+$ & $D^{*-}K^+K^-\pi^+$ \\
    $D^{*-}\pi^+\pi^+$ & \\
    \bottomrule
  \end{tabular}
\end{table}
\begin{table}[htbp]
  \caption{ Decay channels of $D^0$, $D^+$ and $D_s^+$ used in FEI. }
  \label{tab:fei_d_chan}
  \centering
  \begin{tabular}{@{}lll@{}} \toprule
    $D^0\to$ & $D^+\to$ & $D_s^+\to$ \\ \midrule
    $K^-\pi^+$ & $K^-\pi^+\pi^+$ & $K^+K^-\pi^+$ \\
    $K^-\pi^+\pi^0$ & $K^-\pi^+\pi^+\pi^0$ & $K^+\ks$ \\
    $K^-\pi^+\pi^+\pi^-$ & $\ks\,\pi^+$ & $K^+K^-\pi^+\pi^0$ \\
    $\ks\,\pi^+\pi^-$ & $\ks\,\pi^+\pi^0$ & $K^+\ks\,\pi^+\pi^-$ \\
    $\ks\,\pi^+\pi^-\pi^0$ & $\ks\,\pi^+\pi^+\pi^-$ & $K^-\ks\,\pi^+\pi^+$ \\
    $K^+K^-$ & $K^+K^-\pi^+$ & $K^+K^-\pi^+\pi^+\pi^-$ \\
    $K^+K^-\ks\,$ &  & $K^+\pi^+\pi^-$ \\
    & & $\pi^+\pi^+\pi^-$ \\
    \bottomrule
  \end{tabular} 
\end{table}

\section{\boldmath Angular analysis at $\Uf$}
\label{app:angular}

\begin{figure}[htbp]
  \centering
  \includegraphics[width=0.45\textwidth]{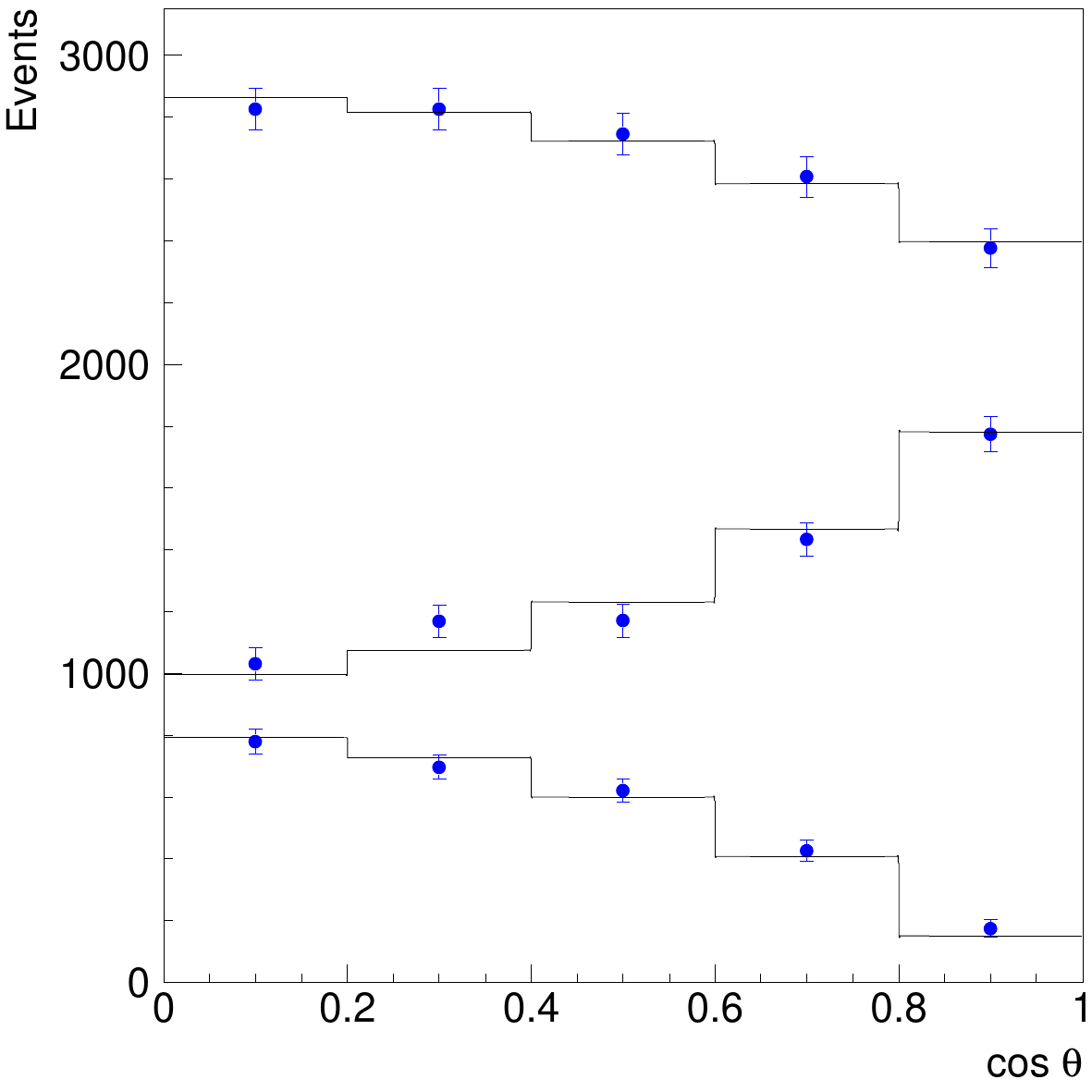}
  \caption{The yield of $\bstbst$ (top set of points), $\bbst$
    (middle) and $\bb$ (bottom set of points) as a function of the
    polar angle of the $B$ meson. The histograms show the fit
    results. }
  \label{nb_vs_coth_081020}
\end{figure}

As a consistency check, we measure signal yields in various intervals
of the $B$ meson polar angle (Fig.~\ref{nb_vs_coth_081020}).
From the MC simulation we find that the variation of the efficiency
with $\cos\theta$ can be neglected. 
The expected distributions for $\bb$ and $\bbst$ are $\sin^2\theta$
and $1+\cos^2\theta$, respectively. The data agree with these
expectations well, the p-values of the corresponding fits are 69\% and
24\%, respectively.

The $\bstbst$ pairs can be produced in three states: $L=1$, $S=2$;
$L=3$, $S=2$; $L=1$, $S=0$, where $L$ is the orbital angular momentum
and $S$ the total spin of the $\bstbst$ pair. The expected total polar
angle distribution is $1+b\cos^2\theta$, with $-1\leq{b}\leq1$. From
the fit we find $b=-0.20\pm0.03$, the p-value of the fit is 88\%.

We measure the parameter $\ah$ in various intervals of $\cos\theta$
and do not find a significant variation.

Currently available experimental information on the $\ee\to\bstbst$
process is insufficient to determine its production amplitudes. To
determine the polarization, reconstruction of $\gamma$ from the
$B^*\to{B}\gamma$ decay would be necessary.

\section{\boldmath Fits to $\mbc$ distributions at various energies}
\label{app:mbc_fits}

\begin{figure}[htbp]
  \centering
  \includegraphics[width=0.46\textwidth]{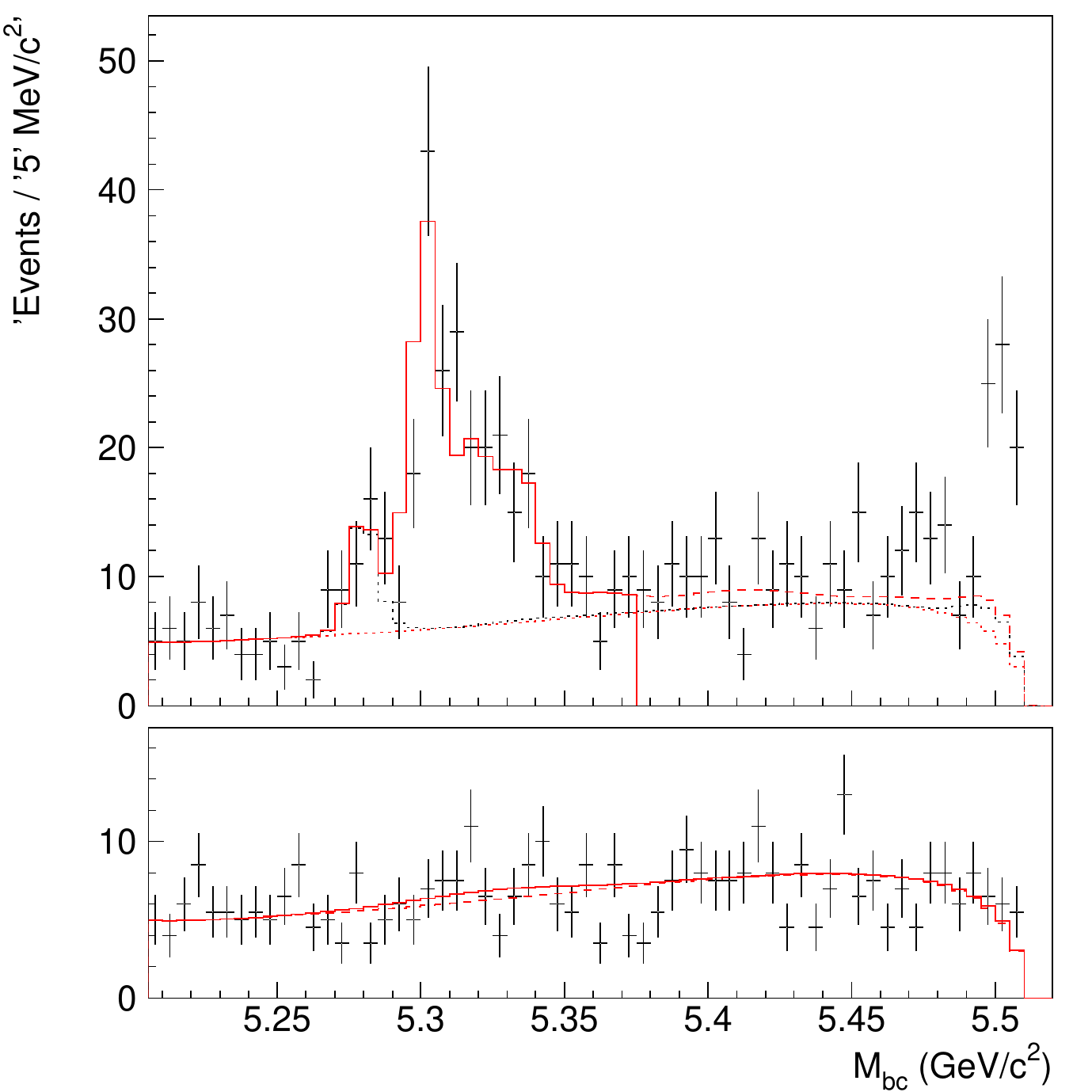}\hfill
  \includegraphics[width=0.46\textwidth]{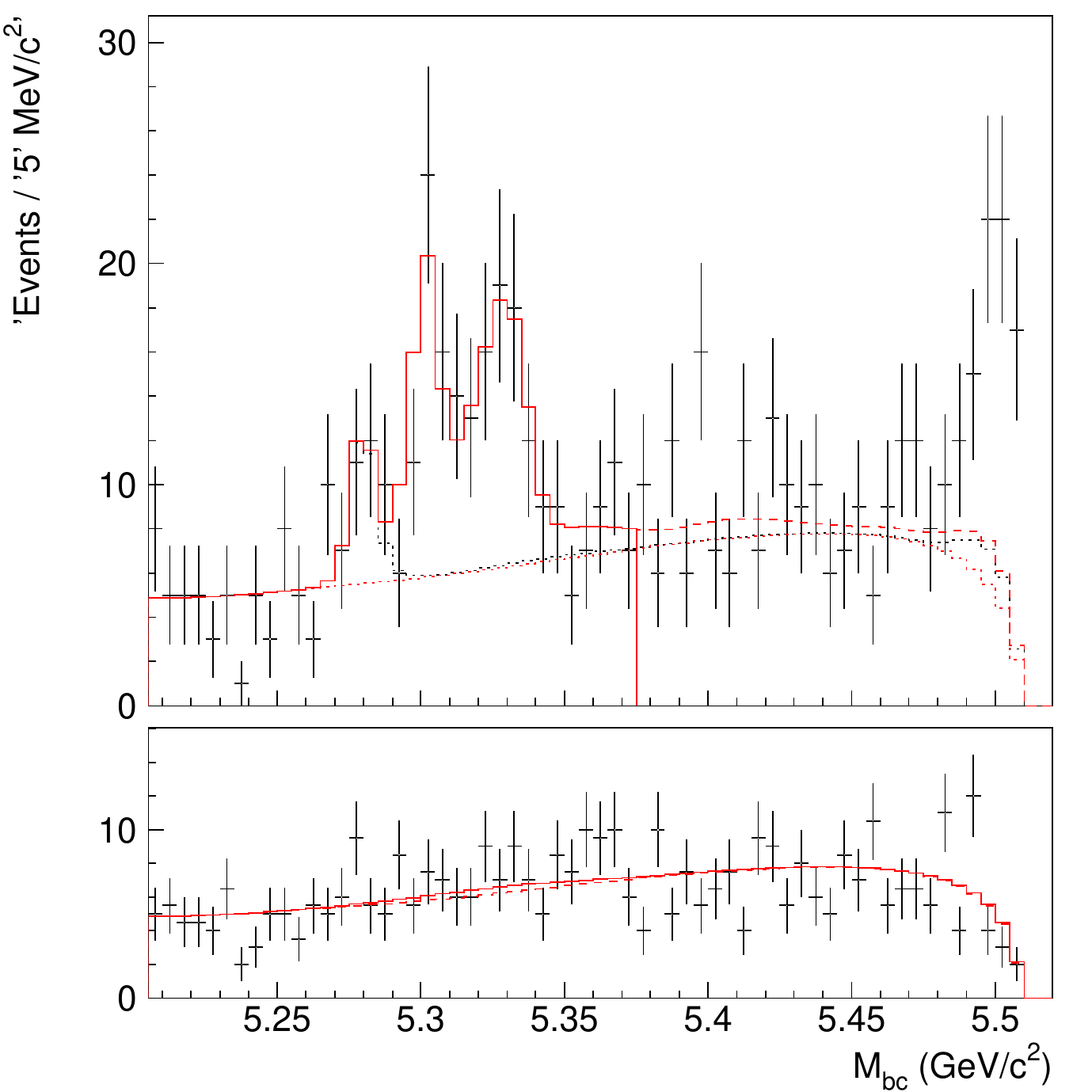}
  \includegraphics[width=0.46\textwidth]{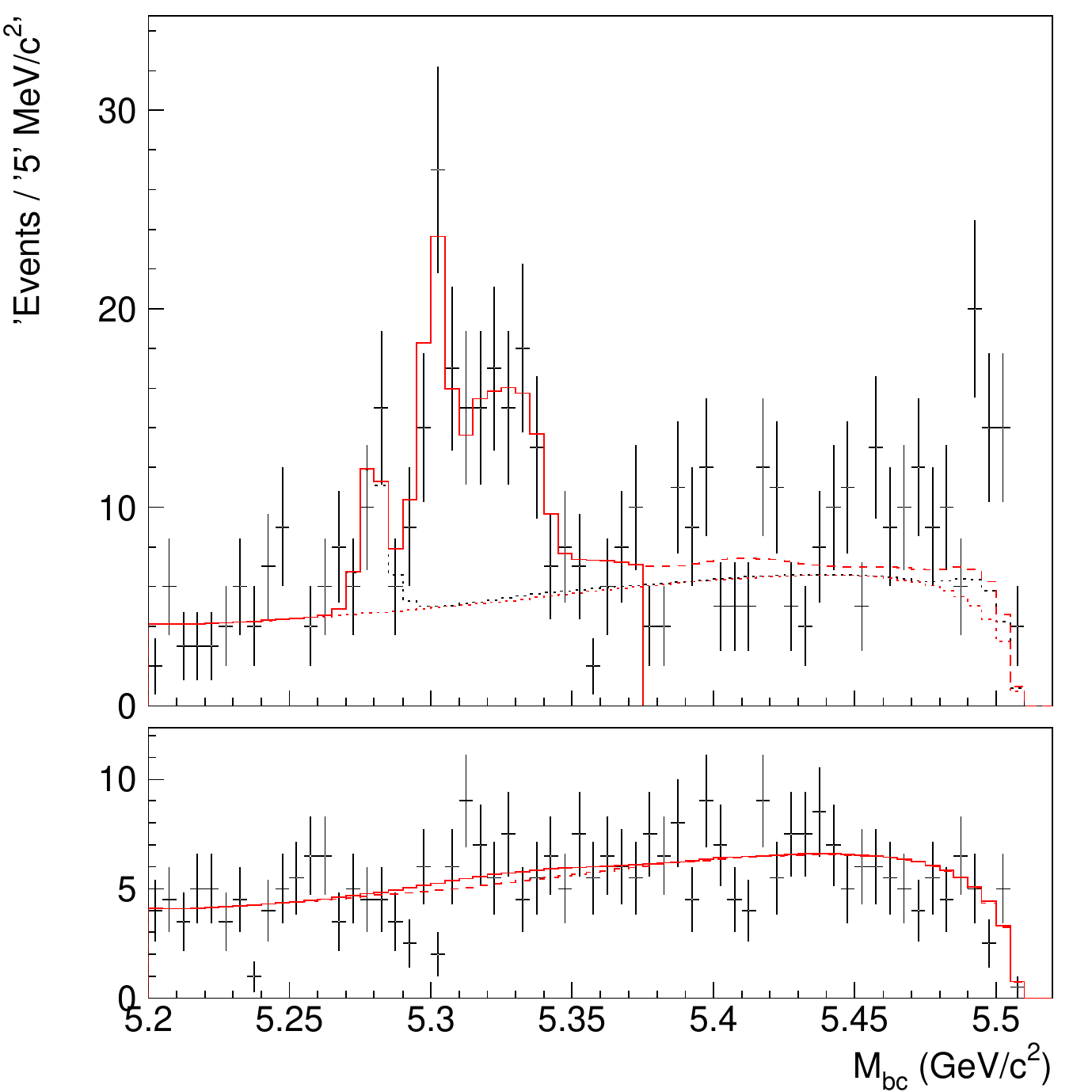}\hfill
  \includegraphics[width=0.46\textwidth]{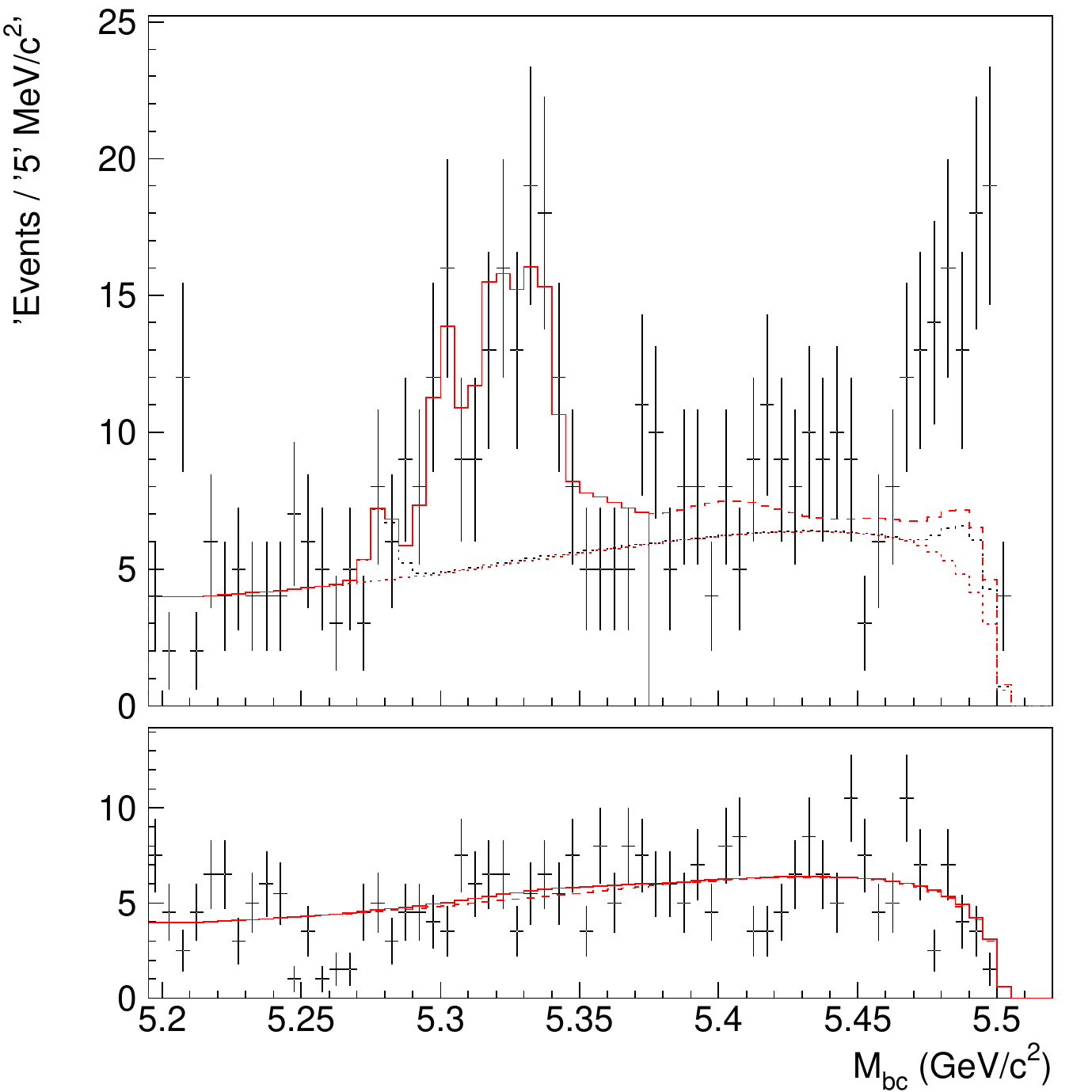}
  \includegraphics[width=0.46\textwidth]{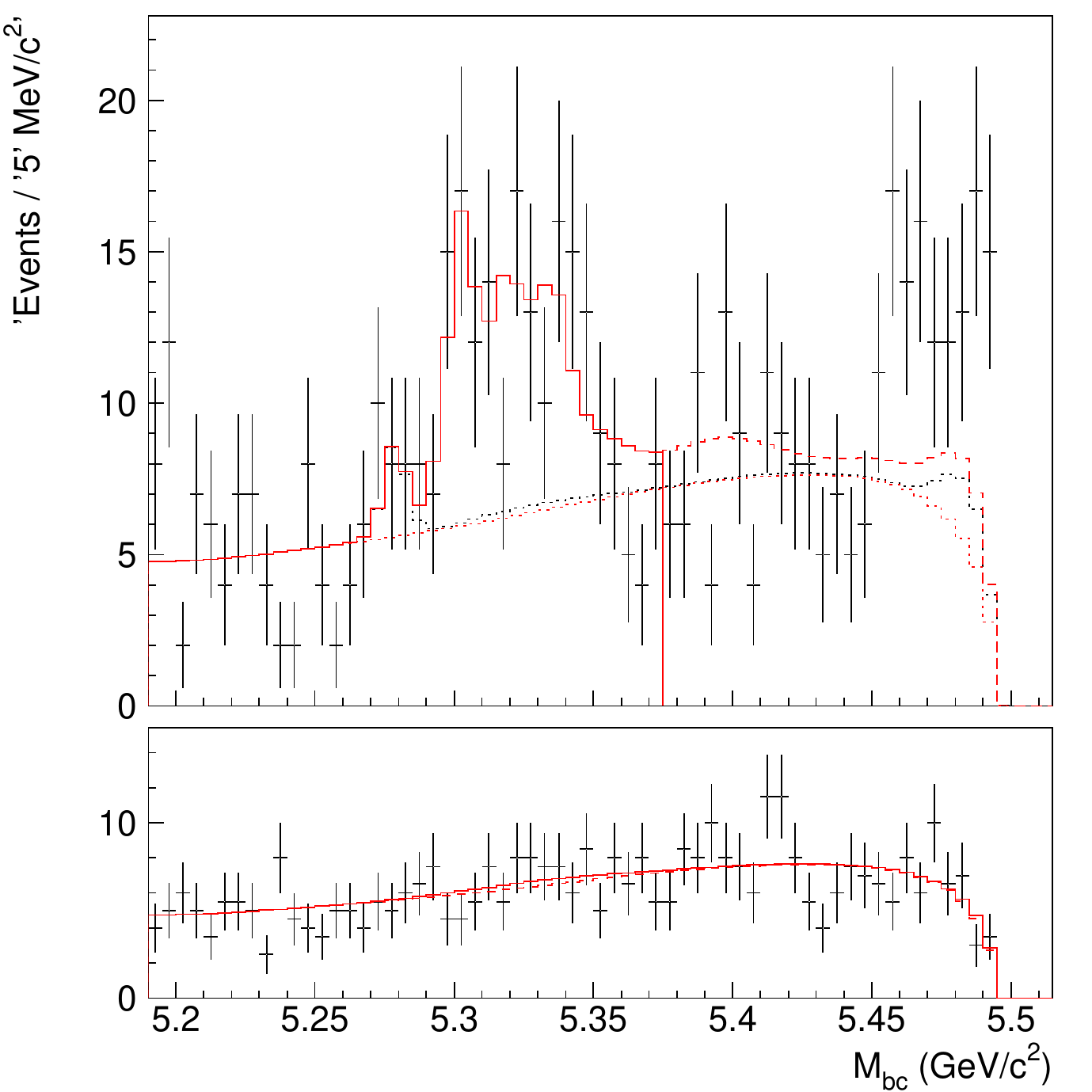}\hfill
  \includegraphics[width=0.46\textwidth]{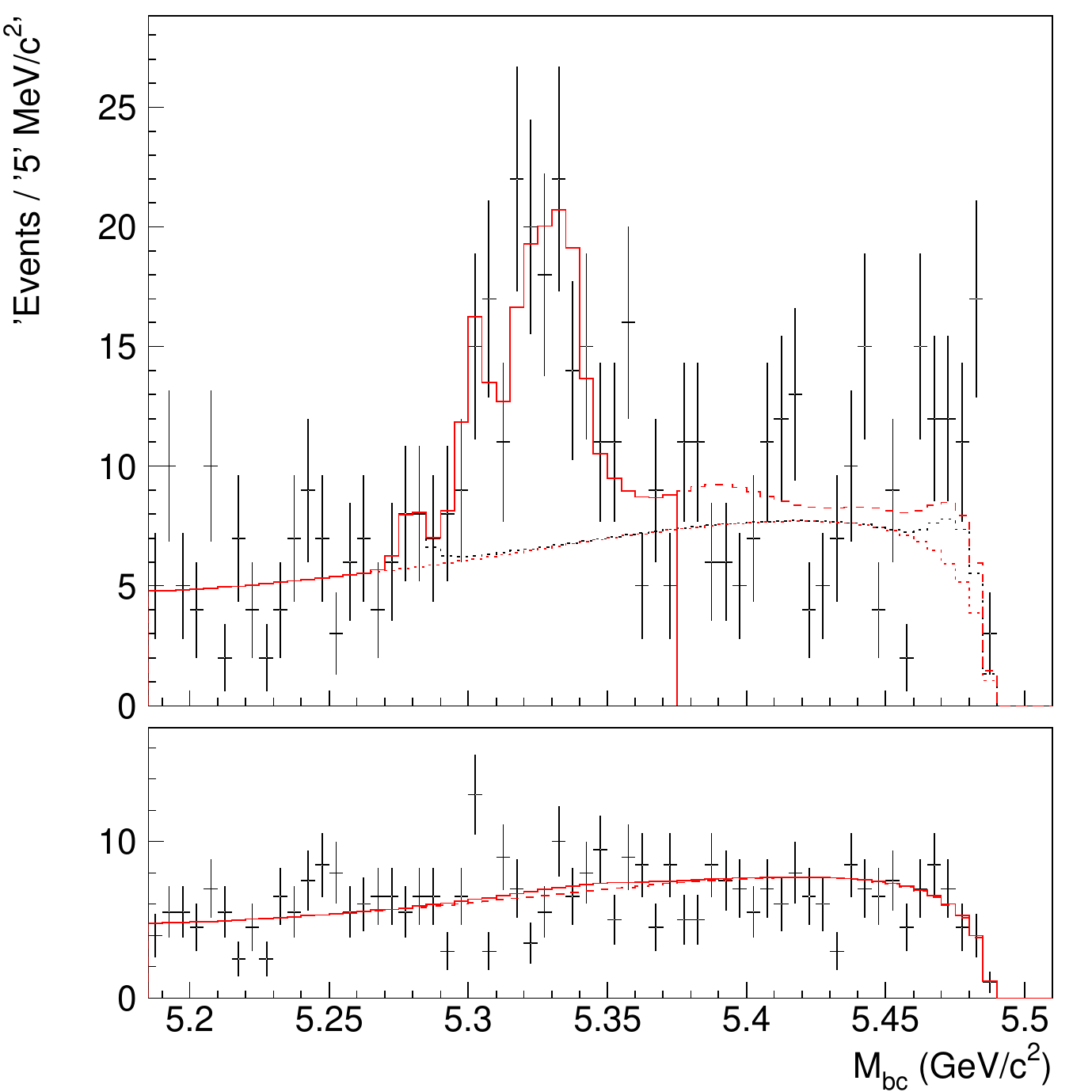}
  \caption{ The $\mbc$ distributions for the points 1 to 6 in
    Table~\ref{tab:xsec_results} (from left to right and from top to
    bottom). Legend is the same as in
    Fig.~\ref{mbc_y5s_it10_140920}. }
  \label{mbc_fit_1}
\end{figure}

\begin{figure}[htbp]
  \centering
  \includegraphics[width=0.46\textwidth]{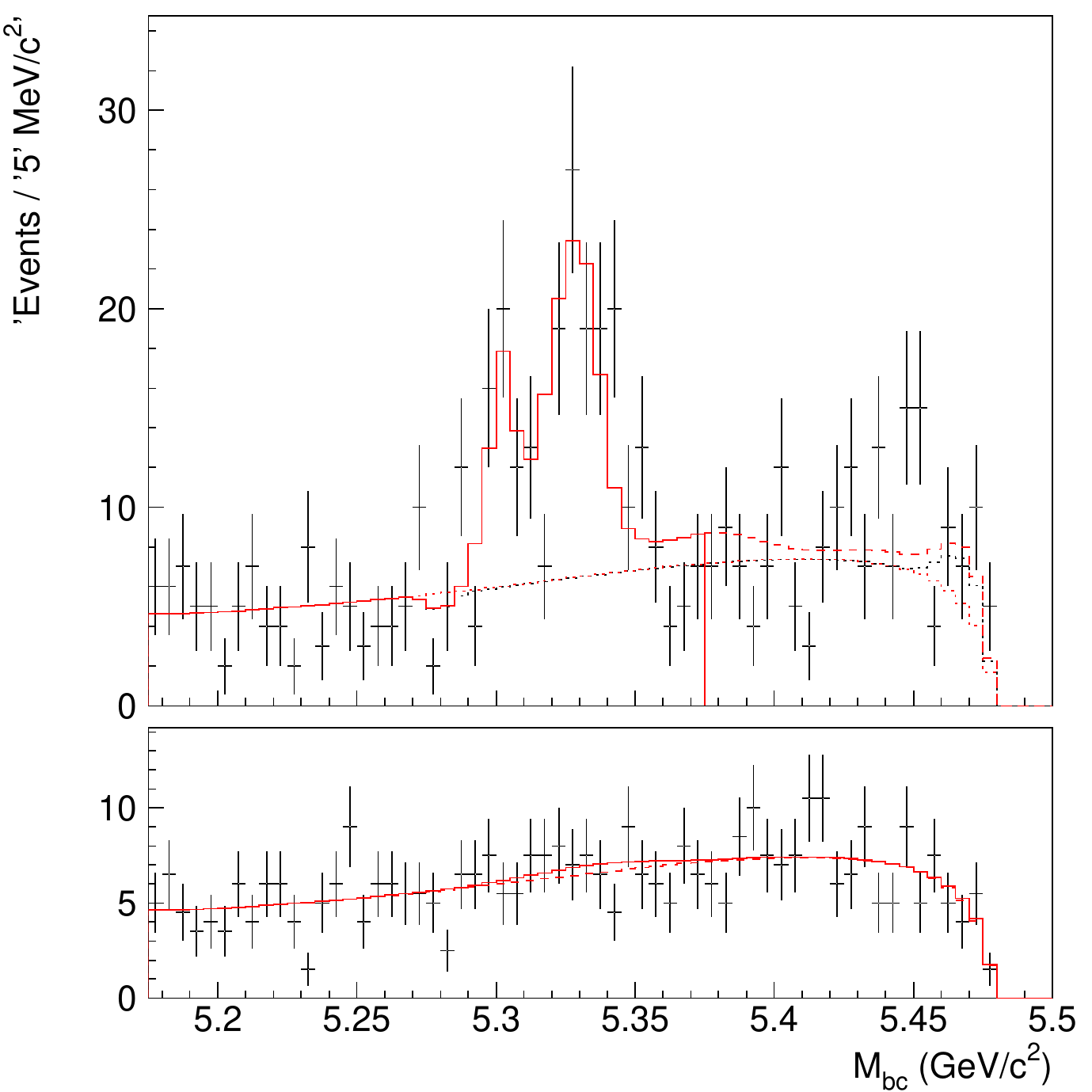}\hfill
  \includegraphics[width=0.46\textwidth]{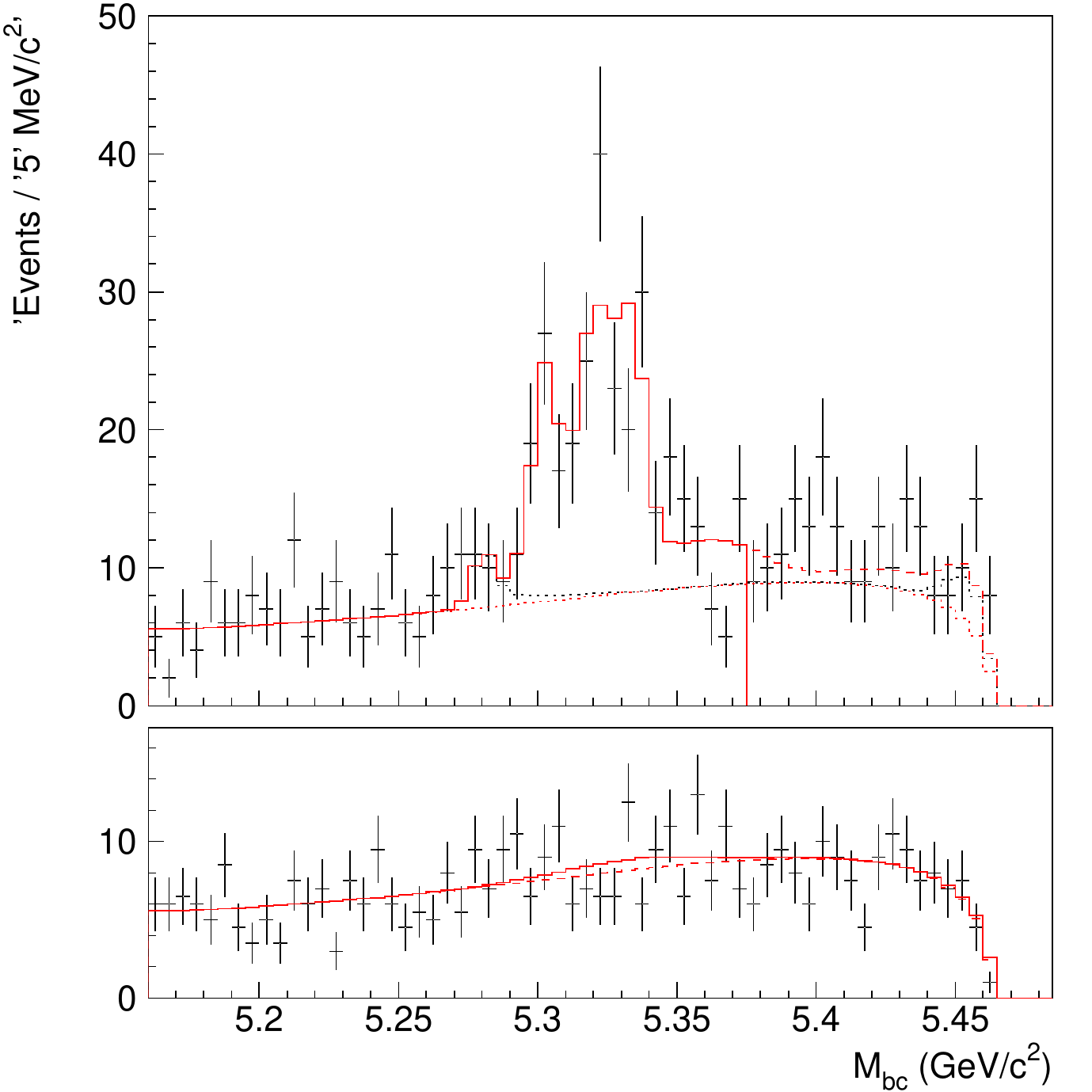}
  \includegraphics[width=0.46\textwidth]{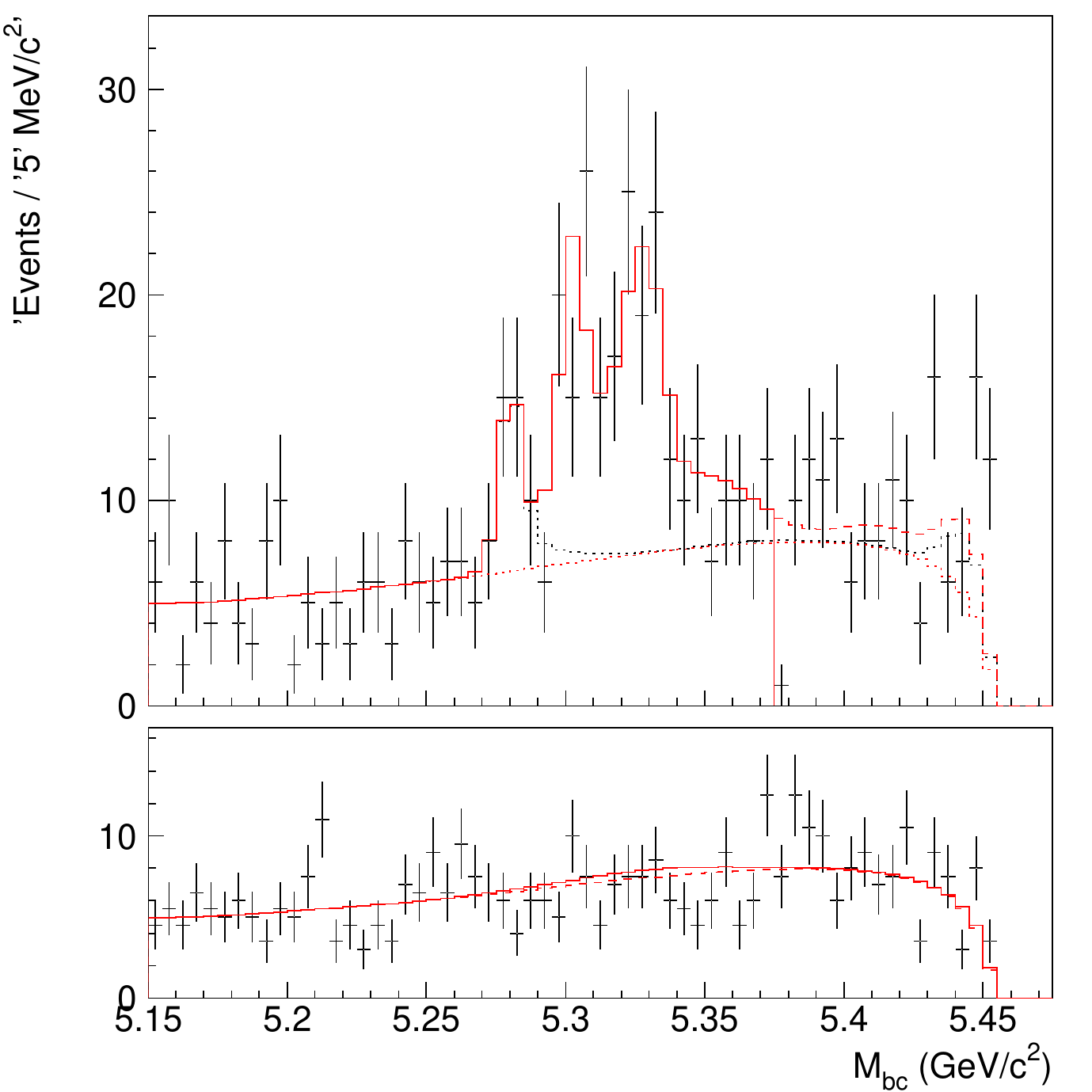}\hfill
  \includegraphics[width=0.46\textwidth]{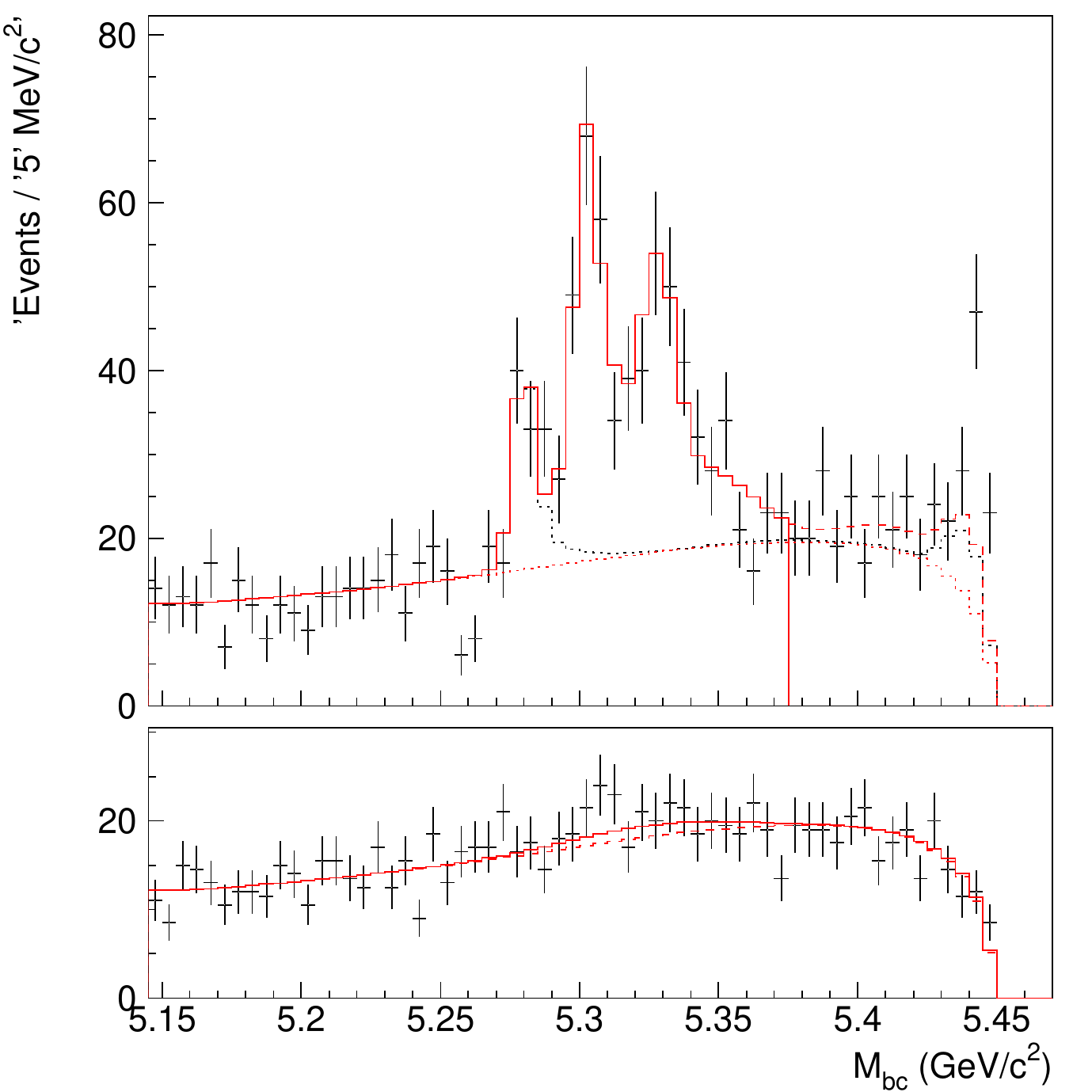}
  \includegraphics[width=0.46\textwidth]{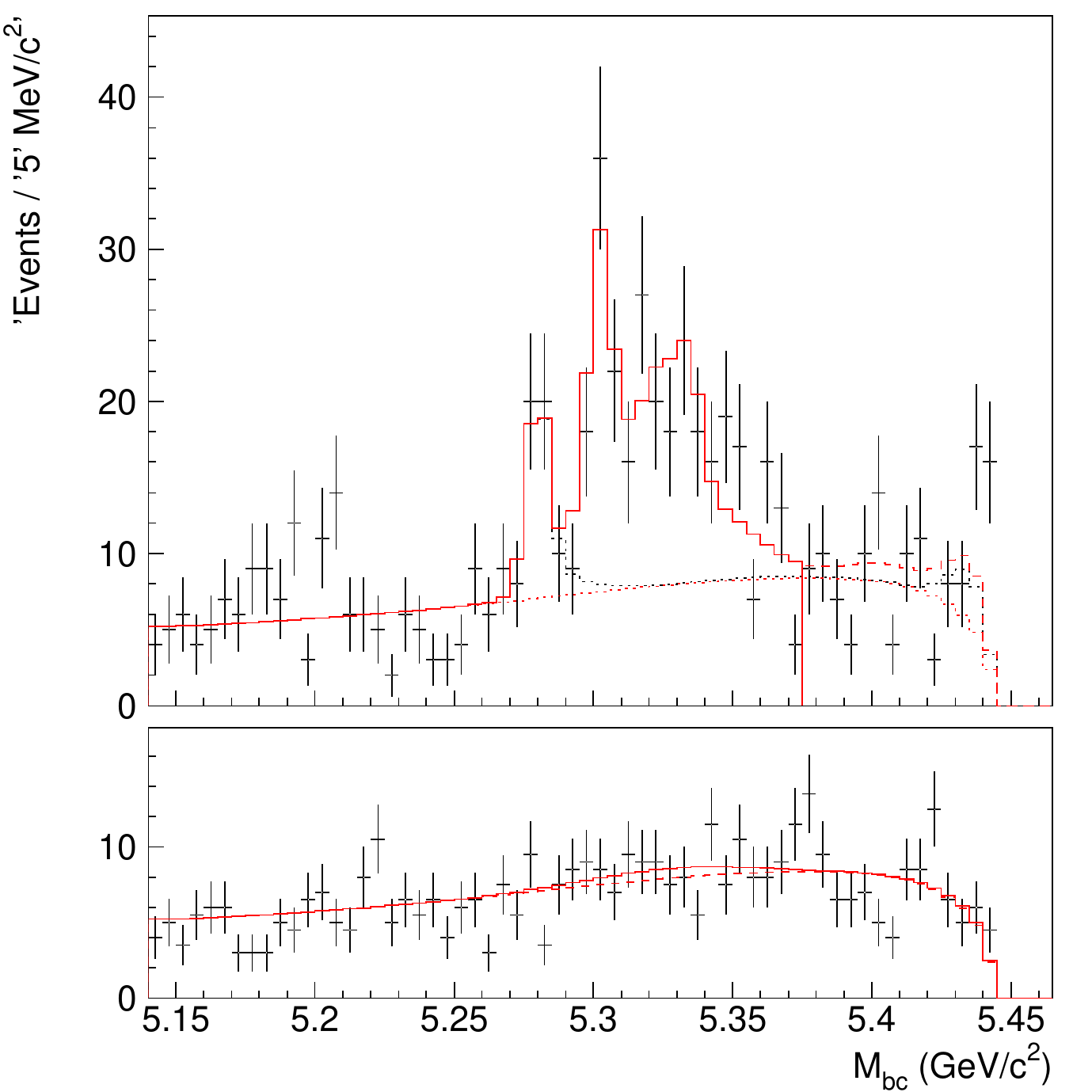}\hfill
  \includegraphics[width=0.46\textwidth]{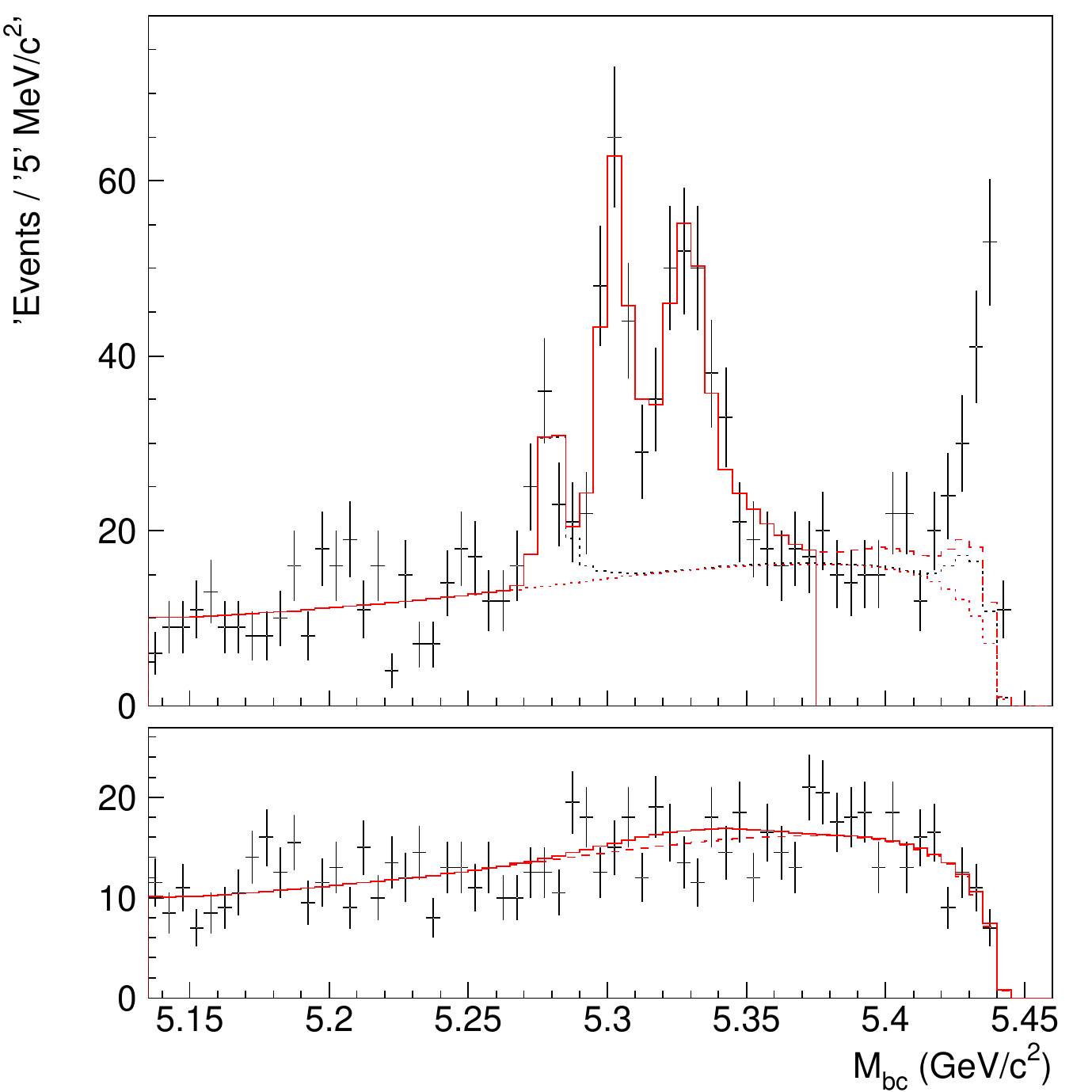}
  \caption{ The $\mbc$ distributions for the points 7 to 12 in
    Table~\ref{tab:xsec_results} (from left to right and from top to
    bottom). Legend is the same as in
    Fig.~\ref{mbc_y5s_it10_140920}. }
  \label{mbc_fit_7}
\end{figure}

\begin{figure}[htbp]
  \centering
  \includegraphics[width=0.46\textwidth]{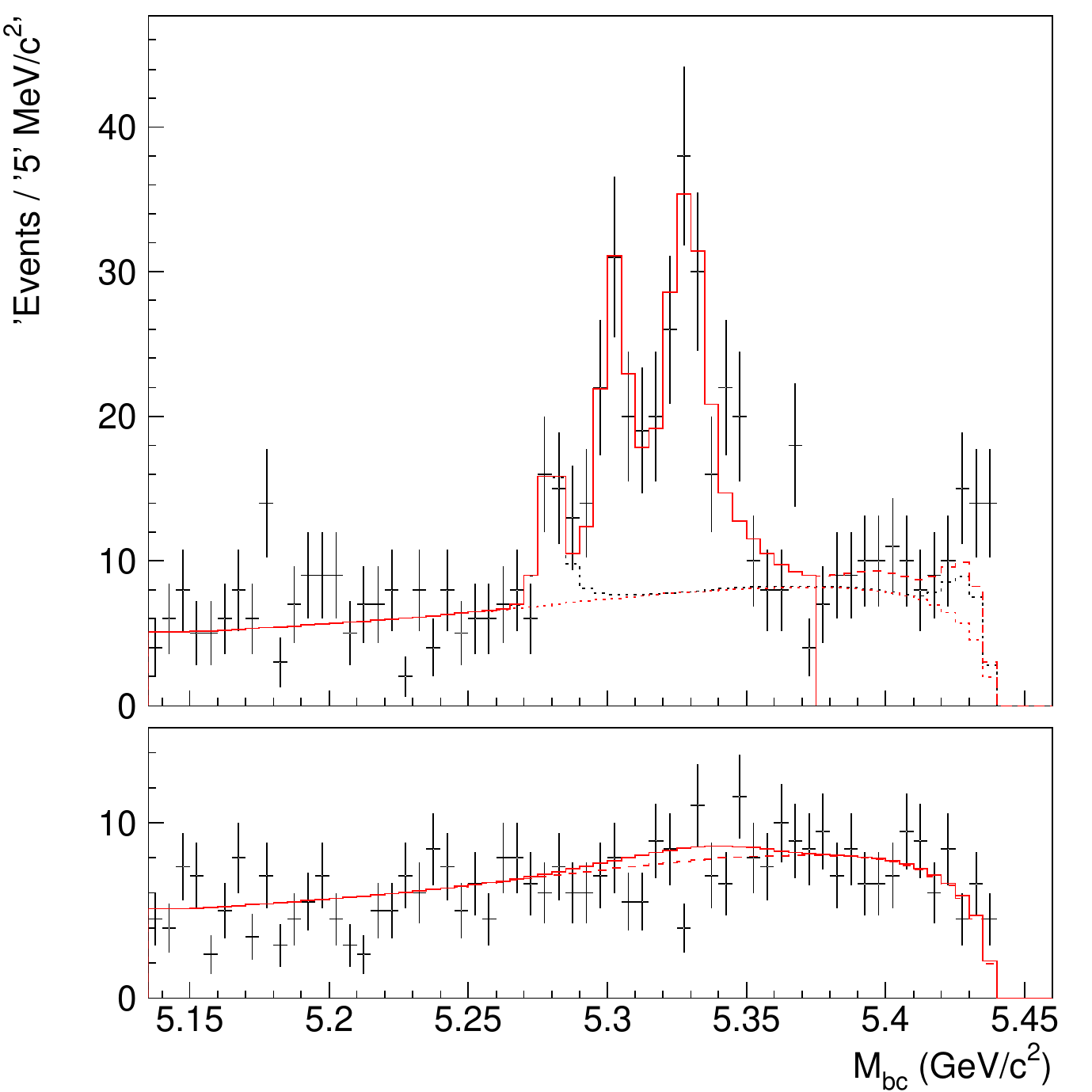}\hfill
  \includegraphics[width=0.46\textwidth]{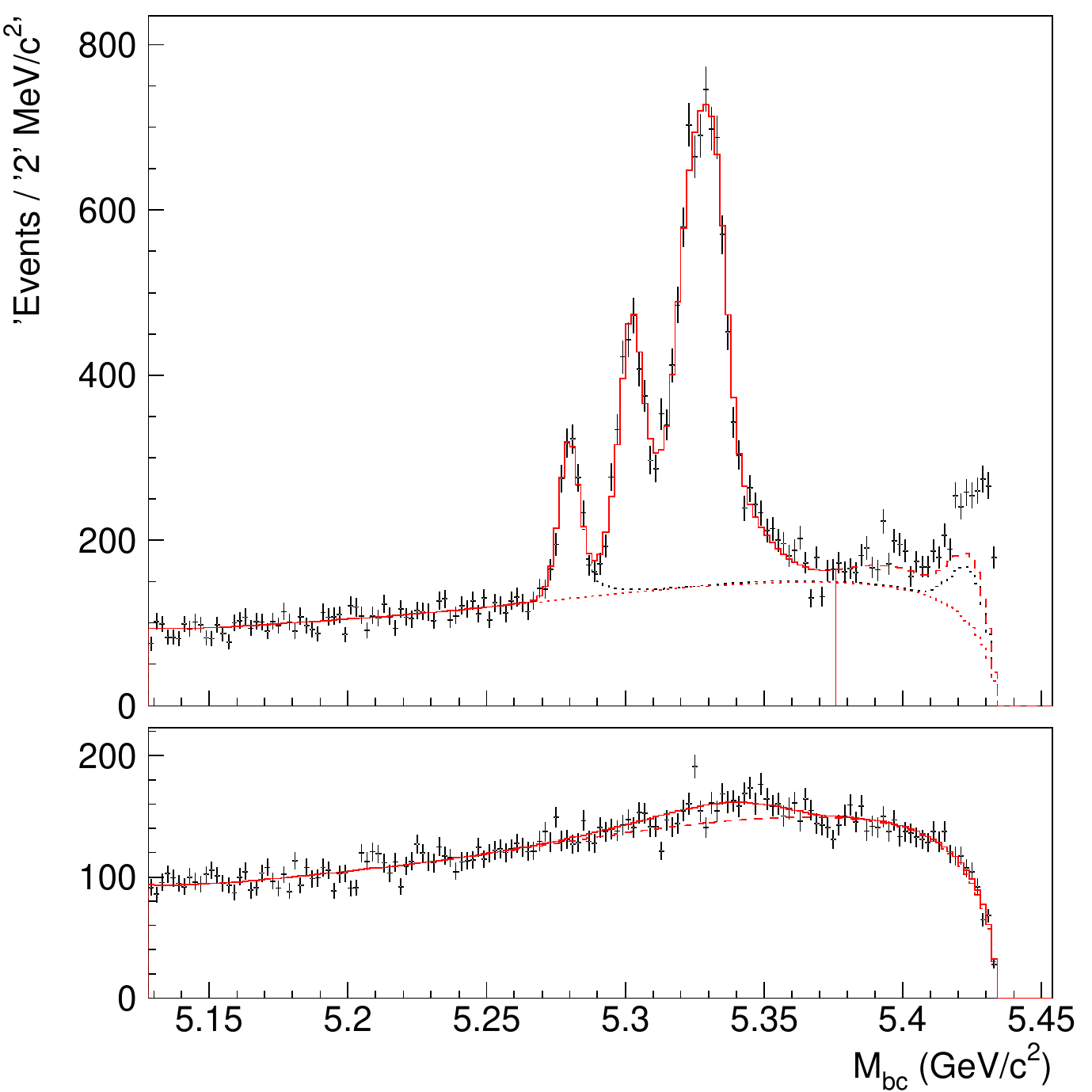}
  \includegraphics[width=0.46\textwidth]{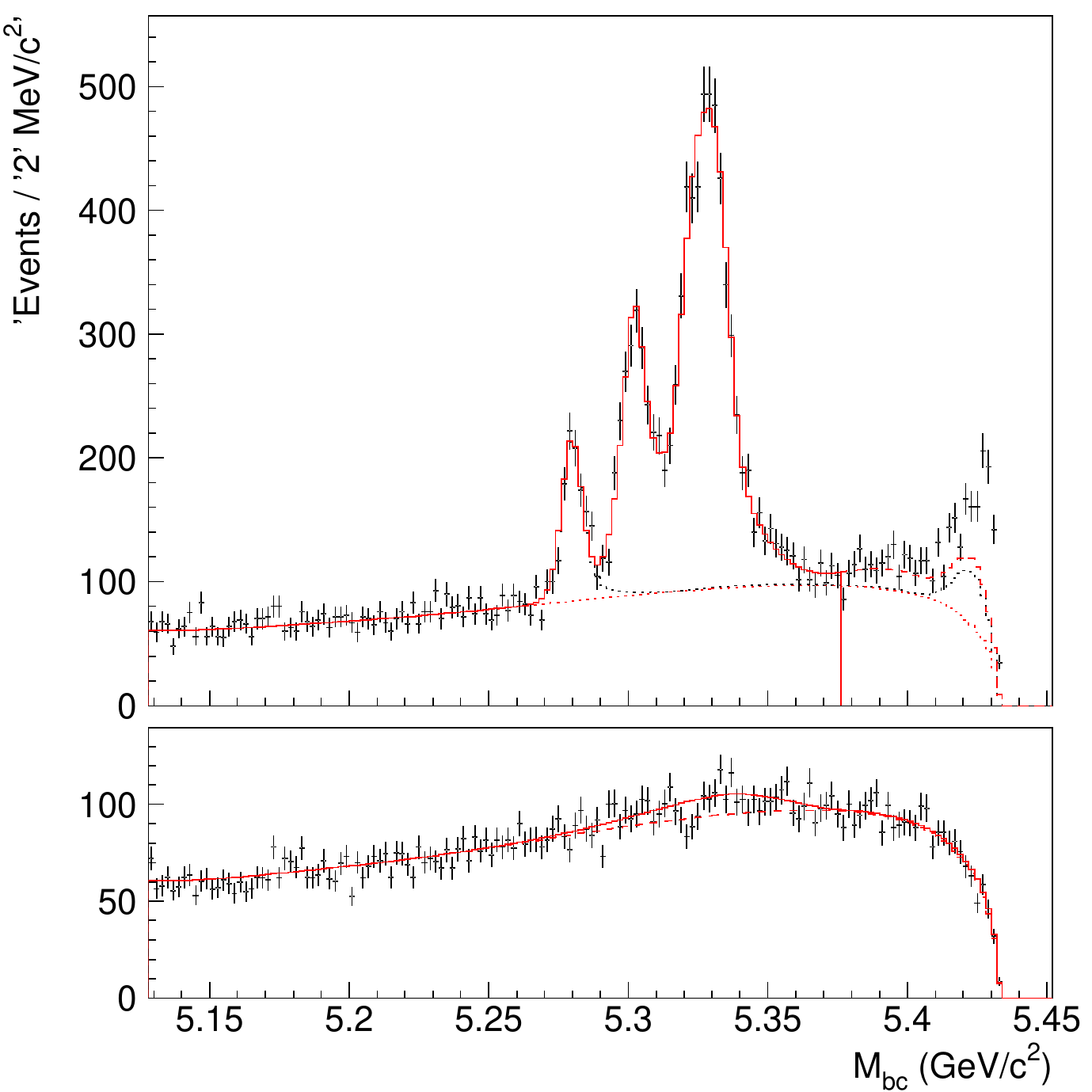}\hfill
  \includegraphics[width=0.46\textwidth]{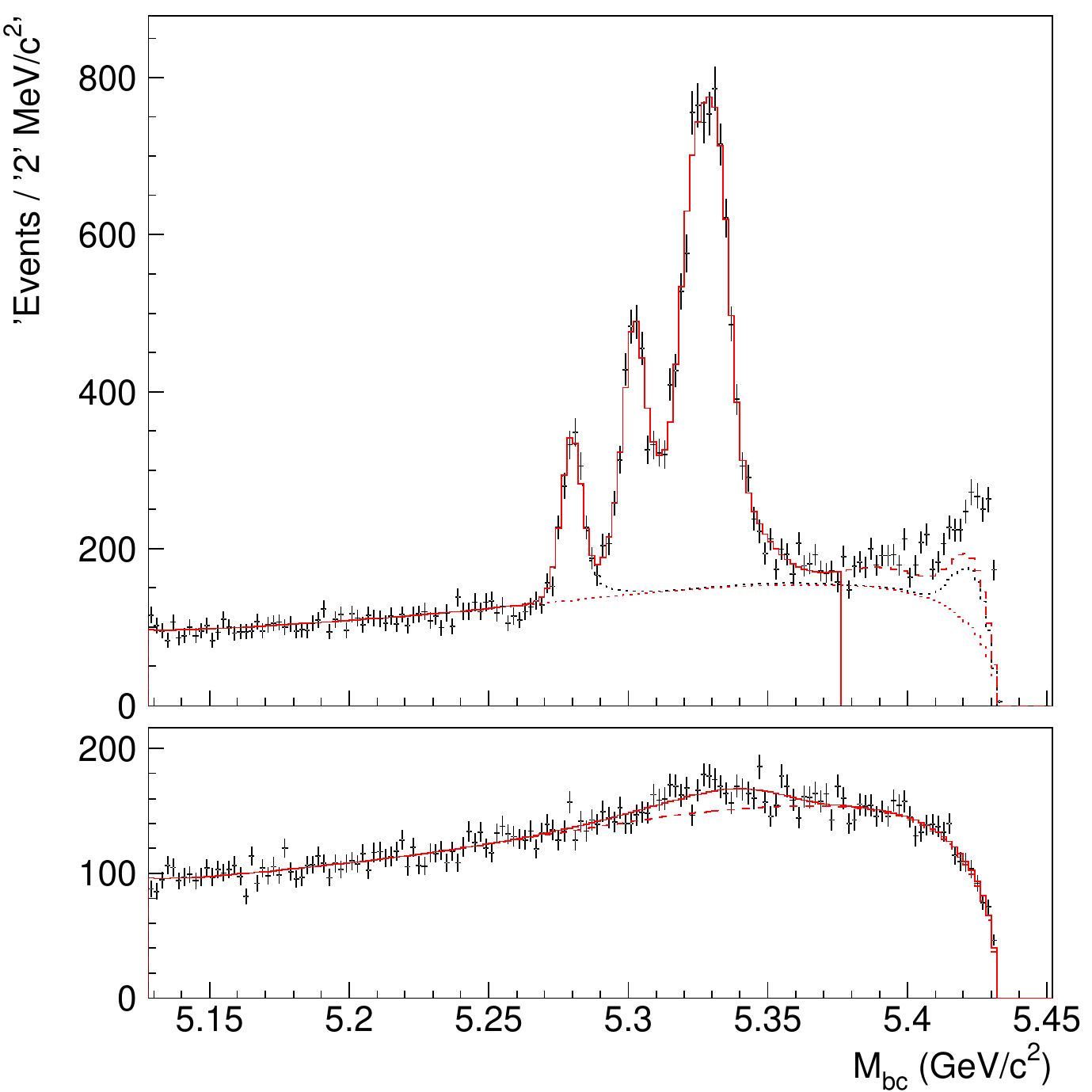}
  \includegraphics[width=0.46\textwidth]{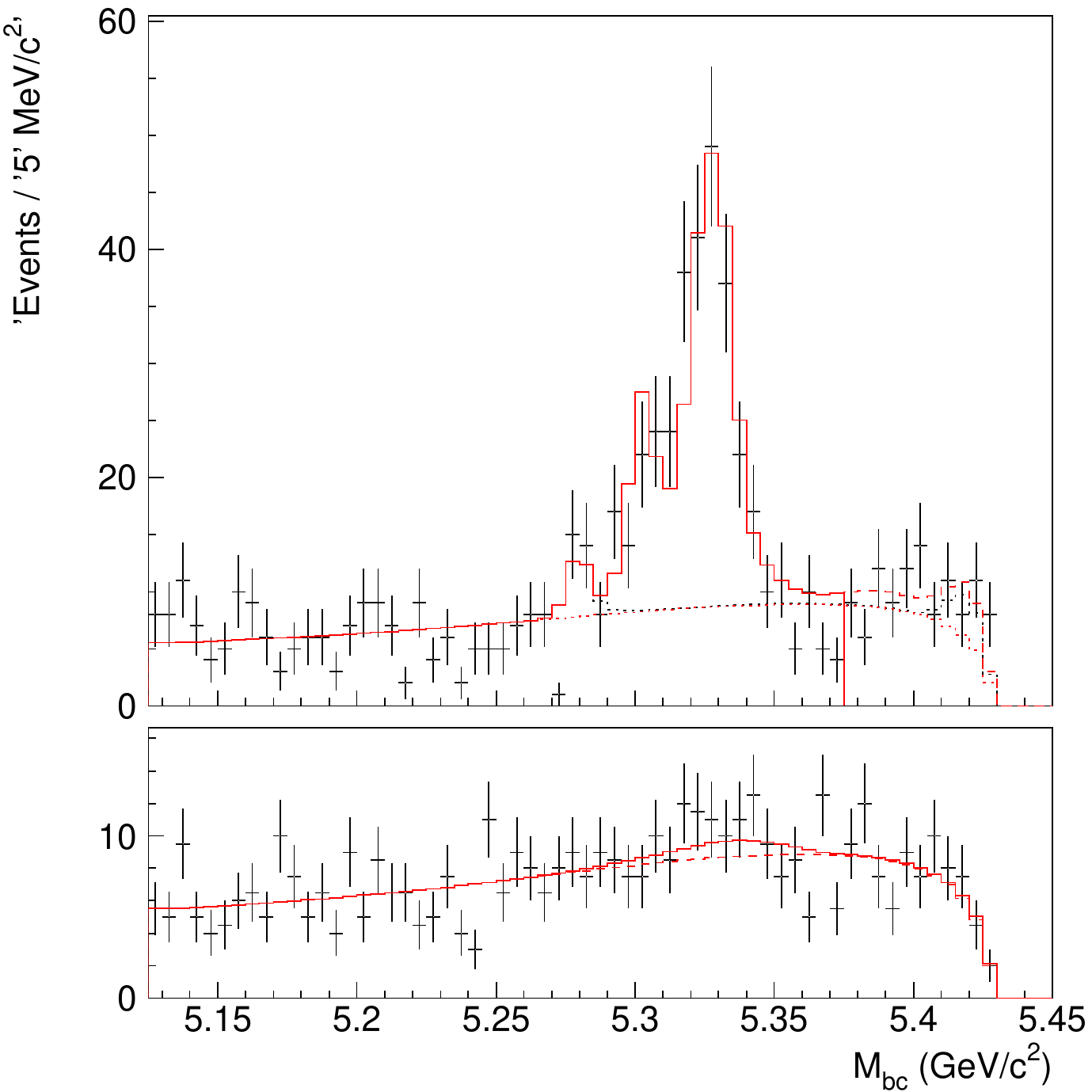}\hfill
  \includegraphics[width=0.46\textwidth]{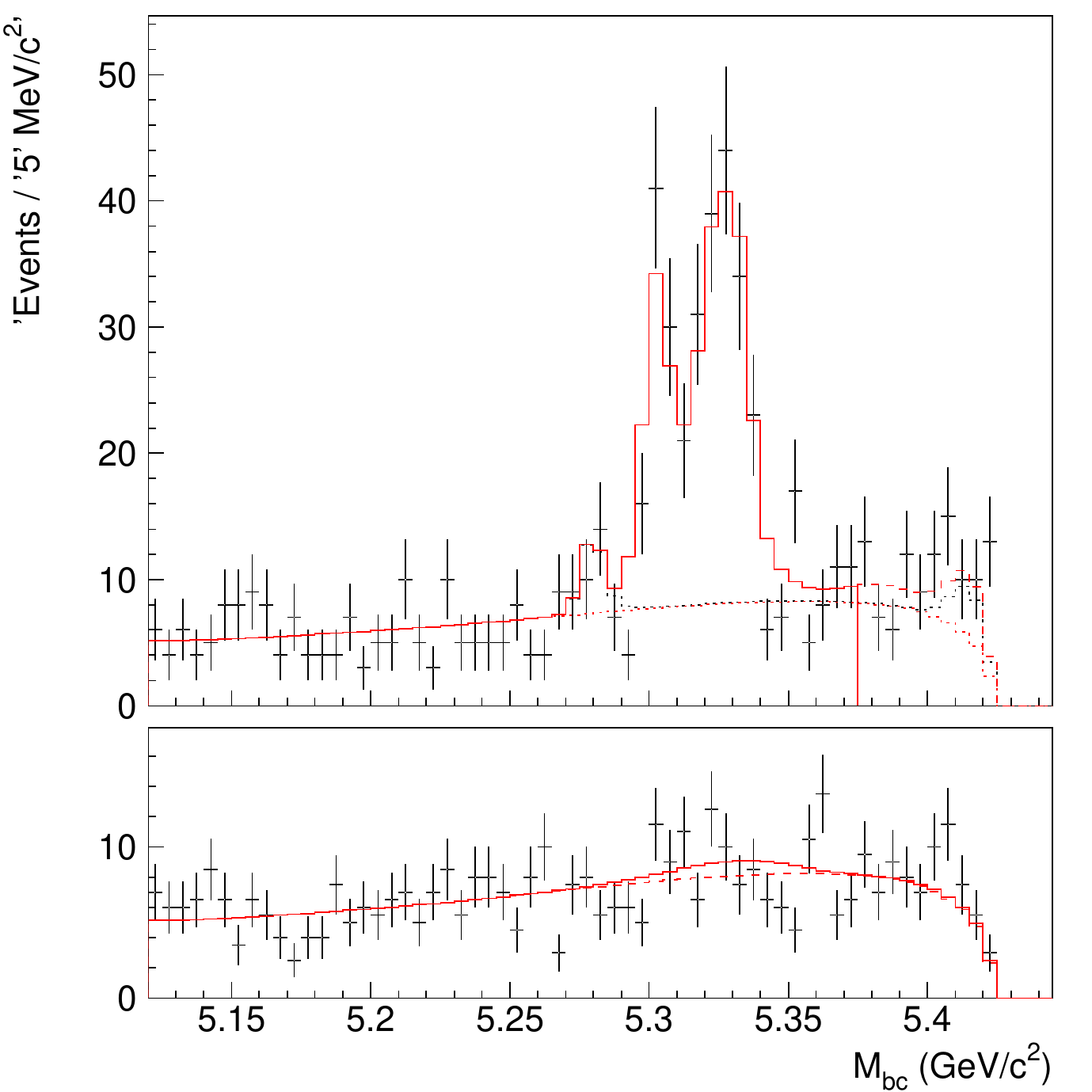}
  \caption{ The $\mbc$ distributions for the points 13 to 18 in
    Table~\ref{tab:xsec_results} (from left to right and from top to
    bottom). Legend is the same as in
    Fig.~\ref{mbc_y5s_it10_140920}. }
  \label{mbc_fit_13}
\end{figure}

\begin{figure}[htbp]
  \centering
  \includegraphics[width=0.46\textwidth]{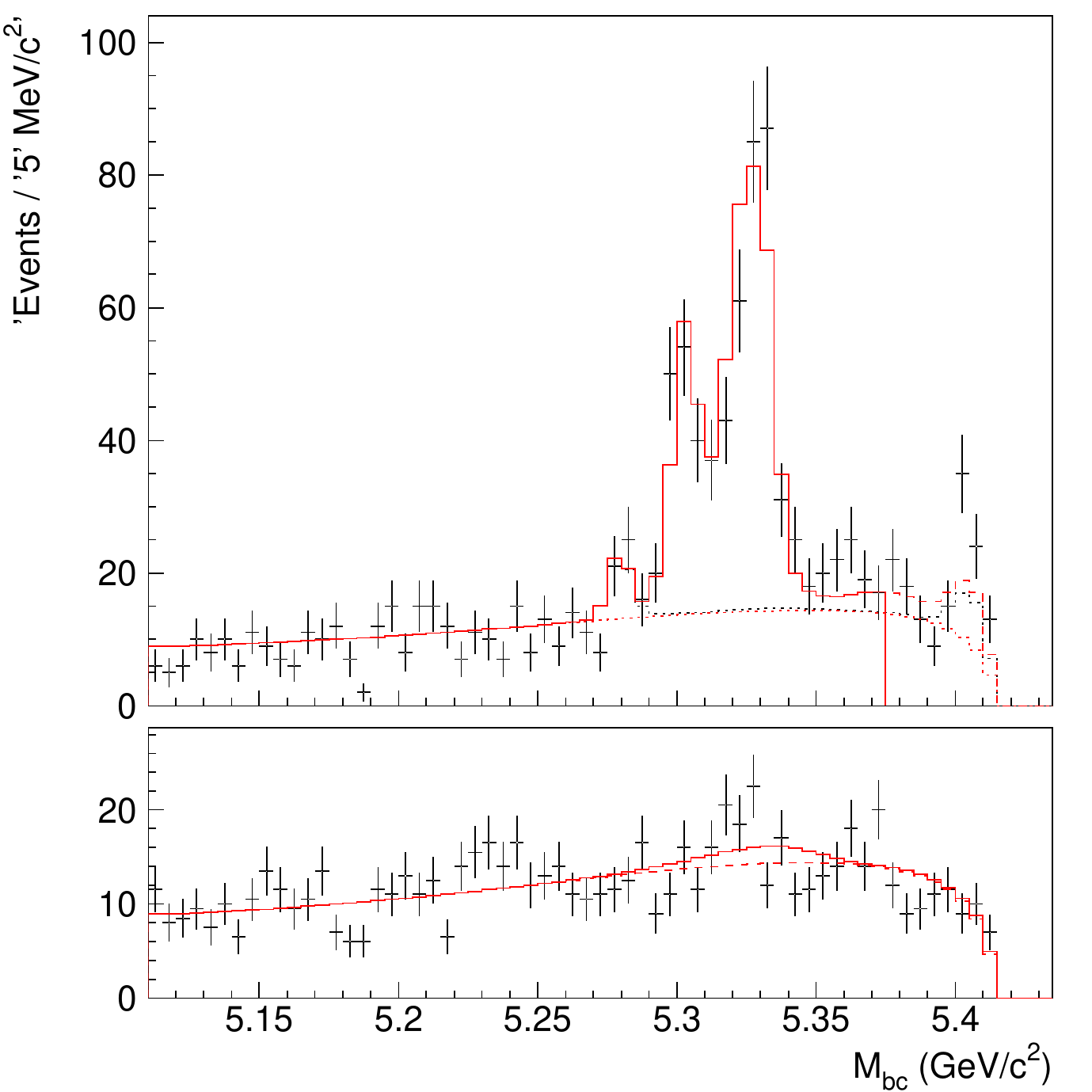}\hfill
  \includegraphics[width=0.46\textwidth]{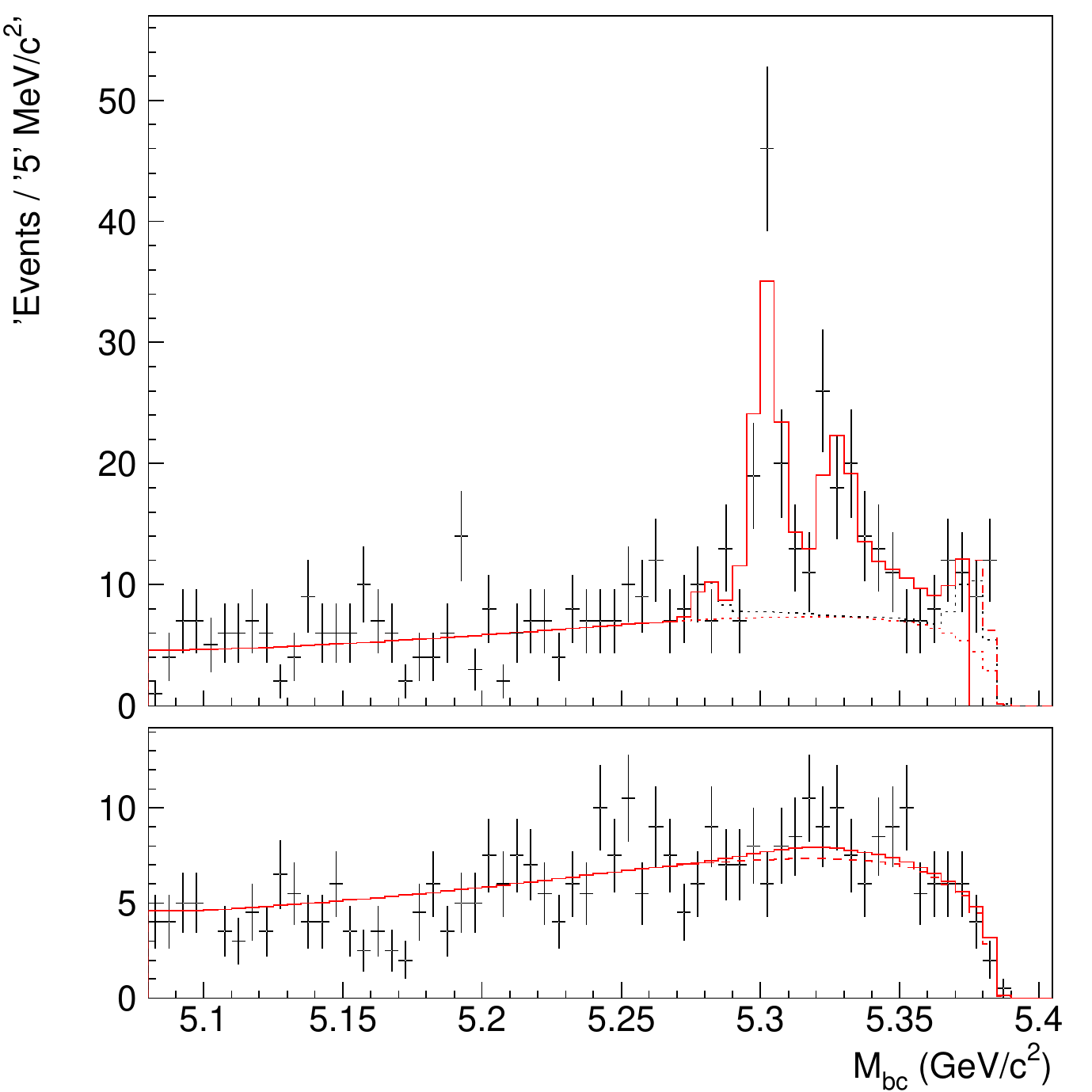}
  \includegraphics[width=0.46\textwidth]{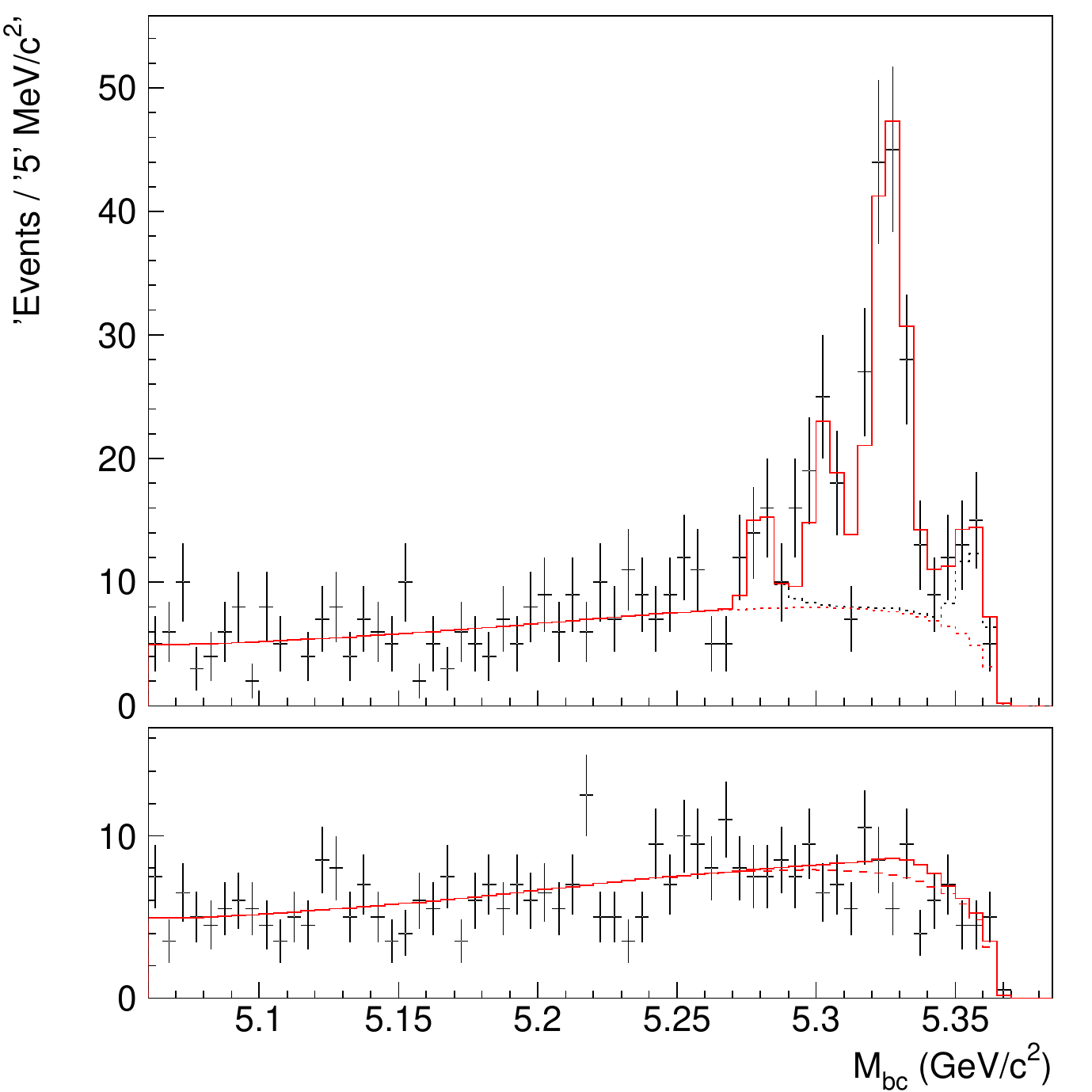}\hfill
  \includegraphics[width=0.46\textwidth]{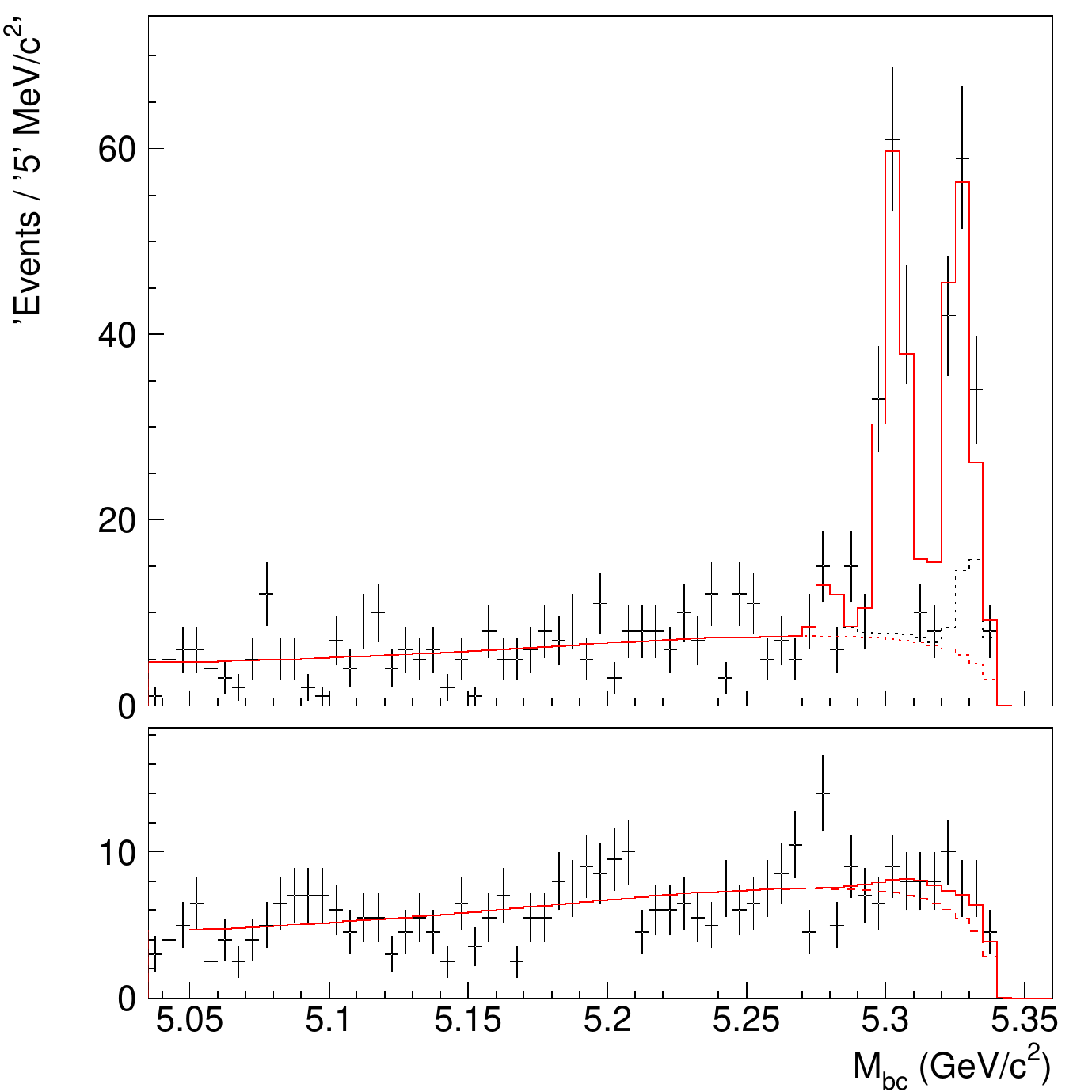}
  \includegraphics[width=0.46\textwidth]{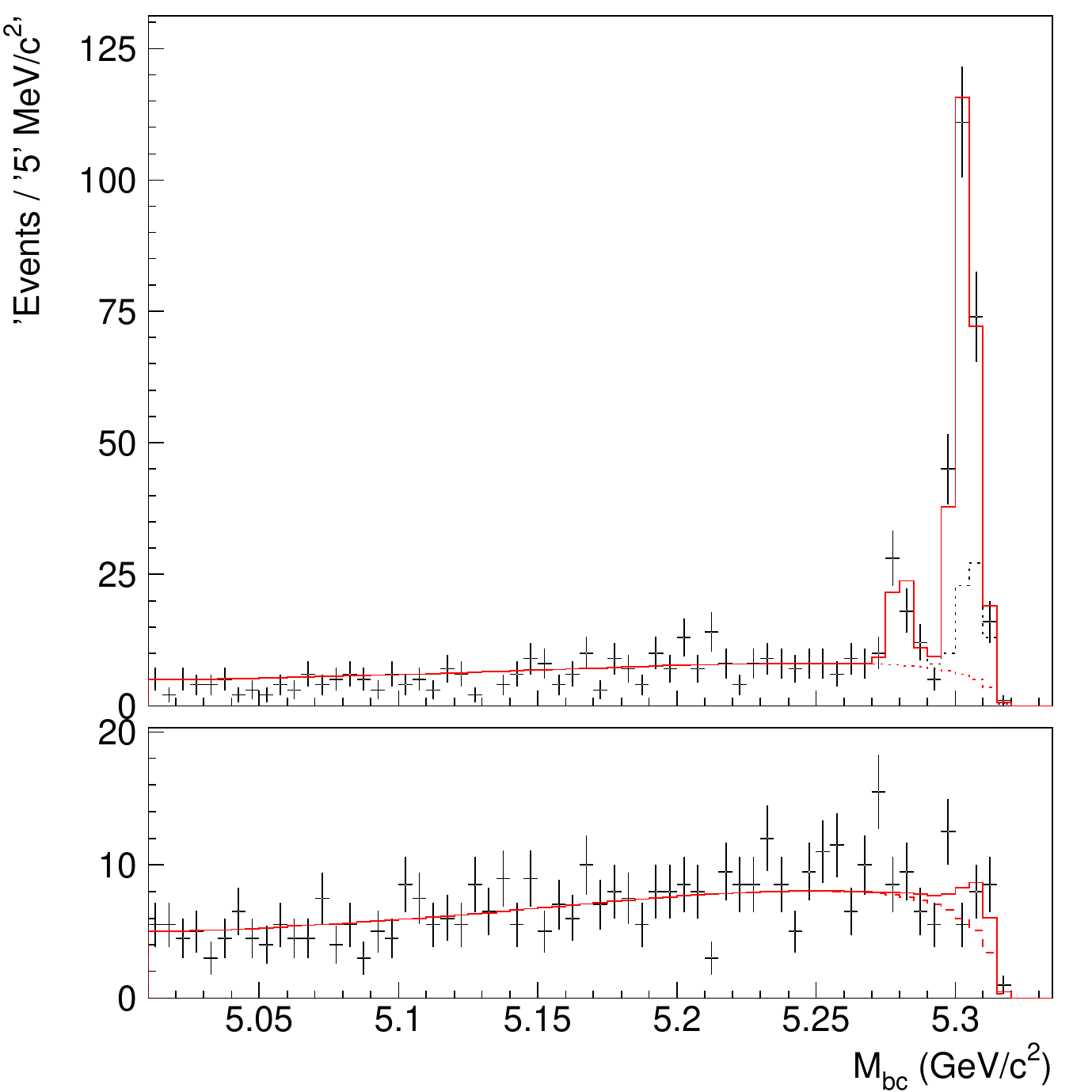}\hfill
  \caption{ The $\mbc$ distributions for the points 19 to 23 in
    Table~\ref{tab:xsec_results} (from left to right and from top to
    bottom). Legend is the same as in
    Fig.~\ref{mbc_y5s_it10_140920}. }
  \label{mbc_fit_19}
\end{figure}

The fits to $\mbc$ distributions at various energies are shown in
Figs.~\ref{mbc_fit_1}--\ref{mbc_fit_19}.

\end{document}